\documentstyle[bezier,epsf,epsfig]{lamuphys}
\newcommand{\ewxy}[2]{\setlength{\epsfxsize}{#2}\epsfbox[10 60 640 570]{#1}}
\makeatletter
\let\chapter\hid@chapter
\makeatother

\begin{document}
\pagenumbering{arabic}

\title{The Heavy Hadron Spectrum}
\author{Christine\,Davies}
\institute{Department of Physics and Astronomy, University of Glasgow, Glasgow,
G12 8QQ, UK}

\maketitle

\begin{abstract}
I discuss the spectrum of hadrons containing heavy quarks ($b$ or $c$),
and how well the experimental results are matched by theoretical ideas.
Useful insights come from 
potential models and applications of Heavy Quark Symmetry and  
these can be compared with new numerical results from
the {\it ab initio} methods of Lattice QCD. 
\end{abstract}

\section{Introduction}

The fact that we cannot study free quarks but only their bound states makes the
prediction of the hadron spectrum a key element in testing Quantum
Chromodynamics as a theory of the strong interactions. This test is by no means
complete many years after QCD was first formulated.

The `everyday' hadrons making up the world around us contain only the light $u$
and $d$ quarks. In these lectures, however, I concentrate on the spectrum of
hadrons containing the heavy quarks $b$ and $c$ (the top quark
is too heavy to have a spectrum of bound states, see for example \cite{top})
 because in many ways
this is better understood than the light hadron spectrum, both experimentally
and theoretically. The heavy hadrons only appear for a tiny fraction of a
second in particle accelerators but they are just as important to our
understanding of fundamental interactions as light hadrons. In fact the
phenomenology of heavy quark systems is becoming very useful; particularly
that of $B$ mesons. The study of $B$ decays and mixing will lead in the next
few years, we hope, to a complete determination of the elements of the
Cabibbo-Kobayashi-Maskawa matrix to test our understanding of CP violation. CKM
elements refer to weak decays from one quark flavour to another but the only
measurable quantity is the decay rate for hadrons containing those quarks.
To extract the CKM element from the experimental decay rate then requires
theoretical predictions for the hadronic matrix element. We cannot expect to
get these right if we have not previously matched the somewhat simpler
theoretical predictions for the spectrum to experiment.

Here I will review the current situation for the spectrum of bound states with
valence heavy quarks alone and bound states with valence heavy quarks and light
quarks. The common thread is, of course, the presence of the heavy quark, but
we will nevertheless find a very rich spectrum with plenty of variety in
theoretical expectations and phenomenology. A lot of the recent theoretical
progress has been made using the {\it ab initio} techniques of Lattice QCD.
These are described elsewhere in this Volume (~\cite{weingarten}) along with
recent results from Lattice QCD for the light hadron spectrum.

Quark model notation for the states in the meson spectrum will prove useful
(baryons will not be discussed until section 3). The valence quark and
anti-quark in the meson have total spin, $S$ = 0 or 1, and relative orbital
angular momentum, $L$. The total angular momentum, which becomes the spin of
the hadron, $\vec{J}$ = $\vec{L}$ + $\vec{S}$. The meson state is then denoted by
$n^{2S+1}L_J$ where $n$ is the radial quantum number. $n$ is 
conventionally given so that the first occurrence of 
that $L$ is labelled by $n$=1 (i.e. $n$+1 is the number 
of radial nodes). $L$ = 0 is given the name $S$, $L$ = 1, the name $P$, etc.  
To give $J^{PC}$ quantum numbers
for the state (the only physical quantum numbers) 
we need the facts that $P = (-1)^{L+1}$ and, for  $C$
eigenstates, $C = (-1)^{L+S}$.
In Table \ref{jpc} a translation between $^{2S+1}L_J$ and $J^{PC}$ is provided.

\begin{table}
\begin{center}
\begin{tabular}{c|c}
$^{2S+1}L_J$ &  $J^{PC}$ \\
\hline \\
$^1S_0$ & $0^{-+}$ \\
$^3S_1$ & $1^{--}$ \\
$^1P_1$ & $1^{+-}$ \\
$^3P_0$ & $0^{++}$ \\
$^3P_1$ & $1^{++}$ \\
$^3P_2$ & $2^{++}$ 
\end{tabular}
\caption{$J^{PC}$ quantum numbers for quark model $S$ and $P$ states}
\label{jpc}
\end{center}
\end{table}

The ordering of levels that we see in the meson spectrum (~\cite{pdg}) is
generally the na\"{\i}ve one i.e. that for a given combination of quark and
anti-quark adding orbital or spin momentum or radial excitation increases the
mass. This is clearer for the heavy hadrons since, because of their masses and
properties, the quark assignments are unambiguous. For heavy hadrons it is also
true, for reasons that I shall discuss, that the splittings between states of
the same $L$ but different $S$ are smaller than the splittings between
different values of $L$ or $n$. To separate this fine structure from radial and
orbital splittings it is convenient to distinguish spin splittings from
spin-independent or spin-averaged splittings. Spin-averaged states are obtained
by summing over masses of a given $L$ and $n$, weighting by the total number of
polarisations i.e $(2J+1)$. Examples are given below - they will be denoted by 
a bar.

In Section 2 I begin with the phenomenology of mesons containing valence heavy
quarks, the heavy-heavy spectrum. I shall discuss potential model approaches to
predicting this spectrum as well as more direct methods recently developed in
Lattice QCD. Section 3 will describe heavy-light mesons and baryons, both from
the viewpoint of Heavy Quark Symmetry ideas and from Lattice QCD, using
techniques successful in the heavy-heavy sector. Section 4 will give
conclusions and the outlook for the future.

\section{The Heavy-heavy Spectrum}

\begin{figure}
\begin{center}
\epsfig{file=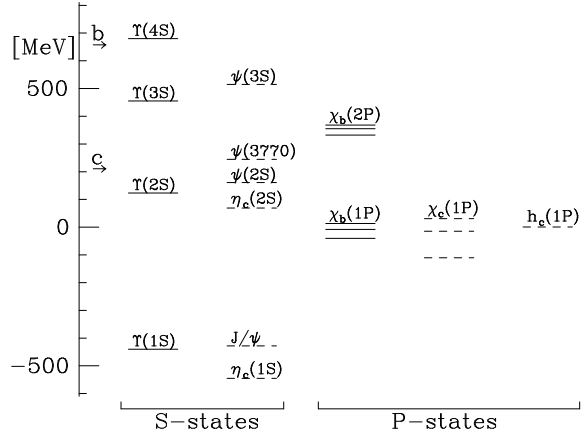, height=7cm, bbllx=10pt, bblly=65pt,
bburx=565pt, bbury=490pt}
\caption[wea]{The experimental heavy-heavy meson spectrum relative to the spin-average
of the $\chi_b(1P)$ and $\chi_c(1P)$ states (\cite{pdg}).}
\label{hhspect}
\end{center}
\end{figure}

Figure \ref{hhspect} shows experimental results for $b\overline{b}$ and
$c\overline{c}$ bound states (\cite{pdg}). They have been fitted on to the same
plot by aligning the spin-average of the
$1 ^3P_{0,1,2}$ states ($\chi_b$ and $\chi_c$). The spin-average $\chi$ mass is
defined by
\begin{equation}
M(\overline{\chi}) = \frac {1} {9} \left[ M(^3P_0) + 3 M(^3P_1) + 5 M(^3P_2)
\right]
\end{equation}
and this has been set to zero in both cases. The overall scale of
$b\overline{b}$ meson masses is much larger than for $c\overline{c}$ but we
see that this
simply reflects the larger mass of the $b$ quarks. The lightest vector state
for $b\overline{b}$ is the $\Upsilon$ produced in $e^{+}e^{-}$ collisions with
a mass of 9.46 GeV. Its radial excitations are known as $\Upsilon^{'}$ or
$\Upsilon(2S)$, $\Upsilon^{''}$ or $\Upsilon(3S)$ and so on. The radial
excitations are separated from the ground state by several hundred MeV. For
$c\overline{c}$ the lightest vector state is the $J/\psi$ or $\psi(1S)$ and
this has a mass of 3.1 GeV. Its radial excitations are the $\psi{'}$ or
$\psi(2S)$ and so on.
Since the scale of Figure 1 spans 1 GeV it is clear that the splittings between
states in both systems are very much smaller than the absolute masses of the
mesons.

It is also clear from Figure 1 that the radial excitations of the vector states
in the two systems match each other very closely. In fact so closely that the
$\psi(3770)$ which has vector quantum numbers cannot be fitted into a scheme of
radial excitations of the $\psi$ system. It is thought to be not an $S$ state
but a $D$ state (\cite{psi-d}). No $b\overline{b}$ $D$ candidates 
have yet been seen).

The matching of radial excitations is even better if we consider spin-averaged
$S$ states,
\begin{equation}
M(\overline{S}) = \frac {1} {4} \left[ M(^1S_0) + 3 M(^3S_1) \right].
\end{equation}
The $^1S_0$ state has only been seen for $c\overline{c}$ and is denoted
$\eta_c$. The $\eta_c(1S)$ lies below the vector state by 117 MeV and so the
spin-average lies one quarter of this below the $J/\psi$. As we shall see, this
spin splitting in the $b\overline{b}$ system
is expected to be much smaller. If we take a reasonable value for $M(\Upsilon)
- M(\eta_b)$ of 40-50 MeV (see later), we would find the $1\overline{S}$ levels on Figure 1 to
be aligned to within 10 MeV despite a difference in overall mass of a factor of
3. Similar arguments apply to the alignment of the $2\overline{S}$ levels,
although the agreement achieved there is not quite as good.

The spin splittings within the $\chi_b(1P)$ states ($\chi_{b0}$, $\chi_{b1}$,
$\chi_{b2}$) are much smaller than those within the
$\chi_c(1P)$ states, so that the spin splittings do depend on the heavy quark
mass, $m_Q$, quite strongly. For example, we can take the ratio for $1P$
levels:
\begin{equation}
\frac {M(\chi_{b2}) - M(\chi_{b0})} {M(\chi_{c2}) - M(\chi_{c0})}  = \frac
{53 \, {\rm MeV}} {141 \, {\rm MeV}} = 0.38(1).
\end{equation}
Na\"{\i}vely this looks very similar to the ratio of $b$ and $c$ quark masses if we
take these to be approximately half the mass of the vector ground states,
$\Upsilon$ and $J/\psi$.
Then $m_c/m_b \approx 0.33$. This might imply a simple $1/m_Q$ dependence for spin
splittings. However, the ratio between $b\overline{b}$ and $c\overline{c}$ does
depend somewhat on the splitting being studied, indicating a more
complicated picture. We have :
\begin{equation}
\frac {M(\chi_{b1}) - M(\chi_{b0})} {M(\chi_{c1}) - M(\chi_{c0})}  = 0.34(2),
\end{equation}
and
\begin{equation}
\frac {M(\chi_{b2}) - M(\chi_{b1})} {M(\chi_{c2}) - M(\chi_{c1})}  = 0.47(2).
\end{equation}

The arrows shown on Figure \ref{hhspect} mark the minimum threshold for decay
into heavy-light mesons, $B\overline{B}$ for $b\overline{b}$ and
$D\overline{D}$ for $c\overline{c}$. Three sets of $S$ states and two 
sets of $P$ states have been seen below
this threshold for $b\overline{b}$ and two sets of $S$ states and one set of $P$
states for $c\overline{c}$.
Another set of $P$ and two sets of $D$ states are expected for 
$b\overline{b}$ (\cite{eich-ups}, \cite{kwong-ups}). 
The states below threshold are very narrow since the
Zweig-allowed decay to heavy-light states (see Fig. 2)
is kinematically forbidden and they must decay by annihilation.
This carries a penalty of powers of the strong coupling 
constant, $\alpha_s(M_Q)$. 
These are then the states that we will concentrate on, because they
can be treated as if they are stable (and none of the 
approaches which we will discuss allow them to decay).
Vector states above threshold
can still be seen in the $e^{+}e^{-}$ cross-section as bumps
but they are much broader and the theoretical understanding
of their masses requires a model for the inclusion of decay channels in the
analysis (\cite{old-cornell}, \cite{ono}).

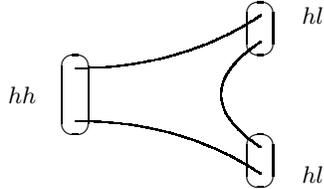
\begin{figure}
\begin{center}
\begin{picture}(130,110)
\qbezier(10,40)(50,40)(80,20)
\qbezier(10,60)(50,60)(80,80)
\qbezier(80,30)(50,50)(80,70)
\put(-10,50){\makebox(0,0){$hh$}}
\put(100,20){\makebox(0,0){$hl$}}
\put(100,80){\makebox(0,0){$hl$}}
\put(10,50){\oval(10,30)}
\put(80,75){\oval(10,20)}
\put(80,25){\oval(10,20)}
\end{picture}
\end{center}
\caption{Decay of a heavy-heavy meson to heavy-light mesons above
threshold.}
\end{figure}

How can we understand the heavy-heavy spectrum below threshold? The fact that
all the
splittings are very much less than the masses, noted above, is critical. It
implies that dynamical scales, such as the kinetic energy of the heavy quarks,
are also very much less than the masses i.e. the quark velocities are
non-relativistic,
$v^2 \ll c^2$. Typical gluon momenta will be of the same order as typical
quark momenta, $m_Qv$. Thus typical gluon energies, $m_Qv$, are very much
larger than typical quark kinetic energies, $m_Qv^2$ (\cite{tl-91}).
The gluon interaction between heavy quarks will then appear `instantaneous'. It
can be modelled using a potential and energies found by solving Schr\"odinger's
equation. In the extreme non-relativistic limit of very heavy quarks the spin
splittings vanish. This was noticed above in the relation between
$b\overline{b}$ and $c\overline{c}$ splittings and will be discussed in more
detail later. In this limit we need only a single spin-independent central potential to solve
for the spin-averaged spectrum of $\overline{S}$ and $\overline{P}$ states
defined above. For a recent review of the history of the heavy quark potential
see \cite{quigg-97}. 

\subsection{The spin-independent heavy quark potential} \label{sipot}

Perturbation theory for QCD gives a flavour-independent central potential based
on 1-gluon exchange, which has a Coulomb-like form,
\begin{equation}
V(r) = - \frac {4} {3} \frac {\alpha_s} {r}
\end{equation}
where $r$ is the radial separation between the two heavy quarks and $\alpha_s$
is the
strong coupling constant. The Coulomb potential cannot be the final answer
because it would allow free quarks to escape. In addition it gives a spectrum
incompatible with experiment in which the $1P$ level is degenerate with $2S$.

For a potential of the form $V \sim r^N$ with $2 > N > 0$ we have a $1P$ level
below $2S$, as we observe, and a $1D$ level above $2S$ (as we see for
charmonium). So, the addition of some positive power of $r$ to the Coulomb
potential can rescue the phenomenology (\cite{martin-rep}). The additional term is usually taken to
be linear in $r$ and thought of
as a `string-like' confining potential.
This gives the simple Cornell potential of
\cite{cornell-pot}:
\begin{equation}
V(r) = - \frac {4} {3} \frac {\alpha_s} {r} + \sigma r
\end{equation}
with $\sigma$ called the `string tension'. This can reproduce the observed
spectrum
reasonably well. Other  successful forms for the heavy quark potential are the
Richardson potential (\cite{richardson-pot}):
\begin{equation}
V(r) = \int d^3q e^{iq\cdot r} \frac {\alpha_s(q^2)} {4\pi q^2},
\end{equation}
in which a running strong coupling constant is included
with non-perturbative behaviour at small $q^2$ (see also \cite{tye-pot}),
and the Martin potential
(\cite{martin-pot}, \cite{power-pot}):
\begin{equation}
V(r) = A r^{\nu}\; {\rm with}\; \nu \approx 0.
\label{martin}
\end{equation}
This last form, essentially a logarithmic potential (\cite{log-pot}),
 has no QCD motivation but is simply observed to work. All three potential
forms can reproduce the $b\overline{b}$ and $c\overline{c}$ spin-averaged
spectra reasonably well if the parameters are chosen appropriately. When this
is done it
is observed that the potentials themselves agree in the region $r \sim 0.1 -
0.8$ fm in which the $\sqrt{\langle r^2 \rangle}$ for the bound
 states sit (\cite{tye-pot}).

It is interesting to compare the dependence of the energies of the states on
the mass of the heavy quark, $m_Q$, in different potentials. 
This can be done for homogeneous
polynomial-type potentials easily
(see, for example \cite{quigg-rep}, 
\cite{close}). 
Schr\"odinger's equation for the wavefunction $\Psi$ is:
\begin{equation}
\left\{ - \frac {\hbar^2 \nabla^2} {2\mu} + V(r) \right\} \Psi(r) = E \Psi(r).
\end{equation}
$E$ is the energy eigenvalue and $\mu$, the reduced mass, $m_Q/2$ for 
the heavy-heavy case. For $V = Ar^N$ we can reproduce
the
same solution at different values of $m_Q$ if we allow for a rescaling
$r \rightarrow \lambda r$. With this rescaling in place
\begin{equation}
\left\{ - \hbar^2 \nabla^2 + A 2 \mu \lambda^{N+2} r^N \right\} \Psi(\lambda r)
=
2 \mu \lambda^2 E \Psi(\lambda r).
\end{equation}
The same solution (with rescaled $r$) will occur for different values of $m_Q$
if
\begin{equation}
\lambda \propto {\mu}^{-1/(2+N)}.
\end{equation}
This gives a solution for $E$ which varies as
\begin{equation}
E \propto m_Q^{-N/(2+N)}
\end{equation}
The values of $E$ (and therefore splittings) will then be 
independent of $m_Q$, as
observed approximately, for $N$ = 0. This corresponds to the Martin potential.
The same result can be achieved by
mixing $N$ = 1 and $N$ = -1 in the Cornell potential.
Note that the Feynman-Hellmann Theorem guarantees that bound 
states fall deeper into the potential as the mass increases, 
$\partial E/ \partial \mu < 0$ (\cite{quigg-rep}). For $N > 0$, $E$ falls
with $m_Q$; for $N < 0$, $E$ increases in the negative direction. 

The Virial Theorem is helpful in extracting some dynamical parameters. It relates the mean kinetic energy to the expectation value of 
a derivative of the potential
(see for example \cite{quigg-rep}): 
\begin{equation}
\langle K \rangle = \frac {1} {2} \langle r \frac {dV} {dr} \rangle
\end{equation}
for homogeneous potentials. For $N \sim 0$ i.e. $V \sim \log r$ we get $K$ = a
constant. Since
\begin{equation}
K = \frac {p^2} {2 \mu}
\end{equation}
this tells us that
\begin{equation}
\langle p^2 \rangle \propto \mu
\end{equation}
and
\begin{equation}
v^2_{m_1} = \frac {2 \langle K \rangle m_2} { (m_1 + m_2) m_1}
\end{equation}
for a meson made of different quarks of masses $m_1$ and $m_2$. $v_{m_i}$
is
the velocity of the quark of mass $m_i$ in the bound state. From fits using
potential
models a value of $\langle K \rangle$ of 0.37 GeV is found (\cite{quigg-rep}), giving
\begin{eqnarray*}
c\; {\rm in}\; \psi, \; \frac {v^2} {c^2} \sim 0.24 \\
b\; {\rm in} \;\Upsilon, \; \frac {v^2} {c^2} \sim 0.07
\end{eqnarray*}
The quarks are non-relativistic as we originally expected. For the
logarithmic potential
we also have the result that $v^2$ is independent of the radial excitation. For a
Coulomb
potential $\langle p^2 \rangle$ decreases with increasing $n$,
whereas for a linearly rising potential, $\langle p^2 \rangle$ increases with
increasing $n$.

\begin{figure}
\centerline{\ewxy{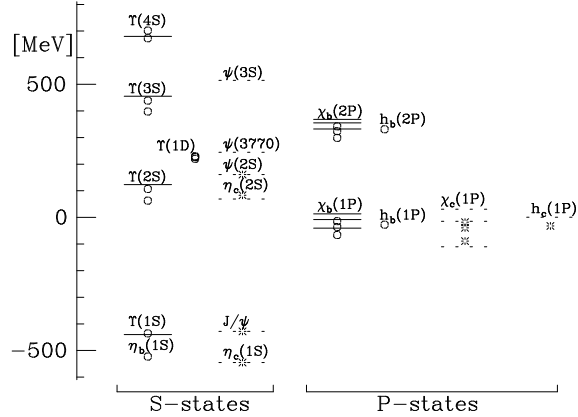}{100mm}}
\caption[hjk]{The heavy-heavy meson spectrum from a recent Richardson potential model
calculation (\cite{equigg}). Circles and bursts show the calculated 
masses relative to the spin average of the $\chi_b(1P)$ and $\chi_c(1P)$ 
states and the solid and dashed lines show experiment results, where
they exist.}
\label{eqspect}
\end{figure}

Potential model calculations of the bottomonium and charmonium
spectra
are reasonably successful. See \cite{equigg} for a 
recent example, whose results are plotted in Figure \ref{eqspect}.
These include not only the central (Richardson) potential discussed here 
but also (perturbative) spin-dependent potentials to be described in 
section \ref{sdpot} to get spin splittings.  
In \cite{equigg} parameters of the potential were fixed from a subset of 
states in the 
experimental $c\overline{c}$ spectrum. Typical deviations from 
experiment for the rest of the $c\overline{c}$ spectrum were 
30 MeV; typical deviations in the $b\overline{b}$ spectrum 
were 25 MeV. Since the $b$ quark is significantly more 
non-relativistic in its bound states than the $c$ quark one 
might expect to get better agreement for the $b\overline{b}$ 
spectrum using fitted parameters from that system. However, agreement
for the $c\overline{c}$ spectrum would then be worse. 
In either case it is necessary to fit the parameters of 
the phenomenological potential from some experimental 
information. 
Instead, the central potential $V(r)$ can be extracted from
first principles
using the techniques of lattice QCD.

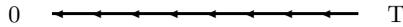
\begin{figure}
\begin{center}
\setlength{\unitlength}{.02in}
\begin{picture}(100,30)
\multiput (10,20)(10,0){8}{\vector(-1,0){10}}
\put(-10,20){\makebox(0,0){0}}
\put(90,20){\makebox(0,0){T}}
\end{picture}
\end{center}
\caption{The world line of a heavy quark on the lattice.}
\label{worldline}
\end{figure}

In the $m_Q \rightarrow \infty$ limit the heavy quark is static. Its world line
in space-time becomes a line of QCD gauge fields in the time direction. In
Lattice QCD we break up space-time into a lattice of points and represent the
gauge field by SU(3) matrices, $U$ (\cite{weingarten}, \cite{montvay}). 
The static quark propagator then becomes a
string of $U$ matrices, as in Figure \ref{worldline}.

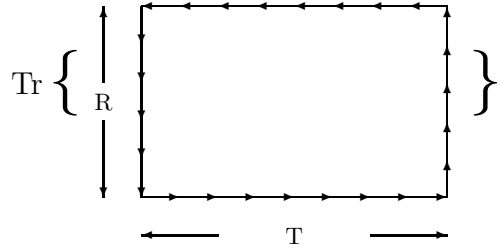
\begin{figure}
\begin{center}
\setlength{\unitlength}{.02in}
\begin{picture}(120,100)(-10,0)
\multiput (20,70)(10,0){8}{\vector(-1,0){10}}
\multiput (10,20)(10,0){8}{\vector(1,0){10}}
\multiput (90,20)(0,10){5}{\vector(0,1){10}}
\multiput(10,30)(0,10){5}{\vector(0, -1){10}}
\put(50,10){\makebox(0,0){T}}
\put(0,45){\makebox(0,0){R}}
\put(30,10){\vector(-1,0){20}}
\put(70,10){\vector(1,0){20}}
\put(0,50){\vector(0,1){20}}
\put(0,40){\vector(0,-1){20}}
\put(-10,50){\makebox(0,0){{\Huge\{}}}
\put(100,50){\makebox(0,0){{\Huge\}}}}
\put(-20,50){\makebox(0,0){{\Large Tr}}}
\end{picture}
\end{center}
\caption{A Wilson loop. The Trace is over colour indices.}
\label{wloop}
\end{figure}

We can put a quark and antiquark together and join them up into a closed, and
therefore
gauge-invariant loop, called a Wilson loop (Figure \ref{wloop}).
 The value of the Wilson loop can be measured on sets of gauge fields $\{U\}$
where a gauge field is defined on
every link of the lattice. These are called configurations. The physically
useful quantity is the matrix element of the Wilson loop between vacuum states
and this is obtained by averaging values of the
Wilson loop over an ensemble of configurations where each configuration has
been chosen as a typical snapshot of the vacuum of QCD. To obtain such an
ensemble we must generate
configurations with a probability weighting $e^{-S_{QCD}}$ and there are
standard techniques to do this (\cite{weingarten}, \cite{montvay}).

The expectation value over such an ensemble of gauge fields of a Wilson loop of
spatial size $R$ is related to the heavy quark potential $V(R)$. This is
because the ensemble average is a Monte Carlo estimate of the path integral
giving the
matrix element of 
the operator which creates and destroys a static heavy quark pair
at separation $R$ on the lattice. The matrix element becomes
exponentially related to the ground state energy of the quark
anti-quark pair as the time extent of the Wilson loop, $T$,
tends to infinity. Since $R$ is fixed, and the quarks
in this picture have no kinetic energy, this is simply the
potential $V(R)$ plus an additive self-energy contribution.
\begin{eqnarray}
\langle \rm{Wilson \, Loop} \rangle &=& \frac {\int {\cal{D}} U {\rm Wilson \,
Loop}(U)
e^{-S_{QCD}} } { \int {\cal{D}} U e^{-S_{QCD}} } \label{wloopv}\\
&=& \langle 0 |
[\psi^{\dagger}(0)\chi^{\dagger}(R)]_{t=0}[\psi(0)\chi(R)]_{t=T} | 0 \rangle 
\nonumber \\
& \stackrel { {\scriptsize T \rightarrow \infty}} {\rightarrow} & | \langle 0 |
\psi(0) \chi(R) | {\rm ground \, state} \rangle |^2 e^{-ET} + {\rm higher \, order \, terms}
\nonumber \\
E \; = \; V_{latt}(R) &=& V(R) + {\rm constant} \nonumber.
\end{eqnarray}

How is the calculation done? Once the ensemble of gauge field
configurations has been generated, Wilson loops of various
different sizes in $R$ and $T$ are measured and
average values of $W(R,T)$ obtained. There is a statistical
error associated with the number of configurations in the
ensemble, i.e. how good an estimate of the path integral has been obtained. For
a fixed $R$, $W(R,T)$ is fitted to the
exponential form above in the large $T$ limit, extracting $E$.
 Away from $T = \infty$ higher order terms should be included
in the fit which are exponentials of excitations of the 
potential. 
There are a number of techniques to improve the
values of $E$ obtained, both the statistical error and any
systematic error from fitting to an exponential form (see, for example,
 \cite{bali}). Several of the techniques are similar to those 
used in direct calculations of the spectrum and are discussed
in section 2.3. 

Once $V_{latt}(R)$ is obtained it can either be used directly or a functional form in
terms of $R$ can be extracted to inform the continuum potential model
approaches
described above. The functional form usually used is
that of the Cornell potential with $e=4\alpha_s/3$ and 
an additive constant, $V_c$:
\begin{equation}
V_{latt}(R) = \sigma R - \frac {e} {R} + V_c. \label{vlatt}
\end{equation}
The fit then yields the parameters $\sigma$, $e$ and $V_c$. 
$e$ is generally taken as a constant, although it is possible 
to determine the running coupling constant $\alpha_s(R)$
from the short distance potential (\cite{UKQCD-alpha}).
Often the running is mimicked by keeping $e$ constant and 
adding an additional term, $f/R^2$. This affects slightly the 
value of $e$ obtained, as does the range of $R$ included in the 
fit. A Martin form plus a constant, equation \ref{martin}, does not fit 
$V_{latt}$ (Bali, private communication).   

It is important to remember that $V_{latt}$, being obtained from the lattice,
is
measured in lattice units. To convert to dimensionful units of GeV we
need to know the lattice spacing, $a$. This requires
one piece of experimental information (see below). The separation, $R$, is
also measured in lattice units, corresponding to a physical distance $r=Ra$ in
fm.
Thus the continuum potential $V$ is obtained by
\begin{equation}
V(r=Ra) = V_{latt}(R) \times a^{-1} \label{vphys}
\end{equation}
However this expression should contain on the r.h.s. only the physical pieces
of $V_{latt}$ and
not $V_c$.

$V_c$ is an unphysical constant which resets the zero of energy.
It arises from corrections to the static quark self-energy induced
by gluon loops sitting around the perimeter of the Wilson loop. These
give a contribution to the logarithm of the Wilson loop proportional
to its perimeter, $V_c(2R+2T)/2$. Thus the term $V_c$ appears as
part of $V_{latt}$. In perturbation theory $V_c$ is a power series
in the coupling constant $\alpha_s$, starting at $\cal{O}$$(\alpha_s)$,
but is otherwise a constant in lattice units. From equation \ref{vphys},
its contribution to the continuum potential diverges on the
approach to the continuum limit, $a \rightarrow 0$, and it should
be subtracted from $V_{latt}$ before equation \ref{vphys} is
applied.  Another way to look at this is to notice that the
heavy quark potential on the lattice is forced to zero
at zero separation, $V_{latt}(0)=0$, when the Wilson loop
collapses to two lines on top of one another. Because the $U$ matrices
are unitary we get $\langle {\rm Wilson Loop} \rangle = 1 = e^{0}$.
However, the continuum potential
diverges in Coulomb fashion at zero separation so that $V(0)a \ne 0$.
The physical pieces of the lattice potential will reproduce the continuum behaviour
so to get $V_{latt}(0)=0$ will require an additive constant
to shift the whole potential upwards. This is $V_c$.

\begin{figure}
\centerline{\ewxy{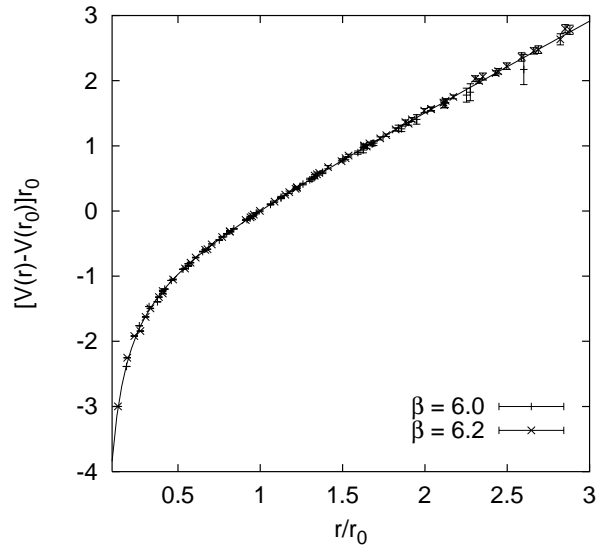}{100mm}}
\caption[eqr]{The heavy quark potential obtained from the lattice 
at two different values of the lattice spacing in the quenched 
approximation. The solid
line is a fit of the form \ref{vlatt} (\cite{bali}). The potential
and separation are given in units of the parameter $r_0$ (see 
text).} 
\label{pot-fit}
\end{figure}

Figure \ref{pot-fit} shows recent results for the lattice potential 
plotted with the fitted
form above, (\ref{vlatt}).
The parameters extracted can be compared to those of phenomenological
potentials.

The Coulomb coefficient, $e$, is dimensionless
and needs no multiplication by powers of the inverse lattice
spacing to get a physical result. $e$ is the coefficient of
the $1/R$ term but the discrete nature of the lattice changes
\begin{equation}
\frac {1} {R} \rightarrow \frac {4 \pi} {L^3} \sum_{q \not= 0} \frac
{e^{iq\cdot R}} {\sum_i \hat{q}_i^2}, \; \hat{q}_i^2 = 2 \sin \frac {q_i} {2}
\end{equation}
Notice that this lattice form of $1/R$ is not rotationally invariant.
At finite lattice spacing there are two alternatives. One is to fit this
modified `lattice' form
of $1/R$; the other is to correct for the discretisation errors in the na\"{\i}ve
lattice action,
$S_{QCD}$ which gave rise to them (\cite{gpl-schladming}).

\begin{figure}
\centerline{\ewxy{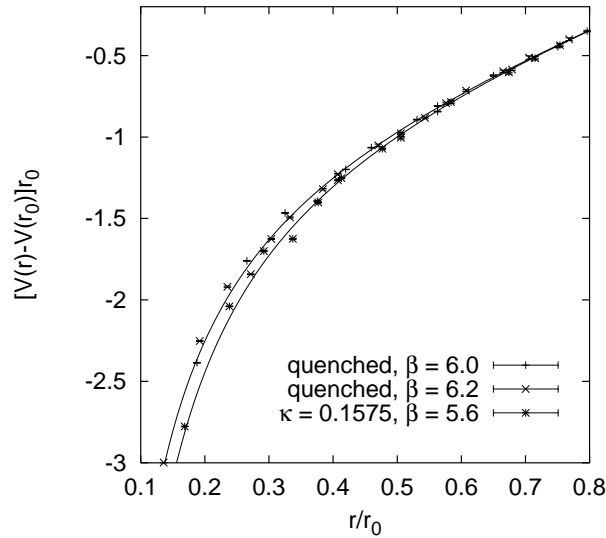}{100mm}}
\caption{A comparison of the short-distance heavy quark potential 
obtained from the lattice on quenched and unquenched 
configurations by the SESAM collaboration.
The potential
and separation are given in units of the parameter $r_0$ (see 
text). Figure provided by Gunnar Bali.} 
\label{pot-dyn}
\end{figure}

Most precision lattice calculations of the heavy quark potential have worked in
the quenched approximation in which only the gluonic terms are included in
$S_{QCD}$. This is equivalent to ignoring quark-antiquark pairs popping in and
out of the vacuum (\cite{weingarten}, \cite{montvay}).
Recent calculations (\cite{bali}, \cite{bali-schilling}, \cite{UKQCD}) have given a value of $e$ around 0.3,
which is rather smaller
than the values that phenomenological potentials have used. For example,
\cite{equigg} use $e$ = 0.54 in their Cornell potential fits. Part of this
discrepancy can
be traced to errors in the quenched approximation. When no $q\overline{q}$
pairs are
available in the vacuum for screening, the strong coupling constant will run to
zero too
fast at small distances. Thus
\begin{eqnarray*}
\alpha_s(r)_{Q.A} < \alpha_s(r)_{\rm full \, theory} \\
V(r)_{Q.A.} > V(r)_{\rm full\, theory}
\end{eqnarray*}
when $V(r)$ is dominated by the Coulomb term. Calculations of the heavy quark
potential that have been done on unquenched configurations (which usually contain
two flavours of degenerate massive
quarks in the vacuum, still not entirely 
simulating the real
world), indicate that $e$ is increased by about 10\%. This gives
a steeper potential at short distances, as in Figure \ref{pot-dyn}.
\cite{SESAM} find 
$e_{Q.A.}$ = 0.289(55) and $e_{unquenched}$ = 0.321(100) without the use 
of the $f/R^2$ term in Equation \ref{vlatt}.
This doesn't then explain all of
the
difference between lattice values of $e$ and phenomenological continuum
values.

\begin{figure}
\centerline{\ewxy{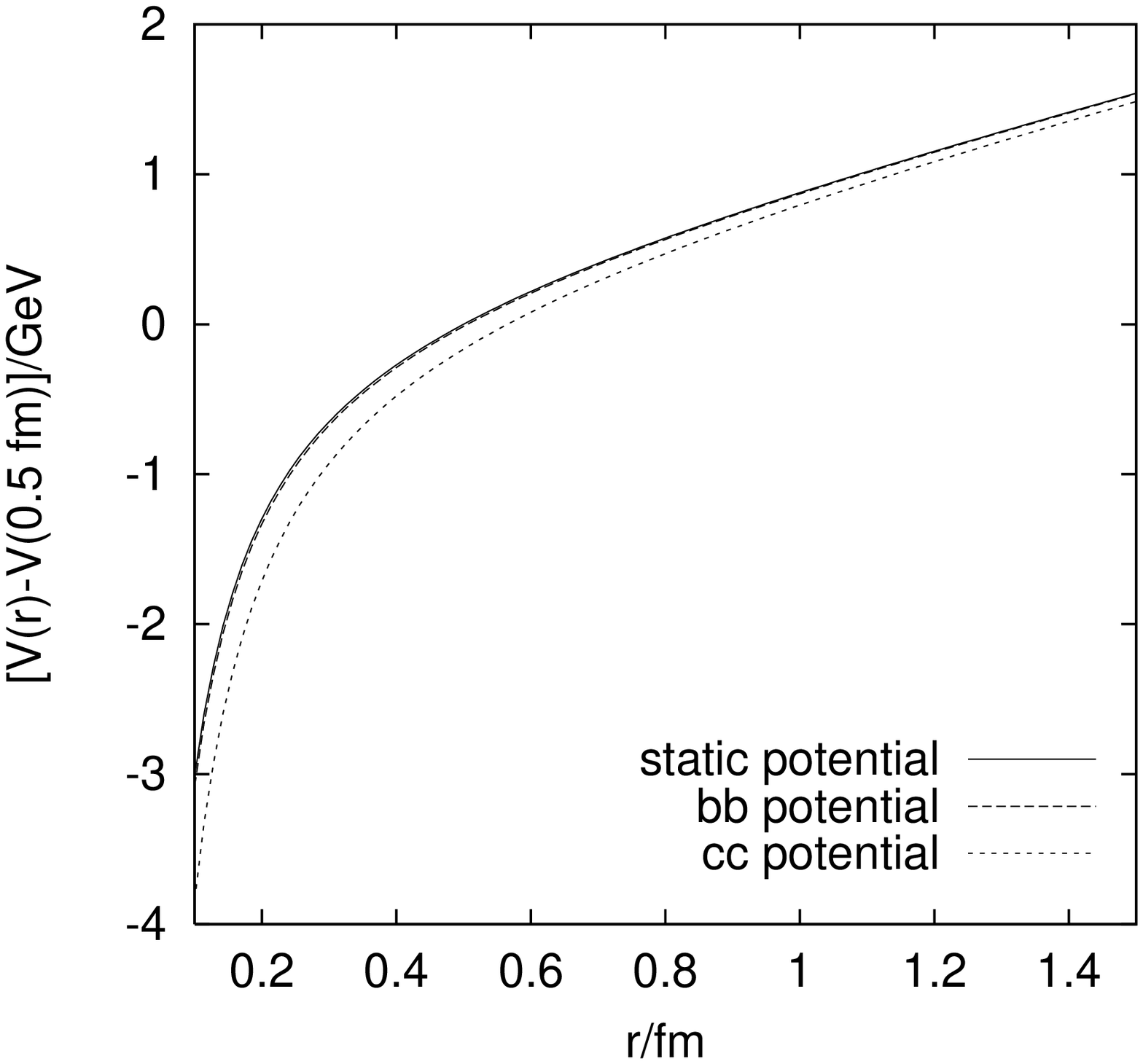}{100mm}}
\caption[kelp]{A comparison of the heavy quark potential 
obtained from the lattice including the first relativistic
corrections which yield an $m_Q$-dependent Coulomb term. 
The static ($m_Q \rightarrow \infty$) potential is 
given by the solid line and those for  $b$ and $c$ 
by dashed lines. Figure provided by Gunnar Bali, 
see \cite{bali}.} 
\label{bc-pot}
\end{figure}

Phenomenological potentials also
implicitly include some relativistic
corrections
to the static picture that can be modelled simply as $r$-dependent additional
potentials. The first such corrections contain a term inversely 
proportional to the square of the heavy quark mass multiplying a
Coulomb potential, and therefore altering the 
effective value of $e$ in an $m_Q$-dependent way.
The coefficient of these corrections can 
be calculated on the lattice (\cite{bali}). It is found that $e$
becomes $e +  b/m_Q^2$ where $b$ = $(0.86(5) {\rm GeV})^2$, giving 
a significant increase (35\%) to the effective value of $e$ 
for charmonium but no change for bottomonium. This is illustrated in Figure
\ref{bc-pot}, and supports the phenomenological use of different
values for $e$ in the two systems as a flavour-dependent dynamical
effect. 

The string tension, $\sigma$, describes the strength of the linearly rising
part
of the potential. It is dimensionful, so $\sigma_{latt} = \sigma a^2$. Using
values
of $\sigma$ from phenomenological potentials ( \cite{equigg} use 
$\sqrt{\sigma} \approx$ 0.43 GeV)
allows us to fix $a$ on the lattice and then convert all other dimensionful
quantities to physical units. However, $\sigma$ and $e$ are anti-correlated
from the
fitted form used in equation \ref{vlatt} and this gives some bias. It is better to use
instead
the value $r_0$ obtained by setting $r^2F(r)$ to a fixed value. $F(r)$ is the
interquark
force, obtained by differentiating the potential,
 and a suitable fixed value is 1.65 which corresponds to $r_0 \approx$ 0.5
fm ($\equiv 2.5 {\rm GeV}^{-1}$ when $\hbar c$ = 1)
 (\cite{sommer}).
 Ensembles at
different values of $a$ are obtained by using different bare coupling constants
in the action,
$S_{QCD}$ (\cite{weingarten}, \cite{montvay}). However, the value of $a$ for a given value of the bare coupling
constant is not known {\it a priori} but has to be obtained by calculating a
dimensionful parameter
and comparing to experiment (or, in the case of $\sigma$ or $r_0$ 
above, to phenomenology). Fixing $a$ is a critical step in a lattice 
calculation and introduces additional systematic and statistical 
errors into the quoted physical results. In the quenched approximation the 
value of $a$ for a given gauge coupling, $\beta$, will depend on the 
experimental quantity chosen to fix it, and so it is 
important to know what quantity was chosen when looking at lattice
results. This point will be discussed further later.   

Given a value for $a$, $V_{latt} - V_c$ can be converted to GeV at separations,
$r$,
in fm ($\equiv {\rm GeV}^{-1}$).  This is then the physical heavy quark
potential and it should be independent of the lattice spacing at which the
calculation was done. Figure \ref{pot-fit} shows that this is true 
for current lattice results. 

Using the fitted form for the potential, the spectrum can be calculated
by solving Schr\"{o}dinger's equation in the continuum (\cite{bali}). 
The heavy quark mass and the overall scale (given by the string 
tension) need to be adjusted to optimise the fit to experiment. Including 
the relativistic corrections to the potential described above 
and adjusting the value of $e$ to mimic an unquenched result, 
yields average deviations from experiment of around 10 MeV for 
bottomonium and rather larger, as expected, 20 MeV for charmonium. 
The remaining systematic errors 
in the lattice potential (see section 2.2) could cause shifts of this size for 
bottomonium and make the 20 MeV deviations for charmonium look 
rather fortuitous.   

\begin{quote}
{\bf Exercise:} Discuss how you would expect the potential appropriate to heavy
baryons to behave. How would you calculate this on the lattice? (\cite{bar-pot}).
\end{quote}

\subsection{The spin-dependent heavy quark potential} \label{sdpot}

As described above, the infinitely massive heavy quark is only a colour
source; it carries no spin. To obtain spin splittings then we must move away
from the static
picture. A useful starting point is a non-relativistic expansion of the Dirac
Lagrangian which is appropriate for heavy quarks
in heavy-heavy systems (\cite{tl-91}). This can be obtained by
a Foldy-Wouthuysen-Tani transformation of the Dirac Lagrangian in Euclidean
space (see for example \cite{itz}):
\begin{eqnarray}
{\cal L} &=&  \psi^{\dag} ( D_t - \frac {\vec{D}^2} {2m_Q} \label{nrqcd} \\
&& - c_1 \frac {(\vec{D}^2)^2} {8 m_Q^3} + c_2 \frac {ig} {8 m_Q^2}
(\vec{D}\cdot\vec{E} - \vec{E}\cdot\vec{D}) \nonumber \\
&& - c_3 \frac {g} {8m_Q^2} \vec{\sigma} \cdot ( \vec{D} \times \vec{E} - \vec
{E} \times \vec{D}) - c_4 \frac {g} {2m_Q} \vec{\sigma}\cdot \vec{B} \ldots )
\psi. \nonumber 
\end{eqnarray}
$\psi$ is a 2-component spinor with heavy quark and anti-quark decoupled. The
mass term $\psi^{\dag}m_Q \psi$ has been dropped. $\vec{D}$ is a covariant derivative
coupling to the gluon field and $\vec{E}$ and $\vec{B}$ are 
chromo-electric and chromo-magnetic fields. The rest of the QCD 
Lagrangian for light quarks and gluons remains as usual.  

The terms in the Lagrangian can be ordered in powers of the squared velocity of
the
heavy quark using the following power counting rules for momentum and kinetic
energy (\cite{nakhleh}, \cite{bodwin}):
\begin{eqnarray*}
\vec{D} \sim  p &\sim& m_Q v \\
K &\sim& m_Q v^2
\end{eqnarray*}
Then from the lowest order field equation
\begin{equation}
( \partial_t - ig A_4 - \frac {\vec{D}^2} {2 m_Q}) \psi = 0
\end{equation}
we have
\begin{eqnarray*}
g A _4 \sim \partial_t \sim K  = m_Q v^2\\
g \vec{E} = [ D_t, \vec{D}] \sim pK = m_Q^2v^3\\
-ig\epsilon_{ijk}B^{k} = [D_i, D_j] \sim K^2 = m_Q^2v^4 .
\label{power}
\end{eqnarray*}

In $\cal{L}$ we then see that the leading order terms on the first line of
equation \ref{nrqcd} are $\cal{O}$$(m_Qv^2)$ and
these give spin-independent splittings in the heavyonium spectrum. On the
second line are spin-independent terms of $\cal{O}$$(m_Qv^4)$ which are
relativistic corrections to the leading terms.
On the third line are spin-dependent terms also of $\cal{O}$$(m_Qv^4)$. These
are the
leading terms as far as spin-splittings are concerned. So, as discussed earlier,
spin splittings should be $\cal{O}$$(v^2)$ times smaller than
spin-independent splittings.
This
is equivalent to $1/m_Q$ behaviour, with a roughly constant
kinetic energy, giving around 120 MeV for
$c\overline{c}$ and 40 MeV for $b\overline{b}$. Note that the $\vec{\sigma}
\cdot(\vec{D} \times \vec{E})$ and $\vec{\sigma} \cdot \vec{B}$ spin-dependent terms 
are of the same order because the chromo-magnetic field is suppressed by one power of $v$
over the chromo-electric field in the power counting. 

Following these power-counting rules the number of operators to be 
included in $\cal{L}$ can be truncated at a fixed order in $v^2/c^2$ and this is
obviously a sensible thing to do if $v^2/c^2 \ll 1$.  
In describing heavy-heavy systems with this Lagrangian, however, we have lost 
the renormalisibility of QCD. To obtain useful results we must put in 
a cut-off, $\Lambda$, to restrict momenta to $p < \Lambda < m_Q$.
The excluded momenta e.g. in gluon loops will 
reappear as a
renormalisation of the coefficients of the non-relativistic operators, the 
$c_i$ of equation \ref{nrqcd}. The $c_i$ can be calculated in perturbation 
theory (since they are dominated by ultra-violet scales
for $\Lambda \gg \Lambda_{QCD}$) 
by matching low energy scattering amplitudes of (\ref{nrqcd}) to full QCD to 
some order in $\alpha_s$ and $p/m_Q$. The $c_i$ are all one at tree-level.     

I will describe two different, but related, approaches to the study of spin splittings in
heavyonium. One is to take $\cal{L}$ of equation \ref{nrqcd} and discretise
it directly on the lattice - this is the NonRelativistic QCD approach
(\cite{tl-91}). The second is
to develop spin-dependent potentials from $\cal{L}$ to add into a
Schr\"odinger equation, $H\psi = E\psi$, and solve for the splittings. 
\begin{equation}
H = \sum_{i=1}^2 \{ \frac {\vec{p}^2} {2m_i} - c_1 \frac {\vec{p}^4} {8 m_i^3} \}
+ V_o(r) + V_{sd}(r, \vec{L}, \vec{S_1}, \vec{S_2})
\label{schroH}
\end{equation}
where $V_0(r)$ is the central potential from section \ref{sipot} and 
$V_{sd}$ includes the spin-dependent potentials. 
Again one can take a
phenomenological approach to the spin-dependent potentials, or extract them
from the lattice. I will describe the potential approach first and then return
to NRQCD in the next subsection.

\begin{figure}
\begin{center}
\setlength{\unitlength}{.02in}
\begin{picture}(80,120)(0,-10)
\put(30,70){\line(1,0){20}}
\put(30,70){\line(0,1){40}}
\put(30,110){\line(1,0){20}}
\put(50,70){\line(0,1){40}}
\put(30,0){\line(1,0){20}}
\put(30,0){\line(0,1){40}}
\put(30,40){\line(1,0){20}}
\put(50,0){\line(0,1){40}}
\put(0,55){\line(1,0){80}}
\put(10,90){\makebox(0,0){{$\Bigg<$ }}}
\put(70,90){\makebox(0,0){{$\Bigg>$ }}}
\put(10,20){\makebox(0,0){{$\Bigg<$ }}}
\put(70,20){\makebox(0,0){{$\Bigg>$ }}}
\put(30,80){\circle*{2.0}}
\put(50,100){\circle*{2.0}}
\put(40,80){\makebox(0,0){{$F_1$ }}}
\put(40,100){\makebox(0,0){{$F_2$ }}}
\put(60,90){\makebox(0,0){{$t$ }}}
\put(55,95){\vector(0,1){5}}
\put(55,85){\vector(0,-1){5}}
\put(20,105){\makebox(0,0){{$\Delta$ }}}
\put(25,107){\vector(0,1){3}}
\put(25,103){\vector(0,-1){3}}
\put(20,75){\makebox(0,0){{$\Delta$ }}}
\put(25,77){\vector(0,1){3}}
\put(25,73){\vector(0,-1){3}}
\put(40,-10){\makebox(0,0){{$R$ }}}
\put(45,-5){\vector(1,0){5}}
\put(35,-5){\vector(-1,0){5}}
\put(20,20){\makebox(0,0){{$T$ }}}
\put(25,35){\vector(0,1){5}}
\put(25,5){\vector(0,-1){5}}
\end{picture}
\end{center}
\caption{The ratio of expectation values required for the 
spin-dependent potentials. $F_1$ and $F_2$ represent 
insertions of $E$ or $B$ as required for that potential. 
For some potentials these will be on the same side of the Wilson
loop. The distance $\Delta$ should be large and $t$ is 
summed over (see text).}
\label{wloop-ratio}
\end{figure}
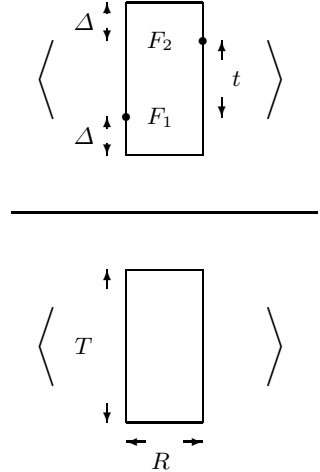

To extract spin-dependent potentials from QCD we start from the Wilson
loop
which represents a static quark anti-quark pair at separation $R$
(\cite{efein}, \cite{peskin}).
As discussed earlier, the heavy quark propagator in this 
case is simply a line in the time direction, from the simplest 
possible heavy quark Lagrangian, 
$\psi^{\dag}D_t \psi$.
 Imagine
adding a
perturbation $\vec{\sigma} \cdot \vec{B}/ 2 m_Q$ to the quark or anti-quark,
such as would come from relativistic corrections to the propagator using 
the Lagrangian of equation \ref{nrqcd}. On
one leg
alone, zero is obtained by symmetry. If the perturbation is added to both legs
and we sum over the time separations, $t$, between the two additions a
new contribution
to the potential is obtained of the form $ \vec{S}_1 \vec{S}_2 \Delta
V / m_{Q1} m_{Q2}$. 
\begin{equation}
\Delta V = 2 \lim_{\tau \rightarrow \infty} \int_0^{\tau} dt \langle \langle
B(\vec{0},0)
B(\vec{R}, t) \rangle \rangle_W
\label{latt-sdpot}
\end{equation}
where $\langle \langle \rangle \rangle_W$ means the expectation value in the
presence of the Wilson loop i.e. the ratio of the expectation value of 
the Wilson loop with the $B$ field insertions to that without.
This is easy to calculate using the methods of Lattice QCD (\cite{mich-rak},
\cite{def-s}).
Figure \ref{wloop-ratio} illustrates this ratio for one value of $t$.
An integration over $t$ is required and this is approximated on the 
lattice by a sum (see \cite{bali} for a recent description of the 
techniques used). The time separations of the insertion points from 
the ends of the Wilson loop, $\Delta$, must be kept large to 
ensure that the spin-dependent contribution to the static 
propagation of a $Q\overline{Q}$ pair is obtained in the ground 
state of the central potential; excited states must have time to decay away. 

The complete spin-dependent potential is given by (\cite{efein}, \cite{chen}):
\begin{eqnarray}
V_{sd} &=& \frac {1} {2r} \left( \frac {\vec{S}_1} {m_{Q1}^2} + \frac {\vec{S}_2}
{m_{Q2}^2} \right) \cdot \vec{L} \left[ d_0 V_0^{'}(r) + 2 d_1 V_1^{'}(r) \right] \label{v-sd}\\
&& \mbox{}+ \frac {1} {r} \left( \frac {\vec{S}_1 + \vec{S}_2} {m_{Q1} m_{Q2}}
\right) \cdot \vec{L} \; d_2 \; V_2^{'}(r) \nonumber \\
&& \mbox{}+ \left( \frac {\vec{S}_1\cdot \vec{r} \vec{S}_2\cdot\vec{r}} {m_{Q1}
m_{Q2} r^2} - \frac {1} {3} \frac {\vec{S}_1\cdot\vec{S}_2} {m_{Q1} m_{Q2}} \right)
d_3 V_3(r) \nonumber \\
&& \mbox{} + \frac {\vec{S}_1\cdot\vec{S}_2} {3 m_{Q1} m_{Q2}} d_4 V_4(r) \nonumber \\
&& \mbox{} + \frac {1} {r} \left( \frac {\vec{S}_1} {m_{Q1}^2} - \frac {\vec{S}_2}
{m_{Q2}^2} \right) \cdot \vec{L} \; \tilde{d_0} \left[ V_0^{'}(r) + V_1^{'}(r) \right] \nonumber \\
&& \mbox{}+ \frac {1} {r} \left( \frac {\vec{S}_1 - \vec{S}_2} {m_{Q1} m_{Q2}}
\right) \cdot \vec{L} \; \tilde{d_2} V_2^{'}(r) \nonumber 
\end{eqnarray}

The primes indicate differentiation with respect to the argument $r$ of 
the different potentials. Note that the last two terms appear
only for the unequal mass case. The $d_i$ and $\tilde{d_i}$ coefficients will be 
discussed below. In perturbation theory the $d_i$ coefficients 
appear at $\cal{O}$$(1)$, the $\tilde{d_i}$ only at $\cal{O}$$(\alpha_s)$
and only for $m_{Q1} \ne m_{Q2}$. $V_0$ is the central potential, discussed 
in section \ref{sipot}, and $V_1$, $V_2$, $V_3$, $V_4$ are obtained
on the lattice by calculating the following expectation values: 

\begin{equation}
\hspace{30mm} \frac {R_k} {R} V_1^{\prime}(R) = 2 \varepsilon_{ijk} \lim_{\tau \rightarrow \infty} \int_0^{\tau} dt \, t \Big\langle 
\setlength{\unitlength}{.02in}
\begin{picture}(30,20)(25,90)
\put(30,70){\line(1,0){20}}
\put(30,70){\line(0,1){40}}
\put(30,110){\line(1,0){20}}
\put(50,70){\line(0,1){40}}
\put(50,80){\circle*{2.0}}
\put(50,100){\circle*{2.0}}
\put(40,80){\makebox(0,0){{$B_i$ }}}
\put(40,100){\makebox(0,0){{$E_j$ }}}
\end{picture}
 \Big\rangle  / Z_W
\label{sameside}
\end{equation}

\vspace{6mm}

\begin{equation}
\hspace{30mm} \frac {R_k} {R} V_2^{\prime}(R) = \varepsilon_{ijk} \lim_{\tau \rightarrow \infty} \int_0^{\tau} dt \, t \Big\langle 
\setlength{\unitlength}{.02in}
\begin{picture}(30,20)(25,90)
\put(30,70){\line(1,0){20}}
\put(30,70){\line(0,1){40}}
\put(30,110){\line(1,0){20}}
\put(50,70){\line(0,1){40}}
\put(30,80){\circle*{2.0}}
\put(50,100){\circle*{2.0}}
\put(40,80){\makebox(0,0){{$B_i$ }}}
\put(40,100){\makebox(0,0){{$E_j$ }}}
\end{picture}
 \Big\rangle  / Z_W
\end{equation}

\vspace{6mm}

\begin{equation}
[\hat{R}_i \hat{R}_j - \frac {1} {3} \delta_{ij}] V_3(R) + \frac {1} {3} 
\delta_{ij} V_4(R)
= 2 \lim_{\tau \rightarrow \infty} \int_0^{\tau} dt \Big\langle 
\setlength{\unitlength}{.02in}
\begin{picture}(30,20)(25,90)
\put(30,70){\line(1,0){20}}
\put(30,70){\line(0,1){40}}
\put(30,110){\line(1,0){20}}
\put(50,70){\line(0,1){40}}
\put(30,80){\circle*{2.0}}
\put(50,100){\circle*{2.0}}
\put(40,80){\makebox(0,0){{$B_i$ }}}
\put(40,100){\makebox(0,0){{$B_j$ }}}
\end{picture}
 \Big\rangle  / Z_W
\end{equation}

\vspace{6mm}

As in Figure \ref{wloop-ratio}, the denominator $Z_W$ 
is the expectation of the Wilson loop without $E$ or $B$ field
insertions. 

\begin{figure}
\begin{center}
\setlength{\unitlength}{.02in}
\begin{picture}(80,50)(25,25)
\put(50,50){\circle*{2.0}}
\put(55,55){\vector(1,0){20}}
\put(75,75){\vector(-1,0){20}}
\put(75,55){\vector(0,1){20}}
\put(55,75){\vector(0,-1){20}}
\put(55,25){\vector(1,0){20}}
\put(75,45){\vector(-1,0){20}}
\put(75,25){\vector(0,1){20}}
\put(55,45){\vector(0,-1){20}}
\put(25,55){\vector(1,0){20}}
\put(45,75){\vector(-1,0){20}}
\put(45,55){\vector(0,1){20}}
\put(25,75){\vector(0,-1){20}}
\put(25,25){\vector(1,0){20}}
\put(45,45){\vector(-1,0){20}}
\put(45,25){\vector(0,1){20}}
\put(25,45){\vector(0,-1){20}}
\put(95,50){\vector(1,0){10}}
\put(95,50){\vector(0,1){10}}
\put(110,50){\makebox(0,0){{$x$ }}}
\put(95,65){\makebox(0,0){{$y$ }}}
\end{picture}
\end{center}
\caption{The sum of four untraced plaquettes around a
point (the clover-leaf operator) that is used for 
a $B_z$ field insertion in a wilson loop for spin-dependent 
potentials (see text). }
\label{clover}
\end{figure}
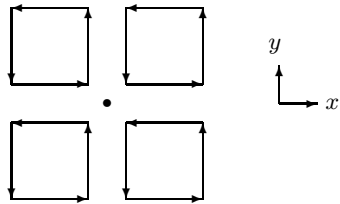

A lattice discretisation of the $E$ and $B$ field strength operators is 
required. The simplest discretisation of $F_{\mu \nu}(\vec{x})$ is
to take the product of four $U$ matrices around a $1 \times 1$ square
in the $\mu, \nu$ plane starting from the corner $\vec{x}$. This 
product is called a plaquette (\cite{weingarten}, \cite{montvay}); its
hermitian conjugate should be subtracted and the resulting SU(3) matrix
made traceless. Note that factors of $g$ that would otherwise appear 
from equation \ref{nrqcd} are absorbed into the lattice version of 
$F_{\mu \nu}$. For the $B$ field, a more symmetric version of this is 
to use, instead of one plaquette, the average of the four plaquettes
around point $\vec{x}$ in the spatial plane perpendicular to $\vec{B}$
(see Figure \ref{clover}). For $\vec{E}$ the spatial average of the  
two plaquettes at a given time is used. See \cite{bali} for details.

The central potential, as calculated on the lattice, is a 
spectral quantity, appearing in the exponent of the exponential 
decay of a correlation function. It can therefore be directly 
interpreted as the continuum potential once converted to 
physical units. The spin-dependent potentials, in contrast, are
calculated from the amplitudes of lattice correlation functions 
and undergo renormalisation when compared to continuum QCD. 
This renormalisation is visible in equation 
\ref{v-sd} as the $d_i$ coefficients. These are functions of  
the $c_i$ coefficients since the potentials are extracted
by perturbing the Wilson loop with operators from 
equation \ref{nrqcd} (\cite{chen}). They reflect the 
matching required between this static/nonrelativistic effective theory
and full QCD. 
In this case it is convenient to do the matching in two stages;
full QCD to continuum effective theory and continuuum effective theory 
to lattice effective theory. 

For the first stage the $c_i$ (and 
therefore $d_i$) have been calculated in leading order
continuum perturbation theory 
(\cite{ehill}, \cite{falk}). They depend logarithmically on 
the quark mass and the cut-off that is applied to the effective 
theory. The spin-dependent potentials also depend on the 
cut-off, but not $m_Q$, so that each term in $V_{sd}$ becomes
\begin{equation}
d_i(\Lambda, m_Q) V_i(\Lambda).
\end{equation}
We can use the $d_i$ calculated in the continuum for the 
lattice calculation if we imagine the continuum effective 
theory at the same cut-off as the lattice cut-off ($1/a$). 

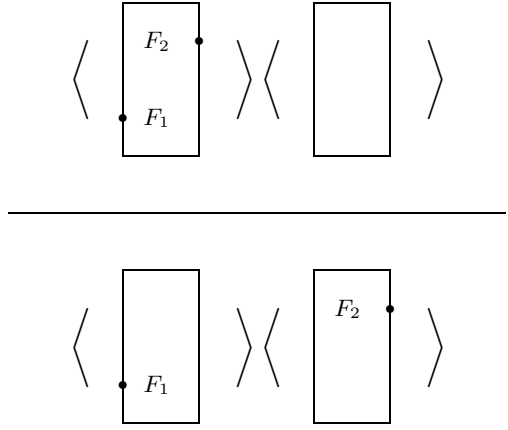
\begin{figure}
\begin{center}
\setlength{\unitlength}{.02in}
\begin{picture}(130,120)(0,-10)
\put(30,70){\line(1,0){20}}
\put(30,70){\line(0,1){40}}
\put(30,110){\line(1,0){20}}
\put(50,70){\line(0,1){40}}
\put(30,0){\line(1,0){20}}
\put(30,0){\line(0,1){40}}
\put(30,40){\line(1,0){20}}
\put(50,0){\line(0,1){40}}
\put(0,55){\line(1,0){130}}
\put(80,70){\line(1,0){20}}
\put(80,70){\line(0,1){40}}
\put(80,110){\line(1,0){20}}
\put(100,70){\line(0,1){40}}
\put(80,0){\line(1,0){20}}
\put(80,0){\line(0,1){40}}
\put(80,40){\line(1,0){20}}
\put(100,0){\line(0,1){40}}
\put(20,90){\makebox(0,0){{$\Bigg<$ }}}
\put(63,90){\makebox(0,0){{$\Bigg>$ }}}
\put(20,20){\makebox(0,0){{$\Bigg<$ }}}
\put(63,20){\makebox(0,0){{$\Bigg>$ }}}
\put(70,90){\makebox(0,0){{$\Bigg<$ }}}
\put(113,90){\makebox(0,0){{$\Bigg>$ }}}
\put(70,20){\makebox(0,0){{$\Bigg<$ }}}
\put(113,20){\makebox(0,0){{$\Bigg>$ }}}
\put(30,80){\circle*{2.0}}
\put(50,100){\circle*{2.0}}
\put(30,10){\circle*{2.0}}
\put(100,30){\circle*{2.0}}
\put(40,80){\makebox(0,0){{$F_1$ }}}
\put(40,100){\makebox(0,0){{$F_2$ }}}
\put(40,10){\makebox(0,0){{$F_1$ }}}
\put(90,30){\makebox(0,0){{$F_2$ }}}
\end{picture}
\end{center}
\caption[kfghf]{The ratio of expectation values used for the 
spin-dependent potentials, taking account of 
renormalisation required to match to the continuum (\cite{hunt-mich}).
 $F_1$ and $F_2$ represent 
insertions of $E$ or $B$ as required for that potential.} 
\label{new-wlrat}
\end{figure}

For the second stage we then match between continuum and lattice 
effective theories at the same cut-off. 
This provides an additional renormalisation which can 
be significant because of the non-linear relationship  
between the continuum and lattice gauge fields
(\cite{lep-mack}). This gives rise to additional 
tadpole diagrams in lattice perturbation theory. 
They have a universal nature and can be thought of 
(even beyond perturbation theory)
as a constant factor multiplying each gauge link. 
Equivalently the renormalisation can be viewed
 as arising from 
the additional perimeter self-energy contributions when the 
$E$ and $B$ field insertion are in place (\cite{def-s}). A 
method to take account of the renormalisation directly on the 
lattice involves calculating, instead of the ratio in 
Figure \ref{wloop-ratio}, the product of ratios in Figure \ref{new-wlrat}
(\cite{hunt-mich}). The additional perimeter/tadpole terms from the 
insertions are thereby cancelled out, and it is hoped that 
any remaining lattice renormalisation is negligible. 

Once the spin-dependent potentials are calculated from 
Figure \ref{new-wlrat} and multiplied by the appropriate 
$d_i$, they can be inserted into a Schr\"{o}dinger equation
and the energy shifts 
from the spin-independent states can be calculated (\cite{bali}). They depend 
on the functional form of the spin-dependent potentials 
and on 
the expectation value of the spin and orbital angular 
momentum operators of equation \ref{v-sd} for a particular
state. To calculate these the following equations are useful 
(for the last relation see \cite{kwong-ups}):
\begin{eqnarray}
\langle \vec{S}_1 \cdot \vec{S}_2 \rangle &=& \frac {1} {2} \left[ S(S+1) -
\frac {3} {2} \right] \label{sl-calc}\\
\langle \vec{L} \cdot \vec{S}_1 \rangle &=& \langle \vec{L} \cdot \vec{S}_2
\rangle = \frac {1} {2}\langle \vec{L} \cdot \vec{S}\rangle \nonumber \\
\langle \vec{L} \cdot \vec{S}\rangle &=& \frac {1} {2} \left[ J(J+1) - L(L+1) -
S(S+1) \right] \nonumber \\
\langle S_{ij} \rangle &=& 4 \langle
3(\vec{S}_i\cdot\vec{\hat{n}})(\vec{S}_j\cdot\vec{\hat{n}}) - \vec{S}_i\cdot\vec{S}_j \rangle
\nonumber \\
&=& 2 \langle
3(\vec{S}\cdot\vec{\hat{n}})(\vec{S}\cdot\vec{\hat{n}}) - \vec{S}^2 \rangle
\nonumber \\
&=& - \frac{ \left[ 12 \langle \vec{L}\cdot\vec{S}\rangle^2 + 6 \langle
\vec{L}\cdot\vec{S} \rangle - 4 S(S+1)L(L+1) \right] }
{ (2L-1) (2L+3) } \nonumber 
\end{eqnarray}

The 
results for the expectation values are tabulated for $S$ and $P$
states in Table \ref{sl-tab}. It is clear from this table that 
the only potential contributing to the hyperfine splitting between
the $^3S_1$ and $^1S_0$ states ($M(\Upsilon)-M(\eta_b)$, $M(J/\Psi)-
M(\eta_c)$) is $V_4$. For the splittings between $P$ states, all 
the spin-spin and spin-orbit potentials can contribute in principle. 
Notice how the spin-averaging described at the beginning of section 
2 removes all the spin-dependent pieces, to obtain the spin-independent
spectrum. To remove $V_4$ terms from $P$ states the spin-average must 
be taken including the $^1P_1$. 
\begin{table}
\begin{center}
\begin{tabular}{c|cccccc}
& $^1S_0$ & $^3S_1$ & $^1P_1$ & $^3P_0$ & $^3P_1$ & $^3P_2$ \\
\hline \\
$\langle \vec{S}_1 \cdot \vec{S}_2 \rangle$ & -$\frac {3} {4}$ &
$ \frac {1} {4} $ & -$\frac {3} {4} $ & $\frac {1} {4} $ & $\frac {1} {4} $
&$\frac {1} {4} $  \\
$\langle \vec{L}\cdot\vec{S} \rangle $ & 0 & 0 & 0 & -2 & -1 & 1 \\
$\langle S_{ij} \rangle $ &  0 & 0 & 0 & -4 & 2 & -$\frac {2} {5} $ \\
\end{tabular}
\caption{Expectation values for combinations of spin and orbital 
angular momentum operators needed for spin splittings in 
heavy-heavy bound states.}
\label{sl-tab}
\end{center}
\end{table}

What functional form do we expect for the different spin-dependent
potentials? In leading order perturbation theory (one gluon exchange): 
\begin{eqnarray}
V_0 &=& - C_F \frac {\alpha_s} {r} \label{v-pert}\\
V_1^{'} &=& 0 \nonumber \\
V_2^{'} &=& C_F \frac {\alpha_s} {r^2} \nonumber \\
V_3 &=& 3 C_F \frac {\alpha_s} {r^3} \nonumber \\
V_4 &=& 8 \pi C_F \alpha_s \delta^{(3)} (r) \nonumber, 
\end{eqnarray}
with $C_F = 4/3$. The `same-side' (see equation \ref{sameside})
 spin-orbit interaction, $V_1$, is 
absent; the `opposite-side', $V_2$ is simply $V_0$.
The form of $V_4$ implies that it is only effective 
for states with a wavefunction at the origin i.e. $S$ states. 
It gives for the $^3S_1 - ^1S_0$ splitting, 
\begin{equation}
\frac {32 \pi \alpha_s} {9 m_Q^2} |\psi(0)|^2.
\label{hypeq}
\end{equation}
We do not then expect any splitting induced by the $V_4$ term 
between the $^1P_1$ and $^3P_1$ states, so 
the $^1P_1$ mass should be at the spin-average
of the $^3P$ states. 

The following inter-relationships between potentials are 
also useful.
\begin{eqnarray}
V_2^{'} - V_1^{'} &=& V_0^{'} \label{gromes} \\
V_3(r) &=& \frac {V_2^{'}(r)} {r} - V_2^{''}(r) \label{v3} \\
V_4(r) &=& 2 \nabla^2 V_2(r)  \label{v4}
\end{eqnarray}
Equation \ref{gromes} is the Gromes relation (\cite{gromes1}),
 derived from Lorentz invariance and 
as such always true.  
Equations \ref{v3} and \ref{v4} hold for any vector-like exchange 
(such as single gluon) but 
only to leading order; they do not survive renormalisation 
of the potentials when the cut-off on the effective 
Lagrangian of equation \ref{nrqcd} is changed. In particular
$V_1$ and $V_2$ then mix (\cite{chen}).  

A crucial ingredient missed in the perturbative analysis is the confining part 
of the central potential, $V_0$, and this can reappear in
the $V_i$. The nature of this confining term is 
important. General considerations (\cite{gromes3}) show that it 
can only arise from vector and/or scalar exchange, but 
these two possibilities yield quite different accompanying 
spin-dependent potentials. A vector exchange gives rise
to $V_2$, $V_3$ and $V_4$, a scalar exchange only to $V_1$  
(\cite{gromes2}). In both cases, a constant term $\sigma$ appears 
in $V_2^{\prime} - V_1^{\prime}$ from the Gromes relation. 

A useful quantity to study in this respect is the 
ratio of $p$ state splittings:
\begin{equation}
\rho = \frac {M(^3P_2)-M(^3P_1)} {M(^3P_1)-M(^3P_0)}.
\label{rho}
\end{equation}
Experimentally this ratio takes the value 0.48(1) for 
$c\overline{c}(1P)$ and 0.66(2) for $b\overline{b}(1P)$
and 0.58(3) for $b\overline{b}(2P)$. For pure $\vec{L}\cdot
\vec{S}$ interactions $\rho$ is simply related to a  
combination of expectation values of $\vec{L}\cdot\vec{S}$
since, considering the spin-dependent potentials as 
a perturbation on the spin-independent one,
 the expectation value of $V_i$ in all the $P$ states is the 
same. This then gives $\rho$ = 2. Similarly a pure 
tensor $V_3$ interaction gives $\rho$ = -0.4. These are 
clearly wrong; we require a mixture of spin-orbit 
and tensor terms. For the leading order perturbative potentials
in equation \ref{v-pert} we can also calculate $\rho$ 
exactly because all the expectation values of $V_i$ reduce to cancelling 
terms of the form $\langle r^{-3} \rangle$. This gives 
$\rho$ = 0.8, larger than all the experimental values. 
The confining term should then appear in such a 
way as to reduce $\rho$. 

This is possible if we make the assumption that the 
confining term grows linearly with $r$ as $\sigma r$; 
such a rapid rise implies a scalar exchange (\cite{gromes2}). 
$V_1$ = $-\sigma r$ and $V_2$ = $- C_F \alpha_s / r$
will satisfy the Gromes relation.  
$\rho$ becomes
\begin{equation}
\rho = \frac {1} {5} \frac {8 \alpha_s \langle r^{-3} \rangle - 5/2 
\sigma \langle r^{-1} \rangle } 
{2 \alpha_s \langle r^{-3} \rangle - 1/4 \sigma \langle r^{-1} \rangle }
\end{equation}
and for positive expectation values this will be less than 0.8,
in agreement with experiment (\cite{moorhouse}). 
The $\sigma$ term will be more effective for longer range 
wavefunctions such as $c\overline{c}$ and $b\overline{b}(2P)$ 
giving a smaller value of $\rho$ 
than for $b\overline{b}(1P)$. 
A vector confining potential would lead to the term proportional 
to $\sigma$ appearing in $V_2$ with opposite sign as well 
as additional $V_3$ terms, so that $\rho >$ 0.8 (\cite{schnitzer}).
Of course this does not rule out a mixture of long-range 
vector and scalar pieces. 

\begin{figure}
\centerline{\epsfig{file=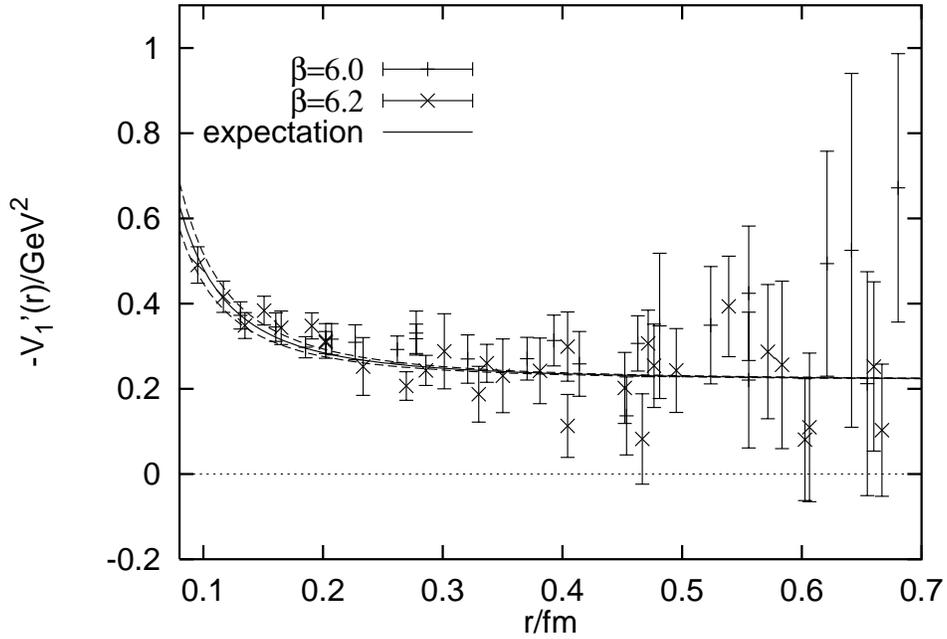,height=90mm,bbllx=50pt,
bblly=50pt,bburx=482pt,bbury=352pt}}
\caption[hjk]{The spin-orbit potential, $ - V_1^{'}$, at 
two different values of the lattice spacing together with 
a fit curve of the form $\sigma + h/R^2$. (\cite{bali}).}
\label{bali-v1}
\end{figure}

The lattice calculation of the spin-dependent potentials confirm 
the behaviour above explicitly, and show (within errors) that 
the long range confining potential is purely scalar (\cite{hunt-mich}). 
$V_3$ and $V_4$ are found to be very short-range with 
$V_3$ showing $1/R^3$ behaviour and $V_4$ approximating a $\delta$ function
on the lattice. $V_1^{\prime}$ is approximately constant at the 
value $-\sigma$ taken from the central potential, whereas
$V_2^{'} \rightarrow 0$ at large $R$. In 
addition $V_1^{'}$ has a small attractive $1/R^2$ piece 
(see Figure \ref{bali-v1}
from \cite{bali}). 
which arises from the mixing between $V_1$ and $V_2$ 
and its size changes as the lattice cut-off ($1/a$) changes,
along with the 
Coulombic $1/R^2$ term present in $V_2^{\prime}$.

\begin{figure}
\centerline{\epsfig{file=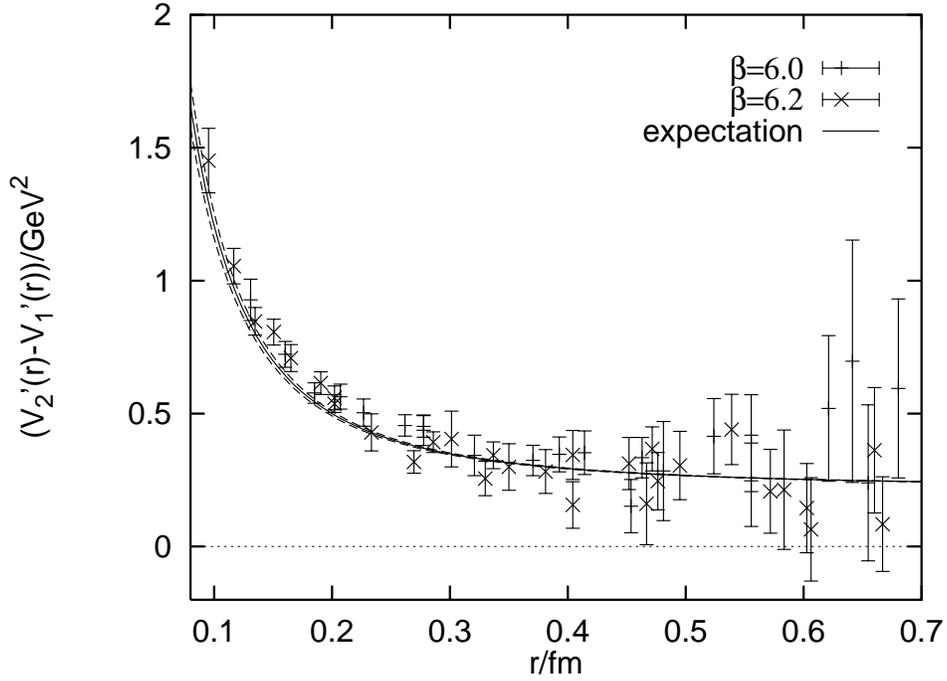,height=90mm,bbllx=50pt,
bblly=50pt,bburx=482pt,bbury=352pt}}
\caption[hjk]{The difference of spin-orbit potentials, 
$V_2^{'}$ and $V_1^{'}$ on the lattice, compared to the expectation from 
the central potential according to the Gromes relation.
(\cite{bali}).}
\label{bali-gromes}
\end{figure}

There is no exact Gromes relation on the lattice (\cite{bali2}), but 
it should be restored in the continuum limit. This relation 
does in fact work well on the lattice at current values of 
the lattice spacing and this is a non-trivial check of the 
lattice renormalisation procedure of Figure \ref{new-wlrat} 
(\cite{hunt-mich}). 
This renormalisation is done for the left hand side of 
equation \ref{gromes} but not for the right. 
See Figure \ref{bali-gromes} from \cite{bali}. 

As discussed earlier, the spectrum from this lattice 
potential yields deviations at the 10 MeV level for 
bottomonium and the 20 MeV level for charmonium. Systematic
errors in the charmonium case are rather larger than this, 
however. The $d_i$ coefficients have large perturbative 
corrections for the lattice cut-off used ($m_ca < 1$)
and so large uncertainties. $c_1$ is set  to 1 in 
equation \ref{schroH}; unknown perturbative corrections 
to that coefficient could induce 50MeV shifts in the 
charmonium spectrum (\cite{bali}). 
As mentioned in section 2.1 there are also relativistic 
corrections to the spin-independent central potential
(\cite{brambilla}). 
These can be calculated from expectation values of 
Wilson loops with $E$ and $B$ insertions in a similar 
way to that for the spin-dependent potentials above. The results 
modify the central potential for charmonium quite strongly, 
again indicating that unknown higher order corrections could 
be significant for that system. 

To go beyond the corrections discussed here in the potential 
model approach is hard; higher order insertions into 
Wilson loops cannot be reduced to the form of an 
instantaneous potential. We need instead more direct 
methods of calculating the spectrum. This must be 
done on the lattice and will be discussed in the 
next subsection.  

\begin{quote}
{\bf Exercise:} Fill out Table \ref{sl-tab} to include $D$ states. 
\end{quote}

\subsection{Direct measurement of the bottomonium spectrum on the lattice}

A direct calculation of the heavyonium spectrum on the lattice
at first sight seems rather hard. There is a large range of 
scales in the problem, all the way from the heavy quark mass
to kinetic energies within bound states ($\approx \Lambda_{QCD}$). 
To cover these properly in a lattice simulation would 
require $a^{-1} \gg m_Q$ and the number of lattice points 
on a side, $L \gg m_Q/\Lambda_{QCD}$. 

As we have seen from the previous sections, however, the quark 
mass itself is not a dynamical scale, simply an overall energy 
shift. We only actually need to simulate accurately the 
important scales for the bound state splittings, $p_Q$ and $K$. 
This leads us to work with a lattice with $a^{-1} < {\cal{O}}(m_Q)$ 
and make use of the non-relativistic effective theory of 
equation \ref{nrqcd}. This Lagrangian can be discretised on the 
lattice (\cite{tl-91}, \cite{nakhleh}) 
and applied using similar techniques to those for 
handling light quarks on the lattice (\cite{weingarten}, \cite{montvay}). 
Details will be discussed below. 

There is an important difference between the NRQCD approach and 
the potential model approach of the section \ref{sdpot}. That 
approach starts from the static theory and so can only 
produce the potential; the missing kinetic energy terms 
are of equal weight (in powers of $v^2/c^2$) in the spectrum 
and they must be added in subsequently in a Schr\"{o}dinger 
equation. The NRQCD calculations, even at lowest order, include both
the $\psi^{\dag}D_t \psi$ term and the $\psi^{\dag}D^2/2m_Q \psi$
terms and yield the spectrum directly; the existence of 
a potential is not invoked at any stage. 
This means that the NRQCD approach can be fully matched to QCD
and handle
the sub-leading 
effects from soft-gluon radiation that eventually cause a 
potential model picture to break down through 
infra-red (long time) divergences (\cite{appelquist}, \cite{tl-91}).
We will find potential models useful for guiding NRQCD 
calculations, nevertheless. 

The NRQCD approach uses the Lagrangian of equation \ref{nrqcd}
as an effective theory on the lattice (\cite{nakhleh}). 
It can reproduce the low energy ($p < 1/a$) behaviour of QCD, 
but the couplings, $c_i$, must be adjusted from their
tree level values of 1 to compensate for neglected high momentum 
interactions. In principle this can be done in perturbation theory 
by matching scattering amplitudes between lattice NRQCD and 
full QCD in the continuum (here a one-stage matching is used). 
The $c_i$ will have an expansion in terms of $\alpha_s(1/a)$. 
They will differ from the $c_i$ of the static approach discussed 
earlier since the $\vec{p}^2/m_Q$ term in the heavy quark propagator
will give additional explicit $1/m_Qa$ terms which diverge 
as $a \rightarrow 0$. 
In this way it is clear that we cannot take a continuum limit in 
NRQCD; we can only demonstrate that results are independent of the lattice 
spacing at non-zero lattice spacing. This is sufficient for 
them to make physical sense, and to be compared to experiment.  

One problem for lattice NRQCD is the possible large 
renormalisations, $c_i$, which come from tadpole diagrams. This 
was discussed earlier in connection with the renormalisation 
of the spin-dependent potentials in the
static case. The tadpoles appear with every occurrence 
of a gluon link field and  can be taken care of by 
renormalising each gauge link by a factor $u_0$ as it is 
read in, 
\begin{equation}    
U_{\mu} \rightarrow \frac {U_{\mu}} {u_0},
\end{equation}
and then using the the renormalised gauge link everywhere
instead of the original. 

$u_0$ represents how far 
the gluon links are from their continuum expectation value 
of 1. The easiest quantity to use to set $u_0$ is 
the plaquette.  Since it contains four links we have:
\begin{equation}
u_0 = u_{0P} = \sqrt[4]{ \frac {1} {3} {\rm Tr} \Big\langle
\setlength{\unitlength}{0.01in}
\begin{picture}(30,15)(-5,10)
\put(0,0){\vector(1,0){20}}
\put(20,20){\vector(-1,0){20}}
\put(20,0){\vector(0,1){20}}
\put(0,20){\vector(0,-1){20}}
\end{picture}
\Big\rangle}
\end{equation} 
A possibly better motivated value is that in which 
we look at a single link field and maximise its value 
by gauge-fixing. This should be most effective at 
isolating (by minimising) the true gauge-independent tadpole 
contribution (\cite{gpl-tsukuba}).
The gauge in which this happens is Landau 
gauge: 
\begin{equation}
u_0 = u_{0L} = \frac {1} {3} {\rm Tr} \langle U_{\mu} \rangle_{{\rm Landau\,gauge}}.
\end{equation}
The difference between the two $u_0$s is small in lattice 
perturbation theory. Measured (non-perturbative) values on 
the lattice differ by a few percent at moderate values of the lattice 
spacing.

\begin{figure}
\centerline{\epsfig{file=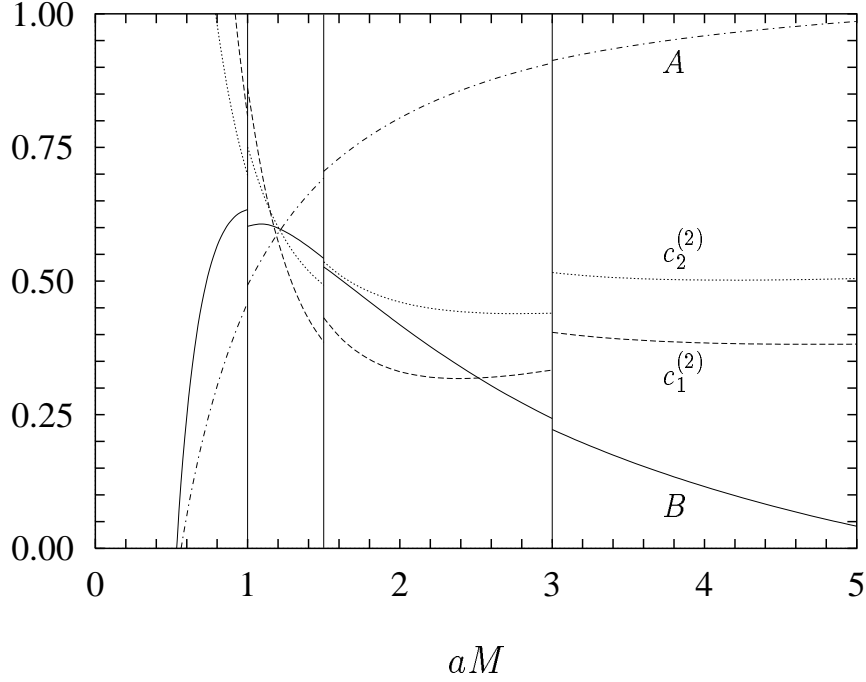,height=100mm,bbllx=83pt,
bblly=367pt,bburx=516pt,bbury=697pt,clip=}}
\caption[hjk]{The $\cal{O}$$(\alpha_s)$ coefficients of 
various terms in the NRQCD Lagrangian, calculated in 
lattice perturbation theory {\it after} tadpole 
improvement. $A$ corresponds to the energy shift and 
$c_1^{(2)}$ to the $D^4$ term. $B$ corresponds to the 
mass renormalisation and $c_2^{(2)}$ to the $D_i^4$ 
term, here denoted $c_5$. The vertical lines 
represent discontinuities when the value of the 
stability parameter, $n$, is changed. (\cite{morning}). } 
\label{colin}
\end{figure}

Once we have renormalised the gauge fields to take account 
of the tadpole contributions (`tadpole-improvement') 
we would hope that 
remaining corrections to the $c_i$ are small. 
They have been  calculated in perturbation theory for those 
terms which contribute to the heavy quark self-energy 
(\cite{morning}). The dispersion relation 
for the heavy quark is required to be
\begin{equation}
E(p) = \alpha_s A + \frac {p^2} {2 m_r} - \frac {p^4} {8 m_r^3}
\end{equation}
with $A$ an energy shift and $m_r$ the renormalised quark
mass, and this fixes the coefficient $c_1$ in the lattice discretised
version of equation 
\ref{nrqcd}. 
Figure \ref{colin} shows that the $\cal{O}$$(\alpha_s)$
coefficient of $c_1$ is small, its magnitude less than 1,
until $m_Qa$ is less than about 0.8, when it starts to diverge.   
This is a sign of the power ultra-violet divergences of 
NRQCD mentioned above; we must stay at values of $a$ where 
$m_Qa > 0.8$. Without tadpole-improvement the $\cal{O}$$(\alpha_s)$
coefficients are all much larger than 1 (\cite{morning}), showing that 
tadpole-improvement has captured most of the renormalisation.   
The results I shall describe here use tadpole-improvement and 
all $c_i$ then set to 1. 

The NRQCD Lagrangian is discretised on the lattice in the 
standard way (\cite{weingarten}, \cite{montvay}). Derivatives 
are replaced by finite differences, and $E$ and $B$ fields 
by clover terms. In the process, all appearances of 
$m_Q$ are replaced by the bare quark mass in lattice units, 
$m_Qa$, and powers of $g$ are absorbed into the lattice fields. 
The lowest order terms in the Lagrangian density (in lattice units) become
\begin{eqnarray}
D_t \psi_t &\rightarrow& U_t \psi_{t+1} - \psi_{t} \label{disc} \\
\frac {\vec{D}^2} {2m_Q} &\rightarrow& \frac {\sum_i U_{x,i} \psi_{x+\hat{i}} + U_{x-\hat{i},i}^{\dag} 
\psi_{x-\hat{i}} - 2 } {2m_Qa}. \nonumber
\end{eqnarray}
Each $U_{\mu}$ field here is understood to have been divided 
by $u_0$ already. 

Then the calculation of the heavy quark propagator is very 
simple. The propagator as a function of spatial indices on 
a given time slice is related to that on previous time slice
by an evolution equation :
\begin{eqnarray}
U_t G_{t+1} - G_t &=& -aH G_t \label{evol}\\
G_{t+1} &=& U_t^{\dag}(1-aH)G_t \nonumber
\end{eqnarray}
where $aH$ is the Hamiltonian, for example, the lowest 
order $\vec{D}^2/2m_Q$ term, discretised on the lattice 
as in equation \ref{disc}.  
This enables the heavy quark propagator to be 
calculated on one pass through the lattice in the 
time direction starting with some source for the 
propagator on time slice 1. This simplicity can be 
traced back to the simple first order time derivative
in equation \ref{nrqcd}; calculations of relativistic 
quark propagators take many sweeps through the lattice 
(\cite{weingarten}, \cite{montvay}).  

One technical problem with (\ref{evol}) is that it 
can become unstable for modes for which $H$ approaches
1. In the free case for the lowest order Hamiltonian 
we have 
\begin{equation}
H_0 = \frac {3 \sum_i 4 \sin^2(p_ia/2)} { 2m_Qa } 
\end{equation}
where $i$ runs over the Fourier modes
and the maximum value for momenta close to 
the lattice cut-off is $6/m_Qa$. This would limit
values of $m_Qa$ to be greater than 6. Instead we 
can stabilise the evolution by adding terms which 
have an effect at the cut-off scale but are not 
important for the physically relevant momenta well 
below the cut-off. This gives rise to an evolution 
equation (\cite{tl-91}):
\begin{equation}
G_{t+1} =  \left( 1 - \frac {aH} {2n} \right)^n U_t^{\dag}
\left( 1 - \frac {aH} {2n} \right)^n G_t \label{newevol}
\end{equation}
and the stability requirement is now $m_Qa > 3/n$ so 
reasonable values of $m_Qa$ of $\cal{O}$$(1)$ can be 
reached for suitable $n$. In fact it is only important
to stabilise the lowest order term $H_0$ like this; 
the higher order terms of the Lagrangian of 
equation \ref{nrqcd} can be added 
in straightforwardly. For example, we can write (\cite{nakhleh}):
\begin{equation} 
G_{t+1} =  (1 - \frac {a\delta H} {2}) \left( 1 - \frac {aH_0} {2n} \right)^n U_t^{\dag}
\left( 1 - \frac {aH_0} {2n} \right)^n (1 - \frac {a\delta H} {2}) G_t \label{newerevol}
\end{equation}
with $a\delta H$ the lattice discretisation of :
\begin{equation}
\delta H =  
- c_1 \frac {(\vec{D}^2)^2} {8 m_Q^3} + c_2 \frac {ig} {8 m_Q^2}
(\vec{D}\cdot\vec{E} - \vec{E}\cdot\vec{D}) 
- c_3 \frac {g} {8m_Q^2} \vec{\sigma} \cdot ( \vec{D} \times \vec{E} - \vec
{E} \times \vec{D}) - c_4 \frac {g} {2m_Q} \vec{\sigma}\cdot \vec{B}. 
\end{equation}

Another technical problem which faces all lattice 
calculations is that of discretisation errors. These 
arise from the use of finite differences on the lattice
to approximate 
continuum derivatives. They mean that even physical 
lattice results, expressed in GeV for example, will 
depend upon the lattice spacing. The dependence will 
be as some power of a typical momentum scale in 
lattice units. Since the momenta inside heavyonium 
systems are quite large, $\cal{O}$(1 GeV), discretisation 
errors will cause a problem if they are not corrected for. 

This is achieved by improving the discretisation of 
derivatives to include higher order terms. For example, 
ignoring gauge fields, 
\begin{eqnarray}
a^2 D_{i,latt}^2 \psi_x &=& \psi_{x+\hat{i}} + \psi_{x-\hat{i}} - 2 \psi_x \\
&=& \left( e^{aD_{i,cont}} - 1 \right)
\left( 1 - e^{-aD_{i,cont}}\right) \psi_x \nonumber \\
&=& \left( a^2 D_{i,cont}^2 + a^4 D_{i,cont}^4 [\frac {2} {6} - \frac {1} {4}]
\ldots \right) \psi_x. \nonumber
\end{eqnarray}
giving $\cal{O}$$(a^2)$ errors relative to the leading term.
A better discretisation is then 
\begin{equation}
a^2 \tilde{D}_{i,latt}^2 = a^2D_{i,latt}^2 - \frac {1} {12} a^4 D_{i,latt}^4
\label{dtilde}
\end{equation}
where $a^2 D_{i,latt}^2$ is given by the na\"{\i}ve finite difference. 
$a^2 \tilde{D}_{i,latt}^2$ has errors at relative $\cal{O}$$(a^4)$.  

The other operator that appears in the leading order terms is
the time derivative operator, $D_t$. Any correction to 
$D_t$ that looks like $D_t^2$ would upset our simple 
evolution equation. Instead the way to correct $D_t$ is 
to require that the time-step operator be 
\begin{equation}
G_{t+1} = e^{-aH}G_t. \label{timestep}
\end{equation}
In fact the modified evolution equation (\ref{newevol}) is
closer to this than (\ref{evol}), and would be correct for 
the kinetic terms in the $n \rightarrow \infty$ limit. The 
gauge potential will appear automatically exponentiated from the 
appearance of $U_t^{\dag}$. The only correction that then needs 
to be made at the next order is to correct for a $H_0^2/n$ term 
that appears when (\ref{newevol}) is expanded out and 
compared to (\ref{timestep}) for $H = H_0$. This correction 
can be made by replacing $H_0$ by 
\begin{equation}
\tilde{H_0} = H_0 - \frac {a H_0^2} {4n}.
\label{h0tilde}
\end{equation}

The discretisation corrections discussed here, 
when added in to $\delta H$ look a lot like relativistic 
corrections. We can apply the same power counting arguments
as before to get an idea of their relative size. 
From (\ref{h0tilde}) we will have a correction of 
$\cal{O}$$(am_Q^2v^4)$. Relative to $H_0$ this is 
$\cal{O}$$(am_Qv^2)$, and for $am_Q \sim 1$ this is 
$\cal{O}$$(v^2)$, the same as the relativistic corrections. 
Similarly for the term from (\ref{dtilde}). Thus it is 
only sensible to correct for the discretisation corrections
up to an order comparable with the order of relativistic 
corrections being included. For the Lagrangian of 
\ref{nrqcd} which has the first spin-independent relativistic
corrections we need only the first spin-independent 
discretisation corrections described above. 
It might be true on coarse lattices, with $m_Qa > 1$, that
higher order discretisation corrections should be kept. 
This can be decided by using potential model
expectation values to better estimate their size (\cite{potmod}). 

There are additional $\cal{O}$$(a^2)$ errors from the 
gluon fields that appear in all the covariant derivatives 
coupling to the heavy quarks, if the gluon fields have 
been generated using the standard Wilson gluon action (\cite
{weingarten}, \cite{montvay}). These errors can be 
treated perturbatively and corrected for at the end of 
the calculation provided they are small (\cite{alpha}). 

The coefficients of the additional terms introduced by 
the discretisation corrections in (\ref{dtilde}) and 
(\ref{h0tilde}) must again be matched to full 
QCD. As before, 
they should be tadpole-improved to remove the largest part of 
the renormalisation of lattice NRQCD, and remaining renormalisations can be 
calculated in lattice perturbation theory. The improvement 
from (\ref{h0tilde}) can be added directly to the 
existing relativistic correction to give the operator
\begin{equation}
\delta H_1 = c_1 \frac {a^4(\vec{D}^2)^2} {8 m_Q^3 a^3} \left( 1 + \frac {m_Qa} {2n} \right).
\end{equation}
The $\cal{O}$$(\alpha_s)$ corrections to $c_1$ were discussed
above and are shown in Figure \ref{colin}. The improvement from 
(\ref{dtilde}) gives
\begin{equation}
\delta H_5 = c_5 \frac {\sum_i a^4 D_i^4} {24 m_Qa}.
\end{equation}
The $\cal{O}$$(\alpha_s)$ corrections to $c_5$ are also shown 
in Figure \ref{colin} and they are confirmed to be 
small after tadpole-improvement (see \cite{morning} and note that 
$c_5$ is there called $c_2$). 

Once the lattice NRQCD Lagrangian (including discretisation corrections)
has been chosen and the quark propagator $G_t$ calculated for a 
given gluon field configuration, then $G_t$ and the anti-quark 
propagator $G_t^{\dag}$ can be put together to make mesons. 
This procedure is identical to that used in lattice calculations 
of the light hadron spectrum. The only difference is that the meson 
operator :
\begin{equation}
\psi^{\dag A}(\vec{x_1})\Omega\phi(\vec{x_1}-\vec{x_2})\chi^{\dag A}(\vec{x_2})
\label{meson}
\end{equation}
has a spin part, $\Omega$, which is only a $2\times 2$ matrix, rather 
than the relativistic $4\times 4$. $\Omega$ is the unit matrix for 
$S=0$ mesons and a Pauli matrix for $S=1$. In addition we have a much better
idea from potential models of what form the spatial operator should take,
than we do in light hadron calculations. This will be discussed below. 
$\psi^{\dag}$ and $\chi^{\dag}$ are the quark and anti-quark 
creation operators respectively, matched in colour, denoted $A$, for 
a colour singlet. 
 
Then the meson correlation function is calculated as
an average over the ensemble of gluon field configurations
(\cite{weingarten}, \cite{montvay}):
\begin{eqnarray}
\langle (\chi \phi^{\dag} \Omega^{\dag} \psi)_T 
(\psi^{\dag} \Omega \phi \chi^{\dag})_0 \rangle 
&=& \langle Tr [ G \Omega^{\dag} \phi^{\dag} G^{\dag} \phi \Omega ] \rangle
\label{energies} \\
& \stackrel { {\scriptsize T \rightarrow \infty}} {\rightarrow} & 
\Phi_1 e^{-E_1T} + \Phi_2 e^{-E_2T} + \ldots . \nonumber
\end{eqnarray}
$E_1$ and $E_2$ are the energies of states in lattice units, $E_{{\rm phys}}a$.
$E_1$ is the energy of the ground state in that $\Omega, \phi$ 
channel, and $E_2$ is the energy of the first radial excitation etc. 
We can project onto different meson momenta at the annihilation time  
point, $T$, to obtain the dispersion relation, $E$ as a function
of $p$ (\cite{ups}). 

Because it is very important in heavyonium physics to calculate 
radial excitation energies, we need to optimise the calculation 
of $E_1$ and $E_2$. The coefficients $\Phi_1$ and $\Phi_2$ in (\ref{energies})
represent the overlap of the mesonic operator used with that state, 
$\Phi_i = \langle 0 | \psi^{\dag} \Omega \phi \chi^{\dag} | i \rangle$, 
see equation \ref{wloopv}. We can then  
adjust $\Phi_1$ and $\Phi_2$ by changing $\phi$ in the mesonic operator.  
For $S$ states we could use an operator in which both 
$\psi$ and $\chi$ appear at the same point but this would have 
overlap with all possible excitations and a poor convergence to 
the ground state. Instead we must $\psi$ and $\chi$ at separated 
points with $\phi$ a `smearing function'.  
Potential model wavefunctions represent a good first approximation 
to the spatial distribution of quark and anti-quark in heavyonium
(\cite{ups}).
To make use of these wavefunctions on the lattice (specifically 
to set $\phi$ equal to the wavefunction) requires us to 
fix a gauge otherwise the meson operator will not be gauge-invariant 
and will vanish in the ensemble average. The best gauge to use 
is Coulomb gauge since this (being the 3-dimensional version of 
the lattice Landau gauge discussed earlier) is the gauge in 
which the spatial gluon field is minimised, and the covariant squared spatial
derivative most like the Schr\"{o}dinger $\vec{p}^2$. 
We then gauge transform the lattice gluon fields to Coulomb 
gauge and use different $\phi$ as simple functions of 
spatial separation for different 
radial and orbital excitations.
For the ground state $\phi$ should 
maximise $\Phi_1$ and minimise $\Phi_2$, $\Phi_3$ etc, and for excited states 
$\phi$ should maximise $\Phi_2$ and minimise $\Phi_1$, $\Phi_3$ etc.   
For each $\phi$ a new quark propagator must be calculated 
in this approach since the fastest way to make the meson 
operator at the initial time slice is to use $\phi(\vec{x})$ as a 
source for the evolution equation for $G_t$. This $G_t$ is 
then combined with a $G_t^{\dag}$ in which a delta 
function at the spatial origin was used as the source. The same source
$\phi$ can be used for the different spin states (to the 
extent that spin-dependent effects on the wavefunction
are relativistic corrections and therefore small) and 
the factors of $\Omega$ at the initial time slice inserted
as $G_t$ and $G_t^{\dag}$ are being combined. 
In this way all $^{2S+1}L_J$ states can be made with as
many radial excitations as required (\cite{ups}). 

It is important to realise that the results for $E_1$ and 
$E_2$ are not affected by the choice of $\phi$; they can 
simply be obtained more efficiently by good choices. 
Methods other than that above have also been used; these
include building meson operators out of a quark and 
anti-quark joined by a string of gauge fields in a 
gauge-invariant way (\cite{manke2}); and calculating 
the propagators from delta function sources at a number of 
spatial points at the initial time and working out 
the optimal $\phi$ at the end using a variational 
method (\cite{mcneile}).  

However good the choice of $\phi$, each meson correlation 
function will contain several exponentials and a 
multi-exponential fit must be performed to extract them. 
This is described with technical details in \cite{ups}.
In general the $n$th exponential is obtained reliably 
from an ${n+1}$-exponential fit. 
In a potential model approach to the spectrum, using orthogonal wavefunctions, 
it is easy to get very precise results for radially excited 
states. In lattice NRQCD it is much harder because the 
ground state will take over exponentially if it is present 
at all in an excited meson correlation function. In addition 
the variance of such a correlation function will be 
dominated by the ground state so that the ratio of the 
signal for the 
excited state compared to noise will fall exponentially
(see, for example \cite{noise}). 

The fits to the zero momentum meson correlation functions 
yield a very accurate set of energies in lattice units but 
these cannot be immediately converted to absolute energies
(although splittings can)
because the zero of energy has been reset by the absence of 
the mass term in equation \ref{nrqcd}. To calculate 
the spectrum we must shift all the lattice energies by a 
constant and then convert to physical units by 
multiplication with $a^{-1}$.

Determining the lattice 
spacing is actually easier in heavyonium than for light 
hadrons. We can make use of the fact, stated before, 
that the radial and orbital splittings are independent 
to a very good approximation of the heavy quark mass. 
This means that we can use one of these splittings, 
e.g. the $1\overline{P}-1\overline{S}$ splitting, to determine $a^{-1}$,
without having necessarily tuned our heavy quark mass 
very well.
In the absence of an experimentally determined spin-average
$S$ state mass for bottomonium we set
\begin{equation}
a^{-1} = \frac {0.44} {aE(\overline{\chi_b}) -
aE(\Upsilon)} {\rm GeV}.
\end{equation}
where the denominator is the difference between the 
lattice energies at zero momentum of the spin 
average of the ground $\chi_b$ states and the $\Upsilon$. 
Given $a^{-1}$ we can now convert all differences in 
energy to splittings to GeV. We can also use 
$E(\Upsilon^{\prime}) - E(\Upsilon)$ to set $a^{-1}$. 

To
tune the bare lattice heavy quark mass, $m_Q$, to the 
appropriate value for the $b$ quark we study the 
dispersion relation for mesons at finite and small momenta,
where the heavy mesons are non-relativistic. 
The absolute meson mass (e.g. for the $\Upsilon$) is given 
not by the energy 
at zero momentum but by the denominator of the kinetic 
energy term :
\begin{equation}
aE_{\Upsilon}(p) = a E_{\Upsilon}(0) + \frac {a^2p^2} {2 a M_{\Upsilon}}
+ \ldots.  
\label{disp}
\end{equation}
Higher order relativistic corrections can also be added to 
this formula.  We adjust $m_Qa$ in the Lagrangian until 
the $\Upsilon$ mass comes out at 9.46 GeV within statistical 
errors, using the $a^{-1}$ determined from the splitting 
above. Now it is clear that if the splittings used for 
determining $a^{-1}$ did depend strongly on $m_Qa$ this 
would be a tricky iterative procedure. It would require 
complete calculations at several different values of 
$m_Qa$, as is generally undertaken in light hadron 
calculations.   

This procedure gives us also the shift of the zero of energy,
 $aM_{\Upsilon} - aE_{\Upsilon}(0)$.
It should be independent of the meson studied 
and so, once calculated, can 
be applied to all mesons. That is, when divided by 2, it 
can be applied as a shift per quark.
This is what allows us to convert
differences in zero momentum energies on the lattice
 directly to splittings in physical
units, given $a^{-1}$. 

The shift obtained can be compared to that calculated in 
lattice perturbation theory from the 
heavy quark self-energy. The 
energy shift is given in lattice units by
\begin{equation}
2 (Z_m m_Qa - E_0a)
\label{pertshift}
\end{equation}
where $Z_m$ is the mass renormalisation. $Z_m$ and $E_0a$ 
are given by perturbative expansions, $Z_m = 1 + \alpha_s B + \ldots$
and $E_0a = \alpha_s A + \ldots$. Again it is clear that $A$ and 
$B$ are smaller when a tadpole-improved lattice Lagrangian 
is used. (see \cite{morning} and Figure \ref{colin}).  
The shifts obtained on the lattice agree well 
with the perturbative estimates (\cite{ups}, \cite{scaling})
when a physical scheme for the lattice coupling 
constant is used (\cite{lep-mack}) and allowance is 
made for unknown higher order terms. See Table \ref{shift}.  

\begin{table}
\begin{center}
\begin{tabular}{c|c|c|c}
$\beta$ & $m_Qa$ & Perturbative shift & Non-perturbative shift \\
\hline 
5.7 & 3.15 & 7.0(6) & 6.54(7) \\
6.0 & 1.71 & 3.5(2) & 3.49(3) \\
6.2 & 1.22 & 2.5(2) & 2.58(3) \\
\end{tabular}
\caption[hfdjgk]{Energy shifts for a heavy quark in lattice NRQCD. Results
are given for the non-perturbative lattice calculation 
of $aM_{\Upsilon} - a E_{\Upsilon}(0)$ and for the 
perturbative shift of equation \ref{pertshift} for three
different values of the lattice spacing, set by $\beta$, and 
for bare quark masses appropriate to the $b$. Errors in the 
perturbative shifts are estimates of unknown higher 
order corrections. (\cite{scaling}). }
\label{shift}
\end{center}
\end{table}

Note that if we take $m_Qa$ to $\infty$ in this 
calculation $aE_0$ will become $V_c/2$ where $V_c$ 
is the unphysical self-energy part of the heavy quark 
potential discussed earlier. Again agreement between 
perturbation theory (\cite{morning}, \cite{eichten-fb}) 
and potential model results (\cite{bali-schilling})
is reasonable given that $aE_0$ starts at $\cal{O}$$(\alpha_s)$
and is only known to this order.
The effect of tadpole-improvement is easy to work out in this 
case. If the Wilson loops were calculated with tadpole-improvement 
(not usually done) then each $U_{\mu}$ would be divided by 
$u_0$ and the loop would pick up a factor $(u_0)^{-2T}$ from 
links in the time direction. Then $V_c \rightarrow V_c +2\ln u_0$. 
Thus to compare the perturbative value of 
$aE_0(m_Qa \rightarrow \infty)$ calculated {\it with} 
tadpole-improvement to the non-perturbative values of $V_c/2$ 
calculated {\it without} tadpole-improvement we must subtract 
$\ln u_0$ from the perturbative calculation.   
It is also true that $B$ for $m_Qa \rightarrow \infty$ vanishes
at $\cal{O}$$(\alpha_s)$ so $Z_m \rightarrow 1$. 
Since the dispersion relation for mesons at finite momentum 
cannot be obtained from potential model approaches, the 
quark mass there has to be fixed in a different way to 
the direct NRQCD method above. If the energies of states
are calculated using the lattice potential including 
$V_c$, $m_Qa$ should be adjusted until
experiment is matched for, say, the $\Upsilon$ on 
applying a shift $2m_Qa - V_c$ (\cite{bali}).

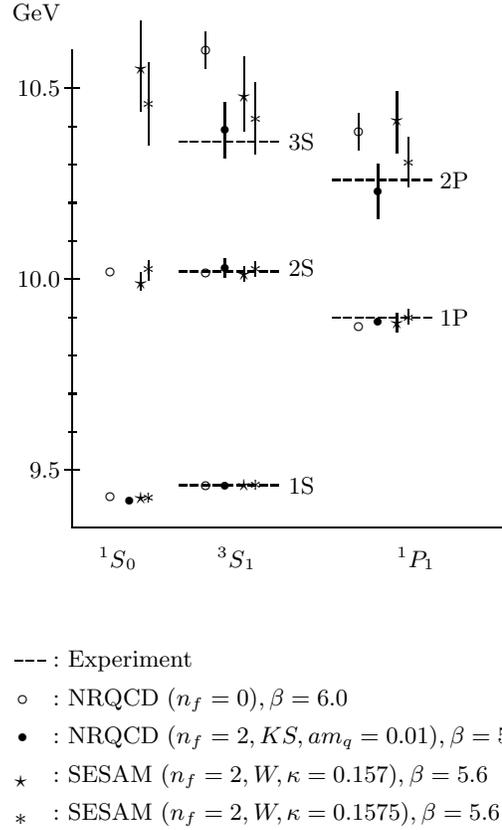
\begin{figure}[h]
\begin{center}
\setlength{\unitlength}{.02in}
\begin{picture}(130,200)(10,860)
\put(15,935){\line(0,1){125}}
\multiput(13,950)(0,50){3}{\line(1,0){4}}
\multiput(14,950)(0,10){10}{\line(1,0){2}}
\put(12,950){\makebox(0,0)[r]{9.5}}
\put(12,1000){\makebox(0,0)[r]{10.0}}
\put(12,1050){\makebox(0,0)[r]{10.5}}
\put(12,1070){\makebox(0,0)[r]{GeV}}
\put(15,935){\line(1,0){115}}


\put(27,930){\makebox(0,0)[t]{${^1S}_0$}}

\put(25,943.1){\circle{2}}
\put(25,1002){\circle{2}}
\put(30,942){\circle*{2}}

\put(33,942.7){\makebox(0,0){$\star$}}
\put(33,942.7){\line(0,1){0.1}}
\put(33,942.7){\line(0,-1){0.1}}
\put(33,999.1){\makebox(0,0){$\star$}}
\put(33,999.1){\line(0,1){2.5}}
\put(33,999.1){\line(0,-1){1.9}}
\put(33,1055.1){\makebox(0,0){$\star$}}
\put(33,1055.1){\line(0,1){12.5}}
\put(33,1055.1){\line(0,-1){11.2}}

\put(35,942.6){\makebox(0,0){$\ast$}}
\put(35,942.6){\line(0,1){0.2}}
\put(35,942.6){\line(0,-1){0.1}}
\put(35,1002.6){\makebox(0,0){$\ast$}}
\put(35,1002.6){\line(0,1){2.4}}
\put(35,1002.6){\line(0,-1){2.8}}
\put(35,1045.7){\makebox(0,0){$\ast$}}
\put(35,1045.7){\line(0,1){11.1}}
\put(35,1045.7){\line(0,-1){10.6}}


\put(58,930){\makebox(0,0)[t]{${^3S}_1$}}
\put(75,946){\makebox(0,0){{\small 1S}}}
\multiput(43,946)(3,0){9}{\line(1,0){2}}
\put(75,1003){\makebox(0,0){{\small 2S}}}
\multiput(43,1002)(3,0){9}{\line(1,0){2}}
\put(75,1036){\makebox(0,0){{\small 3S}}}
\multiput(43,1036)(3,0){9}{\line(1,0){2}}

\put(50,946){\circle{2}}
\put(50,1001.7){\circle{2}}
\put(50,1060){\circle{2}}
\put(50,1060){\line(0,1){4.8}}
\put(50,1060){\line(0,-1){4.8}}

\put(55,946){\circle*{2}}
\put(55,1003){\circle*{2}}
\put(55,1004){\line(0,1){1.4}}
\put(55,1002){\line(0,-1){1.4}}
\put(55,1039.1){\circle*{2}}
\put(55,1039.1){\line(0,1){7.2}}
\put(55,1039.1){\line(0,-1){7.2}}

\put(60,946){\makebox(0,0){$\star$}}
\put(60,1001.3){\makebox(0,0){$\star$}}
\put(60,1001.3){\line(0,1){2.1}}
\put(60,1001.3){\line(0,-1){1.9}}
\put(60,1048.0){\makebox(0,0){$\star$}}
\put(60,1048.0){\line(0,1){10.2}}
\put(60,1048.0){\line(0,-1){9.1}}

\put(63,946){\makebox(0,0){$\ast$}}
\put(63,1002.6){\makebox(0,0){$\ast$}}
\put(63,1002.6){\line(0,1){1.9}}
\put(63,1002.6){\line(0,-1){1.8}}
\put(63,1041.8){\makebox(0,0){$\ast$}}
\put(63,1041.8){\line(0,1){9.6}}
\put(63,1041.8){\line(0,-1){9.1}}


\put(105,930){\makebox(0,0)[t]{$^1P_1$}}

\put(115,990){\makebox(0,0){{\small 1P}}}
\multiput(83,990)(3,0){9}{\line(1,0){2}}
\put(115,1026){\makebox(0,0){{\small 2P}}}
\multiput(83,1026)(3,0){9}{\line(1,0){2}}

\put(90,987.6){\circle{2}}
\put(90,1038.7){\circle{2}}
\put(90,1038.7){\line(0,1){4.8}}
\put(90,1038.7){\line(0,-1){4.8}}

\put(95,989){\circle*{2}}
\put(95,1023){\circle*{2}}
\put(95,1023){\line(0,1){7.2}}
\put(95,1023){\line(0,-1){7.2}}

\put(100, 988.7){\makebox(0,0){$\star$}}
\put(100, 988.7){\line(0,1){2.3}}
\put(100, 988.7){\line(0,-1){2.4}}
\put(100,1041.7){\makebox(0,0){$\star$}}
\put(100,1041.7){\line(0,1){7.4}}
\put(100,1041.7){\line(0,-1){8.6}}

\put(103, 989.8){\makebox(0,0){$\ast$}}
\put(103, 989.8){\line(0,1){2.4}}
\put(103, 989.8){\line(0,-1){1.4}}
\put(103,1030.5){\makebox(0,0){$\ast$}}
\put(103,1030.5){\line(0,1){6.7}}
\put(103,1030.5){\line(0,-1){6.2}}

\multiput(0,900)(3,0){3}{\line(1,0){2}}
\put(10,900){\makebox(0,0)[l]{: {Experiment}}}
\put(2,890){\circle{2}}
\put(10,890){\makebox(0,0)[l]{: {NRQCD $(n_f = 0) , \beta =
      6.0 $}}}
\put(2,880){\circle*{2}}
\put(10,880){\makebox(0,0)[l]{: {NRQCD $(n_f = 2, KS, am_q = 0.01) , 
        \beta = 5.6 $}}}
\put(0,870){\makebox(0,0)[tl]{$\star$}}
\put(10,870){\makebox(0,0)[l]{: {SESAM $(n_f = 2, W, \kappa = 0.157), 
        \beta = 5.6 $}}}
\put(0,860){\makebox(0,0)[tl]{$\ast$}}
\put(10,860){\makebox(0,0)[l]{: {SESAM $(n_f = 2, W, \kappa = 0.1575), 
        \beta = 5.6 $}}}

\end{picture}
\hspace{1.5cm}
\end{center}
\caption[qwqw]{Radial and orbital excitation energies in
the $\Upsilon$ spectrum from lattice NRQCD obtained 
by different groups (\cite{newalpha}, \cite{spitzpub}). 
Errors shown are statistical only. The scale has been set 
by an average of $a^{-1}$ from the $2S-1S$ splitting 
and that from the $1P-1S$.  
Experimental results are given by dashed lines 
and for the $^1P_1$ state is taken as the spin-average
of the $\chi_b$ states. }
\label{sispect}
\end{figure}

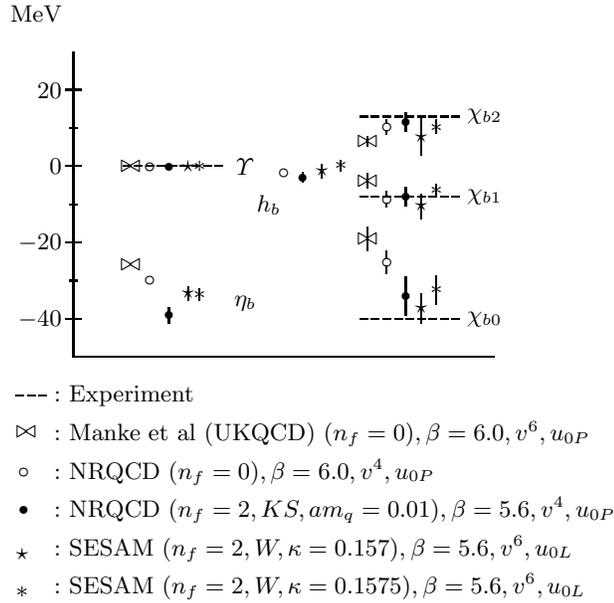
\begin{figure}[h]
\begin{center}
\setlength{\unitlength}{.02in}
\begin{picture}(100,140)(15,-110)
\put(15,-50){\line(0,1){80}}
\multiput(13,-40)(0,20){4}{\line(1,0){4}}
\multiput(14,-40)(0,10){7}{\line(1,0){2}}
\put(12,-40){\makebox(0,0)[r]{$-40$}}
\put(12,-20){\makebox(0,0)[r]{$-20$}}
\put(12,0){\makebox(0,0)[r]{$0$}}
\put(12,20){\makebox(0,0)[r]{$20$}}
\put(12,40){\makebox(0,0)[r]{MeV}}
\put(15,-50){\line(1,0){110}}

\multiput(28,0)(3,0){9}{\line(1,0){2}}
\put(60,2){\makebox(0,0)[t]{$\Upsilon$}}
\put(60,-34){\makebox(0,0)[t]{$\eta_b$}}

\put(35,0){\circle{2}}
\put(35,-29.9){\circle{2}}

\put(30,0){\makebox(0,0){$\bowtie$}}
\put(30,-25.7){\makebox(0,0){$\bowtie$}}

\put(40,0){\circle*{2}}
\put(40,-39){\circle*{2}}
\put(40,-39){\line(0,1){2}}
\put(40,-39){\line(0,-1){2}}

\put(45,0){\makebox(0,0){$\star$}}
\put(45, -33.2){\makebox(0,0){$\star$}}
\put(45, -33.2){\line(0,1){1.5}}
\put(45, -33.2){\line(0,-1){1.9}}

\put(48,0){\makebox(0,0){$\ast$}}
\put(48, -33.7){\makebox(0,0){$\ast$}}
\put(48, -33.7){\line(0,1){1.6}}
\put(48, -33.7){\line(0,-1){1.5}}

\put(63,-10){\makebox(0,0)[l]{$h_b$}}
\put(70,-1.8){\circle{2}}
\put(75,-2.9){\circle*{2}}
\put(75,-2.9){\line(0,1){1.2}}
\put(75,-2.9){\line(0,-1){1.2}}


\put(80,  -1.0){\makebox(0,0){$\star$}}
\put(80,  -1.0){\line(0,1){1.5}}
\put(80,  -1.0){\line(0,-1){2.2}}
\put(85,   0.3){\makebox(0,0){$\ast$}}
\put(85,   0.3){\line(0,1){1.2}}
\put(85,   0.3){\line(0,-1){1.5}}

\multiput(90,-40)(3,0){9}{\line(1,0){2}}
\put(118,-40){\makebox(0,0)[l]{$\chi_{b0}$}}
\multiput(90,-8)(3,0){9}{\line(1,0){2}}
\put(118,-8){\makebox(0,0)[l]{$\chi_{b1}$}}
\multiput(90,13)(3,0){9}{\line(1,0){2}}
\put(118,13){\makebox(0,0)[l]{$\chi_{b2}$}}

\put(97,-25.1){\circle{2}}
\put(97,-26.1){\line(0,-1){2}}
\put(97,-24.1){\line(0,1){2}}
\put(97,-8.6){\circle{2}}
\put(97,-7.6){\line(0,1){1}}
\put(97,-9.6){\line(0,-1){1}}
\put(97,10.2){\circle{2}}
\put(97,11.2){\line(0,1){1}}
\put(97,9.2){\line(0,-1){1}}

\put(102,-34){\circle*{2}}
\put(102,-34){\line(0,1){5}}
\put(102,-34){\line(0,-1){5}}
\put(102,-7.9){\circle*{2}}
\put(102,-7.9){\line(0,1){2.4}}
\put(102,-7.9){\line(0,-1){2.4}}
\put(102,11.5){\circle*{2}}
\put(102,11.5){\line(0,1){2.4}}
\put(102,11.5){\line(0,-1){2.4}}

\put(92,-19){\makebox(0,0){$\bowtie$}}
\put(92,-19){\line(0,1){3}}
\put(92,-19){\line(0,-1){3}}
\put(92,-3.8){\makebox(0,0){$\bowtie$}}
\put(92,-3.8){\line(0,1){2}}
\put(92,-3.8){\line(0,-1){2}}
\put(92,6.5){\makebox(0,0){$\bowtie$}}
\put(92,6.5){\line(0,1){1.3}}
\put(92,6.5){\line(0,-1){1.3}}

\put(106, -36.9){\makebox(0,0){$\star$}}
\put(106, -36.9){\line(0,1){3.5}}
\put(106, -36.9){\line(0,-1){4.2}}
\put(106, -10.1){\makebox(0,0){$\star$}}
\put(106, -10.1){\line(0,1){2.9}}
\put(106, -10.1){\line(0,-1){3.7}}
\put(106,   7.8){\makebox(0,0){$\star$}}
\put(106,   7.8){\line(0,1){4.7}}
\put(106,   7.8){\line(0,-1){4.8}}

\put(110, -32.3){\makebox(0,0){$\ast$}}
\put(110, -32.3){\line(0,1){3.6}}
\put(110, -32.3){\line(0,-1){3.8}}
\put(110,  -6.3){\makebox(0,0){$\ast$}}
\put(110,  -6.3){\line(0,1){1.5}}
\put(110,  -6.3){\line(0,-1){1.7}}
\put(110,  10.2){\makebox(0,0){$\ast$}}
\put(110,  10.2){\line(0,1){1.9}}
\put(110,  10.2){\line(0,-1){1.5}}

\multiput(0,-60)(3,0){3}{\line(1,0){2}}
\put(10,-60){\makebox(0,0)[l]{: {Experiment}}}
\put(0,-68){\makebox(0,0)[tl]{$\bowtie$}}
\put(10,-70){\makebox(0,0)[l]{: {Manke et al (UKQCD)  $(n_f = 0), 
         \beta = 6.0 , v^6 , u_{0P}$}}}
\put(2,-80){\circle{2}}
\put(10,-80){\makebox(0,0)[l]{: {NRQCD $(n_f = 0) , \beta =
      6.0, v^4, u_{0P}$ }}}
\put(2,-90){\circle*{2}}
\put(10,-90){\makebox(0,0)[l]{: {NRQCD $(n_f = 2, KS, am_q=0.01) ,
         \beta = 5.6 , v^4 , u_{0P}$}}}
\put(0,-100){\makebox(0,0)[tl]{$\star$}}
\put(10,-100){\makebox(0,0)[l]{: {SESAM $(n_f = 2, W, \kappa = 0.157), 
         \beta = 5.6 , v^6 , u_{0L}$}}}
\put(0,-110){\makebox(0,0)[tl]{$\ast$}}
\put(10,-110){\makebox(0,0)[l]{: {SESAM $(n_f = 2, W, \kappa = 0.1575), 
         \beta = 5.6 , v^6 , u_{0L}$}}}

\end{picture}
\end{center}
\caption[peo]{Fine structure splittings in 
the ground state $\Upsilon$ spectrum from lattice NRQCD obtained 
by different groups (\cite{manke2}, \cite{newalpha}, \cite{spitzpub}). 
Errors shown are statistical only. 
The scale is set as in Figure \ref{sispect}. 
Experimental results are given by dashed lines. 
The spin-average of the $\chi_b$ states has been set
to zero. 
}
\label{sdspect}
\end{figure}

Figures \ref{sispect} and \ref{sdspect} show recent results 
for the bottomonium spectrum from lattice NRQCD. 
The errors shown on the plot are statistical errors only - it
is clear that they are significantly smaller than those from 
light hadron calculations. The simple form of the 
evolution equation for calculating the heavy quark propagator
means that an average over a very large ensemble of gluon fields 
can be obtained with moderate computing cost. Also several 
different starting points can be used on a single gluon field
configuration. 

The sources of systematic error are also under better control 
than for light hadron calculations. There, one of the most serious 
problems is that of finite volume. A large enough lattice 
is required not to squeeze the mesons under study and 
distort their masses. Because  
the $\Upsilon$ is much 
smaller than, say, the $\rho$, a smaller space-time box is
sufficient for its study. The calculations shown were 
done on lattices of around 1.5fm on a side.
However, radial and orbital excitations 
are larger than ground states and such a lattice may be too small for 
3$S$ and 2$P$ states. Studies on larger volumes should 
be done for these in future. In fact direct calculations of the 
spectrum have worse finite volume errors than calculations 
of the heavy quark potential, because of lattice symmetries
that protect $V(R)$ (\cite{potsym}).  

Discretisation errors are an additional source of systematic 
error and in all the results shown, the leading discretisation
corrections 
have been made as described above. 

Since NRQCD is a non-relativistic expansion, there are 
systematic errors from higher order relativistic terms that 
have been neglected. For the spin-independent terms all groups
have used the lattice-discretised version of the Lagrangian of 
equation \ref{nrqcd}. This includes leading terms of 
$\cal{O}$$(m_Qv^2)$ and corrections of $\cal{O}$$(m_Qv^4)$. 
Errors are therefore at $\cal{O}$$(m_Qv^6)$, giving 
$v^2\times$(a typical kinetic energy) =  0.01 x 400 MeV for 
bottomonium = $\sim$ 4 MeV. This error would be invisible 
on Figure \ref{sispect}, since it is a 1\% error 
in the splittings shown.  For the spin-dependent terms    
a 4 MeV error is more significant since splittings are 
smaller. This is the error estimate if only the leading spin-dependent 
terms of equation \ref{nrqcd} are used, as by the NRQCD 
collaboration (\cite{ups}). The SESAM group (results 
presented at this school by Achim Spitz - see \cite{spitz} and 
\cite{spitzpub}) however has 
used the additional relativistic corrections to the 
spin-dependent terms given in the continuum by (\cite{nakhleh}):
\begin{eqnarray}      
\delta H = &-&c_7 \frac {g} {8m_Q^3} \{ \vec{D}^2, \vec{\sigma}
\cdot \vec{B} \} \label{dhspin} \\
&-&c_8 \frac {3g} {64 m_Q^4} \{ \vec{D}^2, \vec{\sigma}
\cdot (\vec{D}\times\vec{E} - \vec{E}\times\vec{D}) \} \nonumber \\
&-&c_9 \frac {ig} {8m_Q^3} \vec{\sigma}\cdot\vec{E}\times\vec{E}. \nonumber
\end{eqnarray}
This could reduce the relativistic error by another factor of $v^2$
to around 1\% for spin splittings also. The SESAM group tadpole-improve
all the terms above using $u_{0L}$ and set the 
$c_i$ to 1. In principal, however, unknown radiative corrections 
from lower order terms, e.g. $\cal{O}$$(\alpha_s)$ corrections 
to $c_4$ beyond tadpole-improvement, can produce errors at the 
same order as the terms of (\ref{dhspin}) (if we take $\alpha_s \sim
v^2 \sim 0.1$, but see \cite{bodwin}) so this is not 
a complete calculation at the next order. Note that relativistic 
corrections to
spin-dependent terms are not known for a potential 
model.

Figure \ref{sispect} compares radial and orbital splittings 
to experiment. The lattice spacing chosen to set the scale is 
an average of that from the $1P-1S$ splitting and that from the 
$2S-1S$ splitting. One striking feature is the disagreement 
with experiment for the calculation on quenched configurations 
($n_f$ = 0). The results on the partially unquenched configurations 
($n_f$ = 2) give much better agreement. A quantity which 
exposes this error in the quenched approximation is the 
ratio of the $2S-1S$ splitting to that of the $1P-1S$. 
Figure \ref{scal} demonstrates that the fact that this ratio 
is too large is a physical effect - it is not affected substantially by 
lattice discretisation errors.  
We expect such an effect in the quenched approximation because 
$\alpha_s$ runs incorrectly between scales. This means, 
as discussed earlier, that the 
quenched heavy quark potential is too 
high at small values of $R$ and so the $S$ states are pushed
up with respect to $P$ states, making the $1P-1S$ splitting 
too small relative to the $2S-1S$.  The effect may be 
stronger for higher excitations, but they are subject to 
larger lattice errors. 

The error from the quenched approximation is even bigger if 
we look at quantities sensitive to a larger disparity of scales
to maximise the effect of incorrect running. Figure \ref{upsrho}
shows the ratio of the $1P-1S$ splitting in bottomonium 
from the NRQCD collaboration
to the $\rho$ mass from the UKQCD ({\cite{ukqcdhadrons}) 
and GF11 collaborations (\cite{gf11}, \cite{weingarten}). 
The UKQCD $\rho$ mass results have included discretisation 
corrections for the light quarks; the GF11 results have not. 
It is clear
that, although a result independent of lattice spacing is 
obtained when the $\rho$ mass is improved, it is wrong. 

\begin{figure}
\centerline{\ewxy{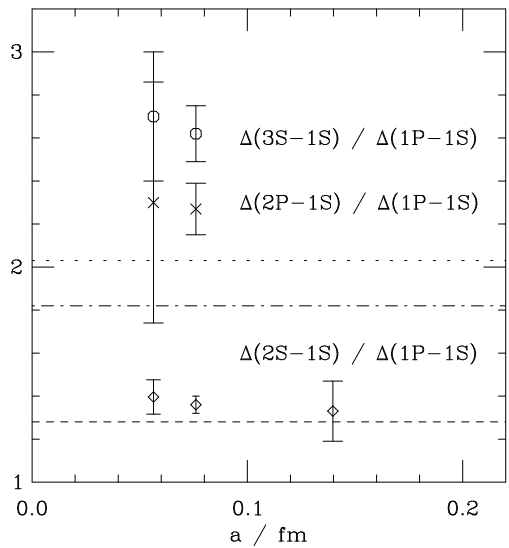}{100mm}}
\caption[hjk]{ Dimensionless ratios of various splittings 
to the $\overline{\chi_b} - \Upsilon$ splitting against the 
lattice spacing in fm, set by the $\overline{\chi_b} -
\Upsilon$ splitting, in the quenched approximation. 
Experimental values are 
indicated by lines. (\cite{scaling}). }
\label{scal}
\end{figure}

\begin{figure}
\centerline{\ewxy{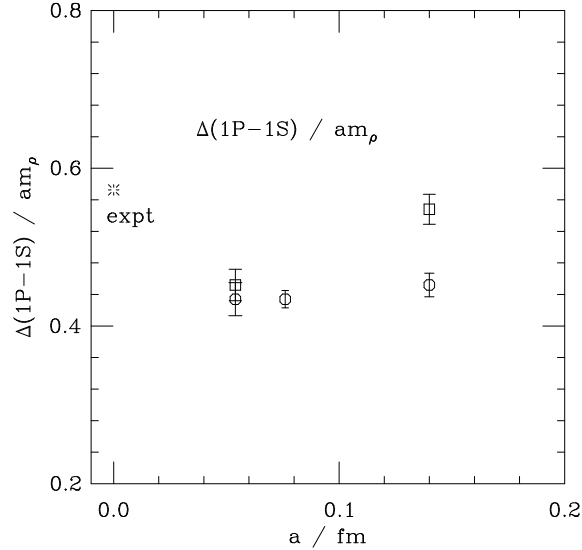}{100mm}}
\caption[hjk]{ Dimensionless ratio of 
the $\overline{\chi_b} - \Upsilon$ splitting to the 
$\rho$ mass against the 
lattice spacing in fm, set by the $\overline{\chi_b} -
\Upsilon$ splitting in the quenched approximation (\cite{scaling}). 
Circles show the 
UKQCD improved $\rho$ mass and the squares the 
GF11 unimproved mass. Experiment is shown by the burst. }
\label{upsrho}
\end{figure}

The fine structure in the spectrum is shown in Figure \ref{sdspect}. 
As already discussed, this is harder to calculate accurately than the spin-independent
spectrum because it only appears as a relativistic correction.  
There is no large leading order term with non-perturbatively
determined coefficient to stabilise the results as there 
is in the spin-independent case ($\vec{D}^2/2m_Q$).
Since the clover discretisation of $E$ and $B$ fields 
each contain four links (see figure \ref{clover}) there are 
several powers of $u_0$ in each term when tadpole-improvement 
is undertaken. This means that spin-splittings 
are affected strongly by the value of $u_0$ and if 
$u_0$ were set to 1 (i.e. no tadpole-improvement) results 
much smaller than experiment would be obtained (\cite{ups}). 
This also means that the spin splittings change when different
definitions of $u_0$ are used, in the absence of a 
perturbative calculation of the remaining radiative corrections 
(for preliminary results on these, see \cite{trottier-97}). 
The difference between $u_{0P}$ and $u_{0L}$ results in 
10-20\% shifts to the splittings at these lattice spacings
(\cite{spitzpub}). The hyperfine splitting is particularly 
sensitive; in leading order perturbation theory it is 
proportional to $c_4^2$ (equivalent to $u_0^{-6}$ when 
$u_0$ factors in the $\vec{D}^2$ term are taken into 
account). The presence or absence of the 
higher order relativistic corrections also affects the 
results at a similar level (\cite{manke2}, \cite{spitzpub}).  
Discretisation corrections, not surprisingly, are 
important as well. Because the fine structure (and particularly 
the hyperfine splitting) is sensitive to short-distance 
scales (consider the functional form of the spin-dependent
potentials), the discretisation errors can be quite 
severe. In the calculation of the NRQCD collaboration
in which only leading order spin terms are included, 
the hyperfine splitting in MeV shows strong dependence 
on the lattice spacing - see Figure \ref{hyprat}.
 This makes a physical result hard to 
determine. Results of the other groups in Figure \ref{sdspect} 
include, along with the relativistic spin-dependent 
corrections, discretisation corrections to the 
leading spin-dependent terms (\cite{nakhleh}). This should reduce 
the lattice spacing dependence of the physical results 
but this analysis is not yet complete (see Figure \ref{hyprat}
and \cite{manke}).   
Finally the spin splittings depend quite strongly on the 
quark mass (particularly again the hyperfine splitting)
and for these the quark mass must be tuned accurately. 
This requires a very accurate determination of the meson kinetic 
mass as well as of the lattice spacing. 

\begin{figure}
\centerline{\ewxy{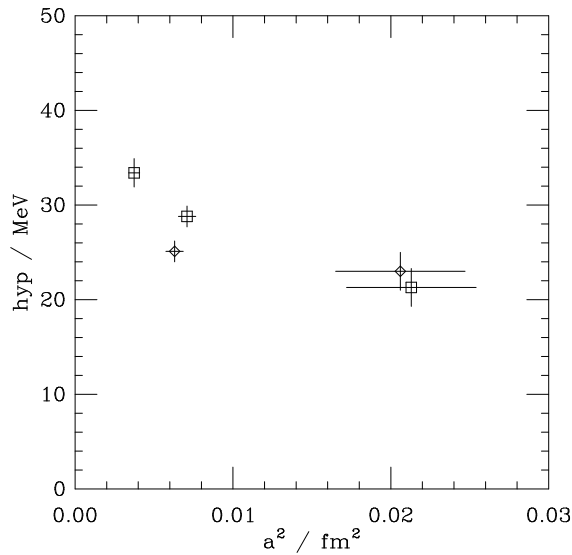}{100mm}}
\caption[hjk]{The $\Upsilon - \eta_b$ splitting in physical 
units vs the square of the lattice spacing in fm in the 
quenched approximation. 
The $2S-1S$ splitting has
been used to set the scale. Squares are results from 
the lowest order action (\cite{scaling}) and diamonds 
from an action which includes spin-dependent relativistic
and discretisation corrections (\cite{manke2}, \cite{manke}). }  
\label{hyprat}
\end{figure}

To compare to the real world we would like results with 
dynamical fermions of appropriate number and mass. To 
estimate how many dynamical fermions are `seen' by the 
bottomonium system, we need to know what the typical 
momenta being exchanged by the heavy quarks are. 
For $\Upsilon$ this $q_{\Upsilon}$ is about 1 GeV, not enough 
to make a $c\overline{c}$ pair, so the effective 
value of $n_f$ should be 3. Almost all available 
sets of gluon field configurations have $n_f$ 
set to 2, so extrapolation is necessary. 
The results will also depend on the light quark 
mass of the dynamical flavours, and this dependence 
at worst should be linear (\cite{grinstein}): 
\begin{equation}
\Delta M \sim \Delta M_0 \bigg( 1 + c \sum_{u,d,s} \frac {m_q}    
{q_{\Upsilon}} \cdots \bigg)
\label{qextrap}
\end{equation}
For $q_{\Upsilon} \gg m_q$ the answer with $u,d$ and $s$ quarks
can be reproduced by 3 degenerate dynamical quark flavours 
of mass $m_s/3$ (ignoring $m_u$ and $m_d$). 
Results should be extrapolated as a function of dynamical 
fermion mass to $m_s/3$ and then extrapolated to $n_f$ = 3
from $n_f$ = 0 and 2. 

In fact no significant $m_Q$ dependence has been seen 
by the two groups, NRQCD and SESAM, who have done calculations 
on dynamical configurations (using different lattice formulations
of the light fermions). The SESAM collaboration
with 3 dynamical quark masses does see a definite trend in $m_q$,
however (\cite{spitzpub}). 
Figure \ref{nfdep} shows the dependence of the ratio of 
the $2S-1S$ to $1P-1S$ splitting on $n_f$ for the two groups. The 
results are consistent with experiment for $n_f$ = 3 
but $n_f$ = 2 cannot be ruled out without better statistics. 
More points at other values of $n_f$ would be useful. 

\begin{figure}
\centerline{\ewxy{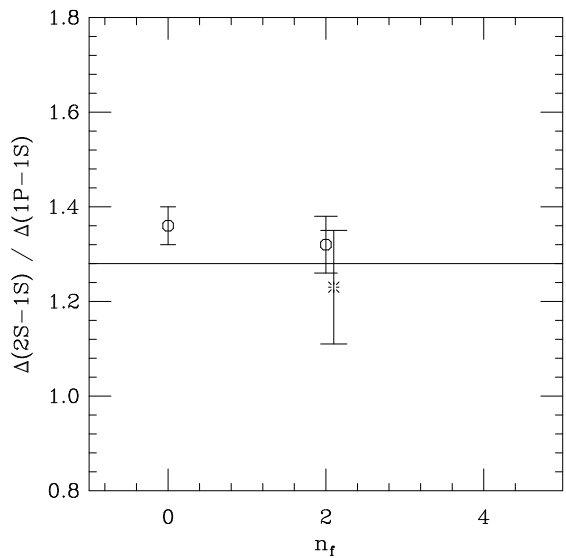}{100mm}}
\caption[hjk]{The ratio of the $\Upsilon^{'} - \Upsilon$ 
splitting to the $\overline{\chi_b} - \Upsilon$ splitting
as a function of the number of dynamical flavours. 
Circles are the results from the 
NRQCD collaboration  (\cite{newalpha})
and the burst from the SESAM collaboration (\cite{spitzpub}).}
\label{nfdep}
\end{figure}

The $n_f$ extrapolation of the fine structure will be 
more difficult, even once physical results at a 
given $n_f$ are obtained. Since the fine structure 
probes much shorter distances it is possible that 
the effective number of flavours that it `sees' is 
higher. Then the challenge will be to find 
appropriate quantities to set the scale for an 
extrapolation to, say, $n_f$ = 4. It will not be 
possible to use spin-independent splittings for which 
the real world $n_f$ value is 3 (\cite{scaling}). 

Figure \ref{sdspect} shows
disagreement with experiment for the 
$P$ fine structure in the quenched approximation, 
both in overall scale and for the ratio $\rho$, equation
\ref{rho}. Agreement is better on unquenched 
configurations, but the systematic errors
described above must be removed before this is clear.
The hyperfine splitting is very
sensitive to the presence of dynamical 
fermions. It increases by $\sim 30\%$ as 
$n_f$ is increased from 0 to 2 for the 
NRQCD results (see Figure \ref{sdspect}).
Extrapolations in $n_f$ using a variety of 
other short-distance quantities to set the 
scale (so the physical results differ 
from those in Figure \ref{sdspect}) 
give a `real world' value for the hyperfine 
splitting of around 40 MeV. The error is very large
at present (25\%) because of the  
inaccuracies in the fine structure. With 
improved calculations this can be reduced to 10\%.

\subsection{Direct measurement of the charmonium spectrum 
on the lattice} 

Unfortunately the NRQCD programme as described for bottomonium 
does not work as well for charmonium. It has been clear
all along that charmonium is much more relativistic; 
with the NRQCD approach we can directly see the effects of 
higher order relativistic corrections to the Lagrangian. 
A calculation with the Lagrangian of equation \ref{nrqcd}
has errors at $\cal{O}$$(m_Qv^6)$ as discussed earlier
(\cite{charm}).
This gives an error of around 30MeV which is 30\% 
for spin splittings. On adding higher order terms these
large corrections to the fine structure become manifest (\cite{trottier})
and are actually rather worse than the naive 30\%. 
An accurate calculation of the $\psi - \eta_c$ splitting, for 
example, would then require a high order in the NRQCD 
expansion, coupled with the determination of radiative 
corrections to the coefficients.    

It seems more useful to treat the $c$ quark as a 
light quark and use standard lattice approaches for 
relativistic quarks (\cite{weingarten}, \cite{montvay}). 
However, the fact that $m_ca \sim 1$ on current lattices
can lead to significant discretisation errors. The heavy 
Wilson approach (\cite{fnal}) is an adaption of 
the standard Wilson light fermion action in which 
higher dimension operators are added to better match 
to continuum QCD by reducing errors of the form 
$(p_Qa)^n$. The coefficients of these operators must
be calculated and in the strict heavy Wilson approach 
they are considered as a perturbative series in $\alpha_s$ 
but to all orders in $m_Qa$. At large $m_Qa$ the Lagrangian 
becomes NRQCD-like (since no symmetry between space 
and time directions is imposed) and at small $m_Qa$ it reduces to the 
form used for the Symanzik improvement of light quarks. In principle it can 
span the region from one extreme to the other. In practise, 
NRQCD is rather simpler and faster to implement for 
really non-relativistic fermions.  

The lowest order heavy Wilson action 
is identical to 
the Sheikosleslami-Wohlert (SW) action for light quarks in 
which a clover term 
\begin{equation}
\frac {ig} {2} c_sw \kappa \overline{\psi}(x) \sigma_{\mu\nu}
F^{\mu\nu} \psi(x)
\end{equation}
 with $m_Qa$-independent
coefficient is added to the Wilson action (\cite{sheik},
\cite{heatlie}):
\begin{equation}
S = \sum_n \overline{\psi}_n \psi_n - \kappa
\sum_{n, \mu} [ \overline{\psi}_n (1-\gamma_{\mu}) U_{n,\mu}
\psi_{n+\hat{\mu}} + \overline{\psi}_{n+\hat{\mu}}
(1+\gamma_{\mu}) U^{\dag}_{n,\mu} \psi_n ].
\end{equation}
The coefficient of the clover term 
is usually taken as 1 after the gauge fields have been 
tadpole-improved, and perturbative calculations 
to $\cal{O}$$(\alpha_s)$ ($m_Qa \rightarrow 0$) indicate no 
large additional radiative corrections (see \cite{luscher} for a 
discussion).
 Non-perturbative  
determinations of this clover coefficient for the 
light quark case are described by \cite{sommer-schlad}. 

The meson energy-momentum relation must also be 
considered carefully for charm systems (\cite{fnal}). For 
the SW action there is 
an energy shift between the energy at zero momentum 
(the pole mass)
and the kinetic meson mass that sits in the 
denominator of the kinetic term, as in the 
NRQCD case (equation \ref{disp}). The shift increases 
as the quark mass increases and for $m_Qa > 0.5$ 
it is important that the physical meson mass is taken 
from the kinetic mass and not from the pole mass
which is used for light hadrons. 
It is possible to remove this shift by adjusting 
coefficients in the full heavy Wilson approach, 
but it is not necessary. 

For the SW action
there is also a problem with non-universality of the shift. 
It should appear simply as a shift per quark and 
therefore twice as big for a meson with two heavy 
quarks as for a meson with one heavy and one 
light quark. However there is a discrepancy between the shift 
per heavy quark in these two cases and it increases 
significantly as the heavy quark mass increases 
(\cite{sara}, \cite{jlqcd}, see Figure \ref{mkin}). The discrepancy arises from 
lattice discretisation errors in relativistic $\vec{D}^4$-type
terms in the heavy quark action affecting the heavy-heavy
mesons. These terms need to be correct in order for the 
binding energy of a meson to be fed into its kinetic mass. Since 
the kinetic mass appears at $\cal{O}$$(m_Qv^2)$ it is 
$\cal{O}$$(m_Qv^4)$ terms which do this, whereas 
the pole mass is $\cal{O}$$(1)$. If the $\cal{O}$$(m_Qv^4)$
terms are incorrect, the binding energy will appear only in 
the pole mass and the shift will then depend on the 
binding energy. For a heavy-light meson the binding energy 
is provided by the light quark and this problem does not 
arise. In the NRQCD case (\cite{mb})
the relativistic terms are correct because the $\vec{D}^4/8m_Q^3$ 
term is added
by hand and discretisation corrections remove $D_i^4$ 
rotational non-invariance. For the  
SW action this isn't true; the $\vec{D}^4$ term has 
an incorrect mass and there is an uncancelled $D_i^4$ term
(\cite{kronfeld}). Both these effects can be corrected for 
in the heavy Wilson approach (\cite{fnal}), but this has 
not been done as yet.  For charm quarks with $m_ca < 0.5$ 
there is not a significant problem, as in clear from Figure \ref{mkin}.  

\begin{figure}
\centerline{\epsfig{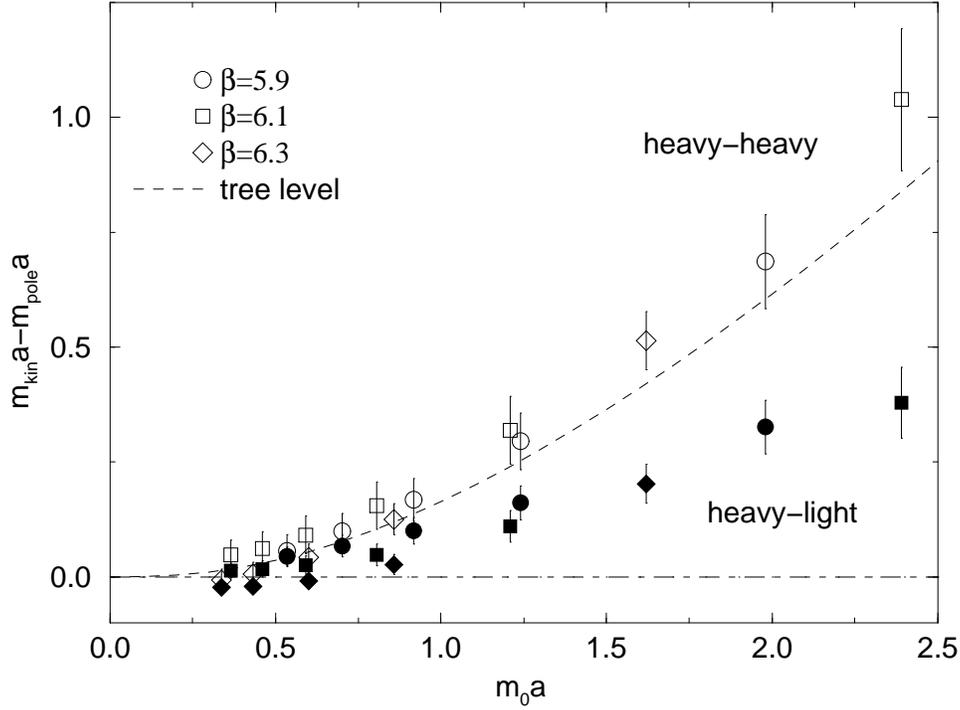}}
\caption[jgjlfk]{The energy shift per heavy quark for the SW action,
as a function of heavy quark mass, $m_0a$. 
Note the difference between results for heavy-heavy mesons (open
symbols) 
and those for heavy-light mesons (filled symbols). (\cite{jlqcd}). }
\label{mkin}
\end{figure}

The calculation of charm quark propagators with the SW action 
proceeds as for conventional light quarks (\cite{weingarten},
\cite{montvay}) with the calculation of rows of the fermion 
matrix inverse by an iterative procedure. This converges 
quite rapidly for heavy quarks, but care must be taken 
to allow enough iterations for the solution to 
propagate over the whole lattice. Meson correlation functions 
are put together using various smearing functions 
and then fitted to multi-exponential 
forms as described for the NRQCD case. The charm quark mass 
is tuned by the kinetic mass method as before.  
 
Figure \ref{peter} shows the spectrum for charmonium 
calculated recently on quenched gluon fields 
with this method and presented at this school 
by Peter Boyle (\cite{boyle}). Previous results (\cite{aida}) 
are in agreement with this, but don't show such 
complete fine structure. 
The hyperfine splitting is clearly underestimated and this could be a 
quenching error, since this splitting increases
with $n_f$, as discussed in the bottomonium case, 
or it could mean that $c_s$ is underestimated. 
The fine structure also shows a discrepancy 
for the ratio $\rho$ (see equation \ref{rho}). 

The systematic errors of the 
SW action for charmonium need some analysis (\cite{aida}) before  
we can extract physical (quenched) results from 
these calculations. Only one calculation on unquenched 
configurations has been done (\cite{sara}) and problems 
with fitting errors made it hard to draw conclusions. 
 
\begin{figure}
\input{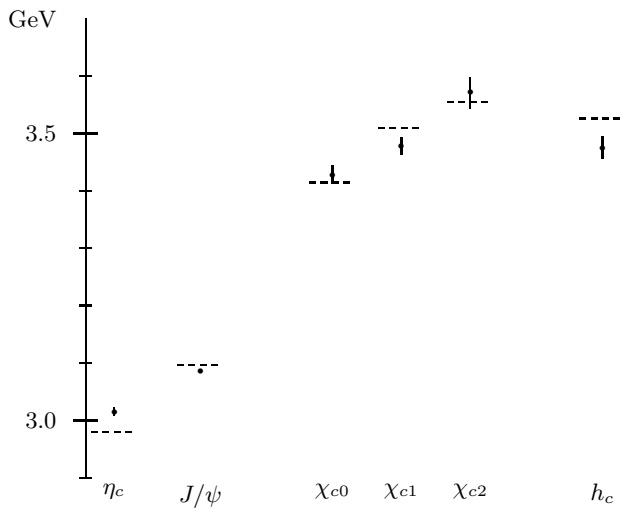}
\renewcommand{\yorg}{-2900}

\setlength{\unitlength}{.003in}
\begin{picture}(300,850)(-120,-10)

\spectrum{800}{3000}{3.0}{3500}{3.5}{3700}{GeV}

\state{\eta_c}{2980}{3015}{7}{7}{50}
\state{J/\psi}{3097}{3086}{2}{2}{200}
\state{\chi_{c0} }{3415}{3428}{15}{15}{430}
\state{\chi_{c1}}{3510}{3478}{14}{14}{550}
\state{\chi_{c2}}{3555}{3572}{26}{28}{670}
\state{h_{c}}{3526}{3475}{19}{19}{900}

\end{picture}
\caption[kdgf]{The charmonium spectrum from quenched lattice QCD using the 
SW action (\cite{boyle}).}
\label{peter}
\end{figure}

Future work will need to investigate the use of actions with higher order terms, 
possibly on anisotropic lattices (\cite{gpl-schladming}). Some preliminary 
work on the charmonium spectrum has been done with these (\cite{alford}).
 A small lattice spacing in the 
time direction is useful for improving exponential fits, particularly
for excited states, and does not need to mean a small lattice spacing 
in the spatial directions. Indeed such an anisotropy is very natural
for non-relativistic systems, as we have seen.

\subsection{Other heavy-heavy states}

There is a lot of interest in the literature in other 
heavy-heavy bound states, which have not yet been seen experimentally. 
The one most likely to be seen in the near future is the 
mixed bound state of bottom and charm quarks; indeed candidates
for the $^1S_0$ ground state, the $B_c$, have been seen 
recently (\cite{delphi}, \cite{aleph}).  

The $b\overline{c}$ system actually has a lot in common with 
the heavy-light systems of the next section, although it is 
classified here as heavy-heavy because of its quark content. 
The charm quark in the $B_c$ will be more tightly bound 
and therefore more relativistic than in charmonium, and 
we have already seen that a non-relativistic approach to 
charmonium is rather inaccurate. In addition, because charge
conjugation is not a good quantum number, the two $1^{+}$ $P$
states will mix to give a different $P$ fine structure to that for 
heavyonium.   

Recent continuum potential model results for the $b\overline{c}$ spectrum 
are given in \cite{equigg} and \cite{russians}. 2 sets of $S$ states,
1 set of $P$ states and 1 set of $D$ states are expected below threshold
for the Zweig allowed decay to $B, D$ (7.14 GeV). Note that the $b\overline{c}$ states
below threshold are particularly stable since the annihilation mode
to gluons is also forbidden.  

First lattice calculations (\cite{lidsey}) 
have used NRQCD for both the $c$ and $b$ quarks. 
Agreement with potential model results was found within 
sizeable systematic uncertainties. Better calculations will 
use NRQCD for the $b$ quark and relativistic formulations for 
the $c$ quark (\cite{shanahan}). However, uncertainties still remain about how to 
fix the bare quark masses in the quenched approximation, because 
of the variations possible in the determination of the scale. 
These problems should become more tractable when complete calculations 
are done 
including the effects of dynamical fermions. Only preliminary
results are available on unquenched configurations 
using NRQCD for $b$ and $c$ (\cite{gorbahn}). 
Figure \ref{fig:bc} shows a comparison of the spectrum for lattice 
and potential model results. 

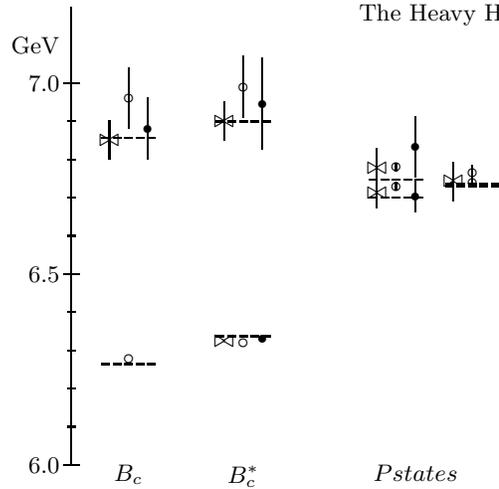
\begin{figure}
\begin{center}
\setlength{\unitlength}{.02in}
\begin{picture}(130,120)(10,580)
\put(15,600){\line(0,1){120}}
\multiput(13,600)(0,50){3}{\line(1,0){4}}
\multiput(14,600)(0,10){11}{\line(1,0){2}}
\put(12,600){\makebox(0,0)[r]{6.0}}
\put(12,650){\makebox(0,0)[r]{6.5}}
\put(12,700){\makebox(0,0)[r]{7.0}}
\put(12,710){\makebox(0,0)[r]{GeV}}

     \put(30,600){\makebox(0,0)[t]{$B_c$}}
     \put(30,628){\circle{2}}
     \multiput(23,626.4)(3,0){5}{\line(1,0){2}}

     \put(30,696){\circle{2}}
     \put(30,696){\line(0,1){8}}
     \put(30,696){\line(0,-1){8}}
     \multiput(23,685.6)(3,0){5}{\line(1,0){2}}
     \put(25, 685){\makebox(0,0){$\bowtie$}}
     \put(25,685){\line(0,1){5}}
     \put(25,685){\line(0,-1){5}}
     \put(35,688){\circle*{2}}
     \put(35,688){\line(0,1){8}}
     \put(35,688){\line(0,-1){8}}

     \put(60,600){\makebox(0,0)[t]{$B^{*}_c$}}
     \put(60,632){\circle{2}}
     \multiput(53,633.7)(3,0){5}{\line(1,0){2}}
     \put(55, 632.3){\makebox(0,0){$\bowtie$}}
     \put(65,633){\circle*{2}}

     \put(60,699){\circle{2}}
     \put(60,699){\line(0,1){8}}
     \put(60,699){\line(0,-1){8}}
     \multiput(53,689.9)(3,0){5}{\line(1,0){2}}
     \put(55, 690){\makebox(0,0){$\bowtie$}}
     \put(55,690){\line(0,1){5}}
     \put(55,690){\line(0,-1){5}}
     \put(65,694.6){\circle*{2}}
     \put(65,694.6){\line(0,1){12}}
     \put(65,694.6){\line(0,-1){12}}


     \put(105,600){\makebox(0,0)[t]{$P states$}}
     \put(100,673){\circle{2}}
     \put(100,673){\line(0,1){1}}
     \put(100,673){\line(0,-1){1}}
     \multiput(93,674.7)(3,0){5}{\line(1,0){2}}
     \put(95, 671.3){\makebox(0,0){$\bowtie$}}
     \put(95,671.3){\line(0,1){4}}
     \put(95,671.3){\line(0,-1){4}}
     \put(105,670.3){\circle*{2}}
     \put(105,670.3){\line(0,1){4}}
     \put(105,670.3){\line(0,-1){4}}

     \put(100,678){\circle{2}}
     \put(100,678){\line(0,1){1}}
     \put(100,678){\line(0,-1){1}}
     \multiput(93,670.0)(3,0){5}{\line(1,0){2}}
     \put(95, 677.8){\makebox(0,0){$\bowtie$}}
     \put(95,677.8){\line(0,1){5}}
     \put(95,677.8){\line(0,-1){5}}
     \put(105,683.3){\circle*{2}}
     \put(105,683.3){\line(0,1){8}}
     \put(105,683.3){\line(0,-1){8}}

     \put(120,674){\circle{2}}
     \put(120,674){\line(0,1){1}}
     \put(120,674){\line(0,-1){1}}
     \put(120,676.5){\circle{2}}
     \put(120,676.5){\line(0,1){2}}
     \put(120,676.5){\line(0,-1){2}}
     \multiput(113,673.0)(3,0){5}{\line(1,0){2}}
     \multiput(113,673.6)(3,0){5}{\line(1,0){2}}
     \put(115, 674.2){\makebox(0,0){$\bowtie$}}
     \put(115,674.2){\line(0,1){5}}
     \put(115,674.2){\line(0,-1){5}}


\end{picture}
\end{center}
\caption[jgfkn]{Lattice results for the spectrum of the
$B_c$ system. Open circles indicate NRQCD results 
on quenched configurations, closed circles those
on partially unquenched configurations from the 
MILC collaboration. Bowties indicate results on 
quenched configurations using NRQCD for the $b$ quark 
and relativistic $c$ quarks. 
Error bars are shown where visible and only indicate statistical
uncertainties. (\cite{lidsey}, \cite{shanahan}, 
\cite{gorbahn}). Dashed lines show results from a recent
potential model calculation (\cite{equigg}).}
\label{fig:bc}
\end{figure}

Other states of a more speculative nature are hybrid states
which include a gluonic valence component; $Q\overline{Q}g$. 
Observation of these states would be a direct confirmation 
of the non-Abelian nature of QCD (see possibly \cite{e852}). 
Hybrids are expected containing 
all quark flavours, but the advantage of studying the heavy-heavy 
hybrids is that the normal $Q\overline{Q}$ spectrum is 
relatively clean, both experimentally and, as discussed here, theoretically. 

There have been several lattice calculations of the hybrid 
potentials for a potential model analysis of these states
(see, for example, \cite{perantonis}, \cite{kuti}). 
Wilson loops are calculated whose spatial ends have the appropriate
symmetries to project onto the different hybrid sectors
(see Figure \ref{hybloop}). 
The hybrid potentials obtained can be compared to 
expectations from excited string models and from bag models. 
From a Schr\"{o}dinger equation, masses for the hybrid states 
can be determined; the particular interest is in `exotic' states
which cannot appear in the usual $Q\overline{Q}$ sector. 
These states have quantum numbers $(J={\rm odd})^{-+}$,
$(J={\rm even})^{+-}$ or $0^{--}$. They are most likely 
to be visible if their energies are below threshold  
for Zweig-allowed decay, but it is not clear where this 
threshold is. Some models expect the hybrid states to 
decay to an $S$-wave heavy-light state and a $P$-wave, 
in which case the threshold is rather higher than for 
conventional heavyonium decay (for a recent review of 
expected hybrid phenomenology see \cite{close-lat97}). 

\begin{figure}
\begin{center}
\setlength{\unitlength}{.02in}
\begin{picture}(140,60)(0,0)
\multiput(0,0)(0,5){10}{\line(0,1){2}}
\multiput(40,0)(0,5){10}{\line(0,1){2}}
\multiput(100,0)(0,5){10}{\line(0,1){2}}
\multiput(140,0)(0,5){10}{\line(0,1){2}}
\put(0,50){\line(5,3){15}}
\put(40,50){\line(5,3){15}}
\put(100,50){\line(-5,-3){15}}
\put(140,50){\line(-5,-3){15}}
\put(15,59){\vector(1,0){40}}
\put(85,41){\vector(1,0){40}}
\put(65,30){\line(1,0){4}}
\end{picture}
\end{center}
\caption{An example of an operator used at the 
end of a Wilson loop in the calculation of the 
hybrid ($\Pi$) potential.} 
\label{hybloop}
\end{figure}
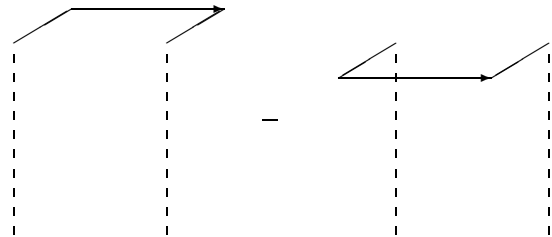

The hybrid potentials obtained (see Figure \ref{hybrid}) 
are very flat, indicating 
broad states, closely packed in energy. The lightest mass 
hybrids from these potentials are close to the 
threshold described above. The same picture is 
obtained by calculating the masses of heavy-heavy hybrids
directly using NRQCD (\cite{collins}, \cite{manke}). 
Further work must be done on the spectrum if the 
states are to be accurately predicted for experimental
searches. 

\begin{figure}
\centerline{\epsfig{file=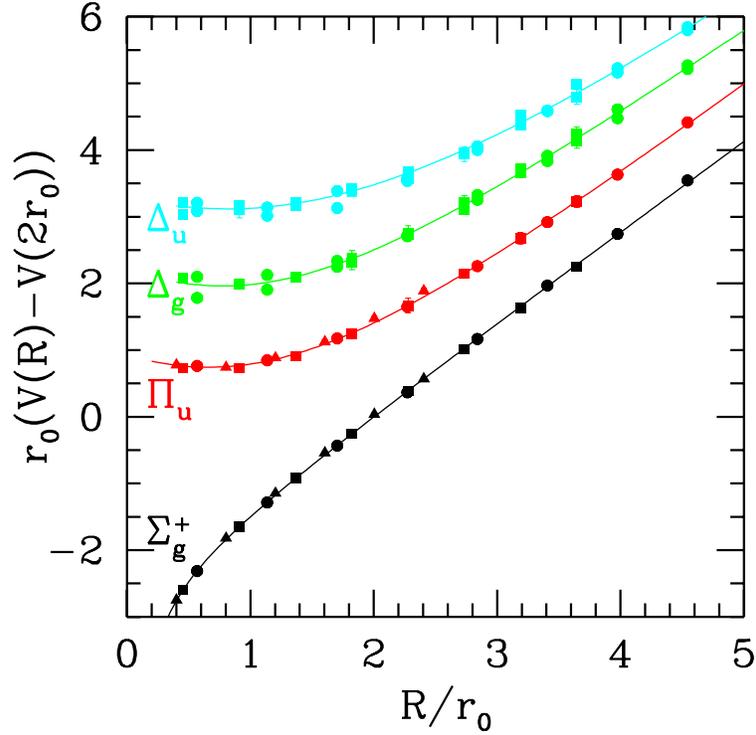,height=100mm,bbllx=18pt,
bblly=145pt,bburx=589pt,bbury=717pt,clip=}}
\caption[hjk]{The heavy hybrid potential from a recent lattice 
calculation (\cite{kuti}). The $\Sigma_g^{+}$ potential is the 
usual central potential; the $\Pi$ potential has a 
gluonic excitation with spin 1 about the $Q\overline{Q}$ 
axis and the $\Delta$ potentials have spin 2. } 
\label{hybrid}
\end{figure}

\section{The heavy-light spectrum}

\subsection{Mesons}
These are bound states with one heavy valence quark or anti-quark and 1 light
anti-quark or quark.
The levels show a similar picture to that for the heavy-heavy spectrum with the
lightest state
the pseudoscalar ($^1S_0$) and close by the vector ($^3S_1$). For charm-light
we
have pseudoscalars $c\overline{d}$ = $D^{+}$, $c\overline{u}$ = $D^{0}$ and
$c\overline{s}$ = $D_s$, and vectors, $D^{*0}$, $D^{*+}$ and $D_s^{*}$. For
bottom-light we have pseudoscalars
$b\overline{d}$ = $\overline{B}^0$, $b\overline{u}$ = $B^{-}$ and
$b\overline{s}$ = $\overline{B}_s$, and vectors again for each. We shall
ignore the distinction (and slight mass difference) between heavy-light mesons
containing $u$ and $d$ quarks and often just refer to
$D$ and $B$. 
Radially excited $S$ states, $D^{\prime}$ and $B^{\prime}$ are
about 500 MeV above the ground states and
below these come a set of positive parity $P$ states, denoted by their spins
$D_0^*/B_0^*$, $D_1/B_1$, $D_2^*/B_2^*$ etc. or more generically $D^{**}$ and $B^{**}$. See
Figure ~\ref{hlspect} and \cite{pdg}.

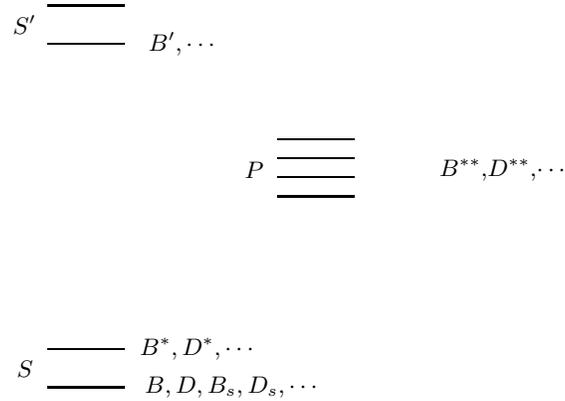
\begin{figure}
\begin{center}
\setlength{\unitlength}{.02in}
\begin{picture}(150,150)(0,0)
\put(10,20){\line(1,0){20}}
\put(10,30){\line(1,0){20}}
\put(10,110){\line(1,0){20}}
\put(10,120){\line(1,0){20}}
\put(5,25){\makebox(0,0){$S$ }}
\put(5,115){\makebox(0,0){$S^{\prime}$ }}
\put(58,20){\makebox(0,0){$B,D,B_s,D_s,\cdots$}}
\put(50,30){\makebox(0,0){$B^{*},D^{*},\cdots$ }}
\put(45,110){\makebox(0,0){$B^{\prime},\cdots$}}
\put(70,70){\line(1,0){20}}
\put(70,75){\line(1,0){20}}
\put(70,80){\line(1,0){20}}
\put(70,85){\line(1,0){20}}
\put(65,77){\makebox(0,0){$P$ }}
\put(130,77){\makebox(0,0){$B^{**}$,$D^{**}$,$\cdots$ }}
\end{picture}
\caption{The spectrum of heavy-light mesons.}
\label{hlspect}
\end{center}
\end{figure}

We do not expect a potential model to work well for the heavy-light spectrum
because the
light quarks are now relativistic. The heavy quarks are still non-relativistic,
however.
Taking $\Lambda_{QCD}$ as a typical QCD momentum scale of 
a few hundred MeV, we have
\begin{eqnarray*}
{\rm Momentum}_Q &\sim& {\rm Momentum}_q \sim \Lambda_{QCD} \\
\frac {v_Q} {c} &\sim& \frac {\Lambda_{QCD}} {m_Q} \\
\end{eqnarray*}
giving $v_Q \sim $ 0.1 for $b$ in $B$ and 0.3 for $c$ in $D$. This is $v_Q$,
not
$v_Q^2$, so the heavy quark is actually more non-relativistic than in
heavy-heavy
systems.

A useful analysis is provided by Heavy Quark Symmetry (see 
\cite{neubert} for a review). In the $m_Q \rightarrow
\infty$ limit QCD has an $[SU(2N_F)]$ symmetry where
$N_F$ is the number of flavours of heavy quark in the theory. This is evident
from the
NRQCD Lagrangian of equation \ref{nrqcd} which, by the 
arguments above, is appropriate for the heavy
quarks here also. As $m_Q \rightarrow \infty$ the Lagrangian becomes
$\psi^{\dag} D_t \psi$ when the quark mass term is removed. Thus the heavy
quarks
become spinless and any distinction between flavours disappears (apart from the
overall energy level set by the missing mass term). The picture of a
heavy-light meson becomes one of a static heavy quark surrounded by a
fuzzy cloud of the light degrees of freedom, known as `brown muck', as in
Figure ~\ref{meson_fig}. 
Interactions that probe only momentum scales appropriate to the 
brown muck will not be able to see details of the heavy quark at the centre.
Notice that this is a completely different physical picture 
to that for heavy-heavy mesons. 

\begin{figure}
\centerline{\epsfig{file=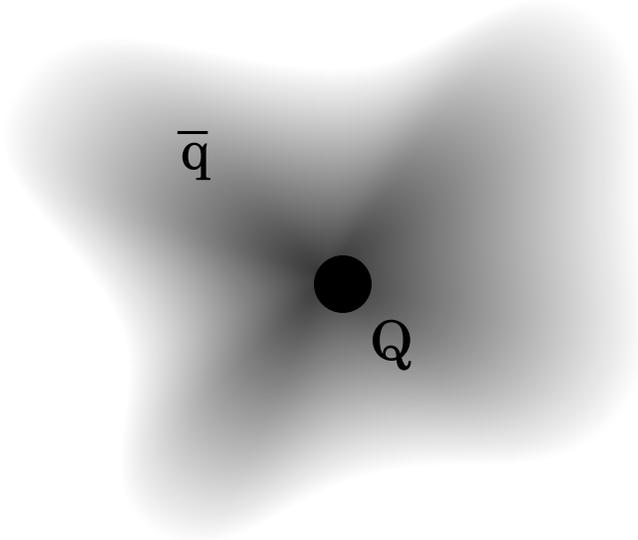,height=80mm,bbllx=85pt,
bblly=526pt,bburx=309pt,bbury=725pt,clip=}}
\caption{A heavy-light meson in the Heavy Quark Symmetry picture.}
\label{meson_fig}
\end{figure}

{}From the picture of Figure \ref{meson_fig} there is a natural distinction between energy shifts in the
spectrum that are caused by something changing for
the light degrees of freedom, e.g. a radial
or orbital excitation, and those caused by something changing for the heavy
quark, such as its flavour or spin. In the first case we expect
that radial and orbital excitation energies should 
be approximately independent of the heavy quark flavour, and 
in the second case we expect much smaller splittings with 
strong heavy quark mass dependence between states of 
different $\vec{S}_Q$.  This hierarchy of splittings is similar 
to that for heavy-heavy mesons but for different reasons. 
The $m_Q$-independence of the $1P-1S$ splitting in 
heavyonium is an accident; in heavy-light mesons it 
is the consequence of a symmetry of the non-relativistic
effective theory as $m_Q \rightarrow \infty$. 

A power-counting analysis of the terms in the NRQCD Lagrangian is useful to
demonstrate this effect (\cite{arifa}). 
\begin{equation}
D_t \sim \vec{D} \sim \Lambda_{QCD}
\end{equation}
from above. Then 
\begin{equation}
\frac {\vec{D}^2} {2 m_Q} \sim \frac {\Lambda_{QCD}^2} {m_Q}. 
\end{equation}
Also 
\begin{equation}
\vec{A} \sim A_t \rightarrow \vec{E},\vec{B} \sim \Lambda_{QCD}^2
\end{equation}
and
\begin{eqnarray}
\frac {\vec{\sigma} \cdot \vec{B}} {m_Q} &\sim& \frac {\Lambda_{QCD}^2}
{m_Q} \label{power2} \\
\frac {\vec{\sigma} \cdot \vec{D} \times \vec{E}} {m_Q^2} 
&\sim& \frac {\Lambda_{QCD}^3} {m_Q^2} \nonumber 
\end{eqnarray}
This shows that, for the heavy-light case, the NRQCD Lagrangian is a $1/m_Q$ 
expansion (unlike the heavy-heavy case where terms at 
different order in $1/m_Q$ appeared at the same order in $v_Q^2$). 
The leading order term is the $D_t$ term and then at the 
next order come two $1/m_Q$ terms - the kinetic energy of the 
heavy quark and the spin coupling to the chromo-magnetic 
field. These are the first two terms to know about the heavy quark 
flavour (mass) and its spin. Any splitting that requires this   
knowledge will appear first at $1/m_Q$ in an expansion in the 
inverse heavy quark mass.  

The heavy quark spin, $\vec{S}_Q$, is a good quantum 
number in the heavy quark limit and so we can classify 
states according to $\vec{j}_l = \vec{J} - \vec{S}_Q$.
Each $\vec{j}_l$ state becomes, on the addition of the 
heavy quark, a doublet with $J = j_l \pm 1/2$ (\cite{isgwise}). 
An analogy can be drawn with atomic physics and the decoupling 
of the nuclear spin as $m_e/m_N \rightarrow 0$. 
The lightest states are the $L$ = 0, $j_l$ = 1/2, 
$^3S_1$ ($D^{*},B^{*}$)/ $^1S_0$ ($D,B$) doublet. 
For heavy-light $P$ states 
the light quark spin, $S_q$, is
coupled to the orbital angular momentum to make states of overall spin, 
$j_l$ = 1/2 ( 2 polarisations ) or $j_l$ = 3/2 ( 4 
polarisations). Coupling $S_Q$ to
$j_l$ = 3/2 gives total J=2 ($B^{*}_2, D^{*}_2$) 
and J=1 ($B_1^{\prime}, D_1^{\prime}$)
(8 states altogether). Coupling $S_Q$ to 
$j_l$ = 1/2 gives J=0 or J=1 (4 states
altogether). Thus in the $jj$ coupled basis we reproduce the same 
12 states as the $LS$ coupled $^1P_1,^3P_{0,1,2}$
multiplet. However, the spin 1 states are a
mixture of the $^1P_1$ and $^3P_1$ (with mixing 
angle $35^{\circ}$) because of a 
lack of charge conjugation. In
the $m_Q \rightarrow \infty$ limit only 
the splittings caused by the light
degrees of freedom remain. The $jj$ basis 
becomes the correct one and all the 
$j_l$ = 3/2 states become degenerate
but split from all the $j_l$ = 1/2 states. 
The $j_l$ = 3/2 states are narrow
(and therefore visible) because of the high
orbital angular momentum required in decays to 
$D^{(*)}\/B^{(*)}\pi$ for J=2.
J=1 can only decay to $D^*/B^* \pi$ but, having 
the same $j_l$ as the J=2
state, has a similar total width (see Figure \ref{pfine} and
\cite{isgwise}).

The difference between the $j_l + 1/2$ and 
$j_l - 1/2$ members of a doublet 
is a spin flip of the heavy quark.
The leading term that gives rise to this in the 
NRQCD Lagrangian is the $\vec{\sigma}_Q \cdot \vec{B}$
term, yielding a splitting behaving 
as $\lambda \times$(spin factors)$\times 1/m_Q$.
$\lambda$ is an expectation value in the light quark degrees of 
freedom so the heavy quark mass dependence of 
the splitting is as $1/m_Q$ in leading order. 
Table \ref{hyp-hl} shows experimental values for the 
vector-pseudoscalar splitting (\cite{pdg}).  
$1/m_Q$ behaviour fits very well if we take $m_c \approx$ 1.5
GeV and $m_b \approx$ 5 GeV. We can also consider 
the strange quark as a heavy
quark rather than a light one and add the value 
for the strange-up/down system,
the $K$ into the Table. This only works moderately 
well, with $m_s \approx$ 0.5
GeV, say. The coefficient of the $1/m_Q$ dependence 
is of order 0.2 ${\rm
GeV}^2$, which is compatible with an expectation value in a light system of
$\Lambda_{QCD}^2$. Note that the variation with light quark mass between $u/d$
and $s$ is very small. It
is clear that $B_c$ cannot be fitted into this heavy-light picture since its
expected hyperfine splitting is much larger (see section 2.5).

\begin{table}
\begin{center}
\begin{tabular}{l|c|c}
Splitting & Experiment/ MeV & `Expected' value / MeV \\
\hline
$K^{*} - K $ & 398 & 457 \\
$D^{*} - D$ & 141 & 152 \\
$D_s^{*} - D_s$ & 144 & \\
$B^{*} - B$ & 46 & 46 \\
$B_s^{*} - B_s$ & 47 & \\
\hline
$K_2^*(1430) - K_1(1270)$ & 154 & \\
$D_{2}^{*} - D_{1}$ & 37 & \\
$D_{s2}^{*} - D_{s1}$ & 38 & \\
\end{tabular}
\caption[kjgp]{ Hyperfine splittings for different heavy-light systems;
the top group for the $L=0, j_l = 1/2$ doublet, the 
lower group for the $L=1, j_l = 3/2$ doublet. The final
column
gives expected values for the $^3S_1 - ^1S_0$ splitting 
rescaling from the $B$ system by the inverse ratio of
quark masses given in the text. (\cite{pdg}).}
\label{hyp-hl}
\end{center}
\end{table}

Table \ref{hyp-hl} also shows results for the $L$ = 1, $j_l$ = 3/2
doublet from the $D$ and $K$ systems (\cite{eich-hl}). 
The ratio of splittings between $D$ and $K$
is rather different for this case to the one above, probably 
showing that the $K$ is stretching the limits of HQS 
arguments. Nevertheless, we expect a splitting 
$B_2^* - B_1$ of $m_c/m_b \times (D_2^* - D_1)$ $\sim$ 12 MeV. 
A value of 26 MeV is given in \cite{delphi95}
for the $B_s$. 
Experimental results for the $L=1$ $j_l=1/2$ doublet 
are not available since these are much broader than the 
$j_l = 3/2$ doublet.  

In contrast there are several splittings that we expect, from the arguments
above, to
be controlled by changes in the light quark degrees of freedom and therefore to
be
independent of the heavy quark mass at leading order. One of these is the
splitting
between the heavy-strange and heavy-up/down mesons. Experimentally this is
satisfied
at the 10\% level (see Table \ref{sd-hl}). Other such splittings are those between
radially excited $S$ states and the ground states for which 
experimental information is very limited ($B^{'} - B$ =
580MeV (\cite{landua}) and $D^{*'} - D^{*}$ = 
630 MeV (\cite{delphi-eps97})), 
and between orbitally excited $P$ states and the
ground $S$ states, which we discuss below.

\begin{table}
\begin{center}
\begin{tabular} {l|c}
Splitting & Experiment / MeV \\
\hline
$D_s - D$ & 99 \\
$B_s - B$ & 90 \\
\end{tabular}
\caption[lpkek]{Experimental values for splittings between heavy-strange and
heavy-up/down systems (\cite{pdg}).}
\label{sd-hl}
\end{center}
\end{table}

\begin{figure}
\setlength{\unitlength}{.02in}
\begin{picture}(300,100)(0,0)
\put(0,60){\line(1,0){20}}
\put(30,20){\line(1,0){20}}
\put(30,50){\line(1,0){20}}
\put(30,90){\line(1,0){20}}
\put(10,65){\makebox(0,0){`$^1P_1$'}}
\put(40,25){\makebox(0,0){$^3P_0$}}
\put(40,55){\makebox(0,0){`$^3P_1$'}}
\put(40,95){\makebox(0,0){$^3P_2$}}
\put(70,50){\vector(1,0){30}}
\put(85,55){\makebox(0,0){$m_Q \rightarrow \infty$}}
\put(140,25){\line(1,0){20}}
\put(140,85){\line(1,0){20}}
\put(130,25){\makebox(0,0){J=1}}
\put(130,85){\makebox(0,0){J=1}}
\put(170,20){\line(1,0){20}}
\put(170,90){\line(1,0){20}}
\put(200,20){\makebox(0,0){J=0}}
\put(200,90){\makebox(0,0){J=2}}
\put(160,100){\makebox(0,0){$j_l=3/2$}}
\put(160,10){\makebox(0,0){$j_l=1/2$}}
\put(165,22){\makebox(0,0){ \}}}
\put(165,88){\makebox(0,0){ \} }}
\put(180,25){\makebox(0,0){$1/m_Q$}}
\put(180,85){\makebox(0,0){$1/m_Q$}}
\end{picture}
\caption{On the left, $P$ fine structure for a degenerate heavy-heavy system. On
the right, $P$ fine structure for a heavy-light system.}
\label{pfine}
\end{figure}
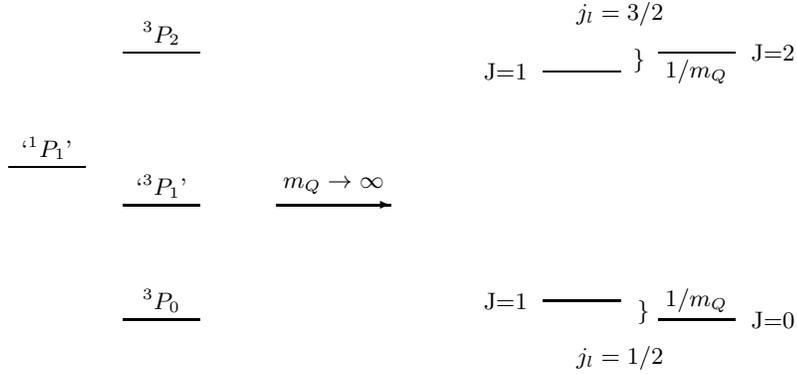

For the orbital splittings between heavy-light $S$ and $P$ states we 
should calculate a splitting between spin-averaged states, to remove 
the spin-dependent $1/m_Q$ effects, and make as 
clear as possible the $m_Q$-independent light quark 
effects. Since $j_l$ = 1/2
states have not been
seen, we compare in Table \ref{sp-hl} the splitting between the spin-average of the
$j_l$ = 3/2 $P$ states and the spin-averaged $S$ states for $D$ and $B$. Good
agreement between the 2 systems is seen. For $B$ states good spin separation of
the $P$ states is not yet available.

\begin{table}
\begin{center}
\begin{tabular}{l|c}
Splitting & Experiment / MeV \\
$\overline{K}_p(1368) - \overline{K}(792)$ & 576 \\
$\overline{D}_p (2445) - \overline{D} (1975)$ & 470 \\
$\overline{D}_{sp} (2559) - \overline{D}_s (2076)$ & 483 \\
$B^{**} (5698) - \overline{B}(5313)$ & 385 \\
$B_s^{**} (5853) - \overline{B}_s(5404)$ & 449 \\
\end{tabular}
\caption[fkjgf]{ Splittings between spin-averaged $j_l$ = 3/2 $P$ states 
(or experimentally unseparated $P$ states) and
spin-averaged
$S$ states for heavy-light systems (\cite{pdg97}).}
\label{sp-hl}
\end{center}
\end{table}

We would also expect the splitting between the $j_l$ = 1/2 and
the $j_l$ = 3/2 states to be 
approximately independent of $m_Q$ (\cite{isg97}), 
although the physical spin 1 states will be a mixture 
of the $jj$ states away from the $m_Q \rightarrow \infty$
limit.
This cannot be
checked experimentally as yet. 

\subsection{Baryons}

There is a huge array of baryon states with one heavy quark and two light
quarks. Again we can make sense of their masses using Heavy Quark Symmetry
arguments.
We view the baryon as a static colour source (for the heavy quark) surrounded
by a fuzzy light quark system which is made of two light quarks this time
instead of a light anti-quark (\cite{falk-hb}).

We will discuss only the case of zero relative orbital angular momentum. For
two different light quarks we can combine the light quark spins to
give a total $S_l$ of 0 or 1. If the light quarks have the same flavour, only
$S_l$ = 1 is possible by Fermi statistics, remembering the overall
anti-symmetry of the colour wavefunction. Coupling the heavy quark spin then
gives the combinations in Table \ref{baryons}, with
overall spin-parity assignments.

\begin{table}
\begin{center}
\begin{tabular}{cccccc}
baryon & $Qqq$ & $S_l$ & $J^P$ & ${\rm mass}_c$/MeV & ${\rm mass}_b$/MeV \\
\hline
$\Lambda$ & $Q[ud]$ & $0^{+}$ & ${\frac {1} {2}}^{+}$ & 2285(1) & 5624(9)\\
$\Sigma$ & $Q\{ud\},uu,dd$ & $1^{+}$ & ${\frac {1} {2}}^{+}$ & 2453(1) & 5797(8) \\
$\Sigma^{*}$ & $\: Q\{ud\},uu,dd \: $ & $1^{+}$ & ${\frac {3} {2}}^{+}$ & 2519(2) & 5853(8) \\
$\Xi$ & $Q[u/ds]$ & $0^{+}$ & ${\frac {1} {2}}^{+}$ & 2468(2) & \\
$\Xi^{\prime}$ & $Q\{u/ds\}$ & $1^{+}$ & ${\frac {1} {2}}^{+}$ & 2568(?) & \\
$\Xi^{*}$ & $Q\{u/ds\}$ & $1^{+}$ & ${\frac {3} {2}}^{+}$ & 2645(2) & \\
$\Omega$ & $Qss$ & $1^{+}$ & ${\frac {1} {2}}^{+}$ & 2704(4) & \\
$\Omega^{*}$ & $Qss$ & $1^{+}$ & ${\frac {3} {2}}^{+}$ && \\
\end{tabular}
\caption[kjkhfl]{$J^{P}$ possibilities for baryons containing one heavy quark along
with two light quarks. The names are given with subscripts $c$ or $b$.
Masses are given in the last two columns, taken from \cite{pdg97},
\cite{delphi95} for $\Sigma_b$ and \cite{werding} for $\Xi^{'}$. }
\end{center}
\label{baryons}
\end{table}

Using HQS arguments we would expect the splitting between the
spin average of $\Sigma$ and $\Sigma^{*}$ states and the
$\Lambda$ to be independent of $m_Q$, since this splitting 
represents a change in $j_l$. We can check this in Table \ref{lamsig},
and it
works well
even when the $s$ quark is considered as a heavy quark. There is in fact very
little
room for sub-leading $1/M_Q$ dependence which can in principle be there
($\Lambda_{QCD}^2/m_c \sim$ 50 MeV). In the
last row is given for comparison the splitting between the 
spin-average of the $\Xi^{*}_c$ and $\Xi^{'}_c$ and the 
$\Xi_c$. This is essentially the same splitting 
except for the different light
quark content. The answer is significantly
different, showing more sensitivity to
light quark
content than for the mesons (\cite{falk-hb}). 
The physical $\Xi^{'}_c$ and $\Xi_c$ will be mixtures
of the HQS states, just like the spin 1 meson $P$ states, 
but this should not be a big effect.  
An equal spacing rule, $\Omega_c - \Xi^{'}_c = 
\Xi^{'}_c - \Sigma_c$ holds well. 

\begin{table}
\begin{center}
\begin{tabular}{c|c}
Splitting & Experiment /MeV \\
$\overline{\Sigma_s} - \Lambda_s$ & 203 \\
$\overline{\Sigma_c} - \Lambda_c$ & 212 \\
$\overline{\Sigma_b} - \Lambda_b$ & 210 \\
\hline
$\overline{\Xi}^{*,\prime}_c - \Xi_c$ & 150 \\
\end{tabular}
\caption[kfjfdl]{ Experimental values for the splitting between the spin-average of
$\Sigma$
states and the $\Lambda$ for different heavy quarks including $s$. In the last
row a comparable splitting is given for the $\Xi_c$.} 
\label{lamsig}
\end{center}
\end{table}

All the fine structure splittings between states of the same $S_l$ but
different J should
behave as $1/m_Q$. Table \ref{hypbar} shows the experimental information on this for the
$\Sigma$ baryons. The $s$ quark fits well into this picture, 
but the experimental $\Sigma^{*}_b - \Sigma$ splitting looks 
significantly different from the expected value (\cite{falk-hb}).
The experimental results need to be confirmed, however. 
The $\Xi^{*} - \Xi^{'}$ splitting agrees well with 
the $\Sigma^{*} - \Sigma$ showing no large $m_s$ effects here. 

\begin{table}
\begin{center}
\begin{tabular}{c|c|c}
Splitting & Experiment / MeV & `Expected' / MeV\\
\hline
$ \Sigma^*_s - \Sigma_s$ & 191 & 198 \\
$ \Sigma^*_c - \Sigma_c$ & 66 & 66 \\
$ \Sigma^*_b - \Sigma_b$ & 56 & 20 \\
\hline
$\Xi^{*}_c - \Xi^{'}_c$ & 80 & 
\end{tabular}
\caption[klgjfl]{Splittings between $\Sigma^*$ and $\Sigma$ states for different heavy
quarks,
including $s$. The last column gives expected values for $1/m_Q$ behaviour
compared to the splitting for $c$. } 
\label{hypbar}
\end{center}
\end{table}

We can also take splittings between baryons and mesons. The simplest splitting
is between the $\Lambda$ baryons and the $S$ state mesons. To remove spurious
$m_Q$ dependence we should take the spin-average of the $^1S_0$ and $^3S_1$
meson states (\cite{martinrich}). Table \ref{bar-mes} shows the experimental results; again Heavy Quark Symmetry
works
much better than might be expected.

\begin{table}
\begin{center}
\begin{tabular}{c|c}
Splitting & Experiment / MeV \\
$\Lambda_s - \overline{K}$ & 323 \\
$\Lambda_c - \overline{D}$ & 310 \\
$\Lambda_b - \overline{B}$ & 310 \\
\end{tabular}
\caption{The splitting between the $\Lambda$ baryon and the spin average of
$S$ state heavy-light mesons for different heavy quarks, including $s$. }
\label{bar-mes}
\end{center}
\end{table}

HQS yields only the $m_Q$ dependence of the splittings; it 
must be combined with a non-perturbative method of determining the 
coefficients of this dependence. QCD sum rules can be invoked 
here (\cite{neubert}); Lattice QCD provides a better {\it ab initio} method. 
We discuss results from lattice QCD in the next subsection. 

\begin{quote}
{\bf Exercise:} Discuss what you would expect for heavy-heavy-light baryons.
Take $Q_1 \ne Q_2$.
\end{quote}

\begin{quote}
{\bf Exercise:} Compare orbitally excited $\Lambda_s$ and 
$\Lambda_c$ baryons from the Particle Data Tables. What 
does this lead you to expect for the orbitally excited 
$\Lambda_b$ ? (\cite{roshb}). 
\end{quote}

\subsection{Direct calculations of the heavy-light spectrum on the lattice}

Following the methods described for the heavyonium spectrum, we can calculate 
the heavy-light spectrum directly using lattice QCD. We must combine a heavy
quark propagator with a light anti-quark propagator (or two light quark propagators)
to make a meson (baryon) correlation function. This we fit as before to a sum 
of exponentials to extract ground and excited state energies and 
masses. 
In principle some of the excited states can undergo strong decays 
upsetting this relation, but this does not happen in current
lattice simulations. 

For hadrons containing a $b$ quark the best method is probably to use NRQCD for 
the heavy quark as described for bottomonium in section 2.3. Because of the 
different power-counting rules for the heavy-light case, the Lagrangian used can be
different to that for heavyonium. For example, a consistent calculation to 
$\cal{O}$$(1/m_Q)$ would include $D_t$, $\vec{D}^2/2m_Q$ and 
$\vec{\sigma}\cdot\vec{B}/2m_Q$ terms (tadpole-improved as before). 
In fact for the heavy-light case the spectrum can be calculated in the static 
limit with simply the $D_t$ term, because the light quark provides the 
kinetic energy. In this case, of course, only states of a given $j_l$ are
obtained with no hyperfine splittings. 
The static limit is very cheap computationally 
but much noisier (\cite{potmod}) than NRQCD even at very large
$m_Q$ and for this reason it may be more accurate to 
obtain static results from the limit of NRQCD calculations. For hadrons 
containing a $c$ quark, we will discuss results using the heavy Wilson (SW)
action. 
 
\begin{figure}
\begin{center}
\setlength{\unitlength}{.02in}
\begin{picture}(130,100)(30,500)

\put(15,500){\line(0,1){100}}
\multiput(13,500)(0,50){3}{\line(1,0){4}}
\multiput(14,500)(0,10){11}{\line(1,0){2}}
\put(12,500){\makebox(0,0)[r]{{\large5.0}}}
\put(12,550){\makebox(0,0)[r]{{\large5.5}}}
\put(12,600){\makebox(0,0)[r]{{\large 6.0}}}
\put(12,570){\makebox(0,0)[r]{{\large GeV}}}
\put(15,500){\line(1,0){160}}


     \put(25,510){\makebox(0,0)[t]{{\large $B$}}}
     \put(23.5,526.9){$\otimes$}
     \multiput(20,527.9)(3,0){4}{\line(1,0){2}}
     \put(23,588.3){\circle*{3}}
     \put(23,588.3){\line(0,1){7.1}}
     \put(23,588.3){\line(0,-1){7.1}}
     \put(37,595){\makebox(0,0)[t]{{\large $(2S)$}}}
     \multiput(20,586)(3,0){4}{\line(1,0){0.5}}
     \put(27,583.7){$\!\!\Box$}
     \put(27,585){\line(0,1){12}}
     \put(27,585){\line(0,-1){12}}

     \put(55,510){\makebox(0,0)[t]{{\large $B^{*}$}}}
     \put(53,530.3){\circle*{3}}
     \multiput(50,532.6)(3,0){4}{\line(1,0){2}}
     \multiput(50,532.4)(3,0){4}{\line(1,0){2}}
\put(57,529.3){$\!\!\Box$}
\put(57,591.8){$\!\!\Box$}
\put(57,593){\line(0,1){12}}
\put(57,593){\line(0,-1){12}}

     \put(80,510){\makebox(0,0)[t]{{\large $B_s$}}}
     \put(76,537){\circle*{3}}
     \put(80,539.0){\circle{3}}
     \multiput(75,538.1)(3,0){4}{\line(1,0){2}}
     \multiput(75,536.9)(3,0){4}{\line(1,0){2}}
     \put(76,592.7){\circle*{3}}
     \put(76,592.7){\line(0,1){4.8}}
     \put(76,592.7){\line(0,-1){4.8}}
     \put(80,593.7){\circle{3}}
     \put(80,593.7){\line(0,1){4.8}}
     \put(80,593.7){\line(0,-1){4.8}}
     \put(92,601){\makebox(0,0)[t]{{\large $(2S)$}}}
     \put(83,538.0){$\!\!\Box$}
     \multiput(75,538.1)(3,0){4}{\line(1,0){2}}
     \multiput(75,536.9)(3,0){4}{\line(1,0){2}}

     \put(105,510){\makebox(0,0)[t]{{\large $B^{*}_s$}}}
     \put(103,539.8){\circle*{3}}
     \put(103,539.8){\line(0,1){1.1}}
     \put(103,539.8){\line(0,-1){1.1}}
     \put(107,541.9){\circle{3}}
     \multiput(100,542.8)(3,0){4}{\line(1,0){2}}
     \multiput(100,541.6)(3,0){4}{\line(1,0){2}}

     \put(145,510){\makebox(0,0)[t]{{\large $P-States$}}}
     \put(155,584.2){\circle*{3}}
     \put(155,584.2){\line(0,1){5}}
     \put(155,584.2){\line(0,-1){5}}
     \put(165,584.2){\makebox(0,0){{\large $(B^*_2)$}}}
     \put(130,561.5){\circle*{3}}
     \put(130,561.5){\line(0,1){4.5}}
     \put(130,561.5){\line(0,-1){4.5}}
     \put(140,564.0){\makebox(0,0)[t]{{\large $(B^*_0)$}}}
     \put(138,571.3){\circle*{3}}
     \put(138,571.3){\line(0,1){4}}
     \put(138,571.3){\line(0,-1){4}}
     \put(142,575.9){\circle*{3}}
     \put(142,575.9){\line(0,1){3.4}}
     \put(142,575.9){\line(0,-1){3.4}}
     \put(145,576.1){$\!\!\Box$}
     \put(145,577.5){\line(0,1){3.4}}
     \put(145,577.5){\line(0,-1){3.4}}
     \put(155,574){\makebox(0,0){{\large $(\overline{B^*_1})$}}}
     \multiput(128,568.6)(3,0){14}{\line(1,0){2}}
     \multiput(128,571.0)(3,0){14}{\line(1,0){2}}

\end{picture}
\end{center}
\caption[jgj]{The $B$ spectrum from lattice QCD using 
NRQCD for the $b$ quark. Circles are in the quenched 
approximation; open circles use $m_s$ from $K$ and closed 
circles, $m_S$ from $K^{*}$. Squares are results on configurations 
with $n_f=2$ dynamical fermions. Experimental results (\cite{pdg97}) are given by
dashed horizontal lines. The $B$ meson mass is fixed 
to its experimental value in all cases (\cite{arifareview}).}
\label{hlmesons} 
\end{figure}
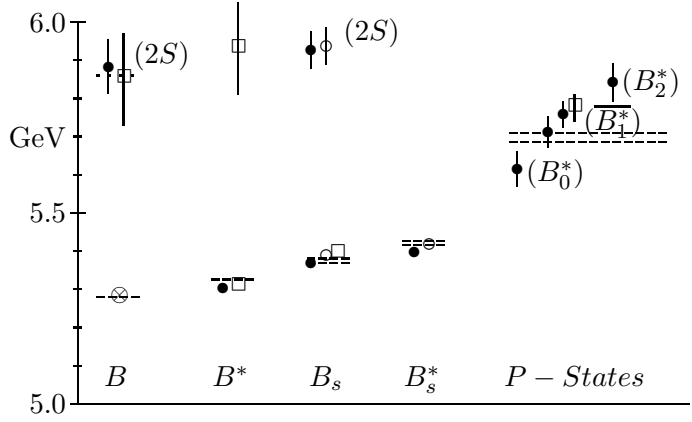

Since we do not have a potential model in principle to guide our intuition, it is more difficult to think of good smearing functions for heavy-light mesons.
 A lot of effort has been put
into this for mesons in the static case (\cite{eichten-fb}, \cite{mcneile}) to ameliorate the noise problems. Again, the smearing does not affect the values of masses obtained, but a good smearing can reduce the errors. In fact potential-model type wavefunctions (much broader than for 
heavyonium) do work reasonably 
well (\cite{eichten-fb}, \cite{arifa}), used as a source for the 
heavy quark, as do gauge-invariant smearings typical of  light hadron calculations (\cite{ukqcdcharm}).
The light anti-quark for the meson is taken to have a delta function 
source. Alternatively both propagators can be smeared, 
and for baryons it is certainly a good 
idea for all the propagators to be smeared (\cite{ukqcdcharm}). 

The meson operators are similar to those for heavyonium - $\psi_Q^{\dag}\Omega\phi\chi_q^{\dag}$. For the NRQCD heavy quark case $\Omega$ is a $2\times2$ matrix in spin space and only 2 components are taken from the 4-component light quark. The colors
of heavy quark and light anti-quark are matched for a colour 
singlet. The baryon operators need an anti-symmetric colour combination, and the light quark propagators 
combined with appropriate spins (\cite{ukqcdcharm}). For 
example the $\Lambda_Q$ operator (with smearing factors, $\phi$ 
suppressed) is:
\begin{equation}
{\cal O} = \epsilon^{ABC} (\chi_{q_1}^{A^T} {\cal C} \gamma_5 \chi_{q_2}^{B})\psi_Q^{\dag C}
\end{equation}
where $\cal{C}$ is the charge conjugation matrix. 

In principle, having calculated the bottomonium spectrum in NRQCD as in section 2.3 on a 
given set of gluon configurations, we can determine $a^{-1}$ and 
the bare $b$ quark mass, $m_b$, and the calculation of 
the $B$ spectrum should have no parameters
to tune. Unfortunately this is not true in the quenched approximation. The disagreement with experiment shown in Figure \ref{upsrho} makes it clear that the 
$a^{-1}$ fixed from the $\Upsilon$ spectrum would be $\sim 20 \%$ different to that 
from $M_{\rho}$, because of the different momentum scales appropriate to the two 
systems. Heavy-light systems are much closer to light hadrons in these terms than to 
heavyonium. For the best quenched results we really need to use a value for $a^{-1}$ 
from the heavy-light system itself, but the lack of experimental information on $P$ 
states makes this hard, since the obvious quantity to use is the $1P-1S$ splitting. 
Usually $a^{-1}$ is taken instead from light hadron spectroscopy. Large statistical 
and systematic uncertainties there then give a rather large error. 

\begin{figure}
\begin{center}
\setlength{\unitlength}{.02in}
\begin{picture}(130,100)(30,500)
\put(15,500){\line(0,1){100}}
\multiput(13,500)(0,50){3}{\line(1,0){4}}
\multiput(14,500)(0,10){11}{\line(1,0){2}}
\put(12,500){\makebox(0,0)[r]{{\large5.0}}}
\put(12,550){\makebox(0,0)[r]{{\large5.5}}}
\put(12,600){\makebox(0,0)[r]{{\large 6.0}}}
\put(12,570){\makebox(0,0)[r]{{\large GeV}}}
\put(15,500){\line(1,0){160}}


     \put(20,510){\makebox(0,0)[t]{{\large $B$}}}
     \put(20,526.9){$\!\!\otimes$}
     \multiput(15,527.9)(3,0){4}{\line(1,0){2}}

     \put(50,510){\makebox(0,0)[t]{{\large $\Lambda_b$}}}
     \put(43,565.1){\circle*{3}}
     \put(43,565.1){\line(0,1){12.6}}
     \put(43,565.1){\line(0,-1){12.6}}
     \multiput(39,562.4)(3,0){6}{\line(1,0){2}}
     \put(45.5,563){$\diamondsuit$}
     \put(48,564){\line(0,1){6}}
     \put(48,564){\line(0,-1){6}}
     \put(50.3,571.8){$\triangle$}
     \put(53,572.8){\line(0,1){14.5}}
     \put(53,572.8){\line(0,-1){14.5}}
     \put(43,581.8){$\!\!\Box$}
     \put(43,583){\line(0,1){4}}
     \put(43,583){\line(0,-1){4}}

     \put(100,510){\makebox(0,0)[t]{{\large $\Sigma_b$}}}
     \put(98,584.3){\circle*{3}}
     \put(98,584.3){\line(0,1){4.3}}
     \put(98,584.3){\line(0,-1){4.3}}
     \multiput(95,579.9)(3,0){5}{\line(1,0){0.5}}
     \put(100.5,576){$\diamondsuit$}
     \put(103,577){\line(0,1){7}}
     \put(103,577){\line(0,-1){7}}

     \put(150,510){\makebox(0,0)[t]{{\large $\Sigma_b^*$}}}
     \put(148,586.2){\circle*{3}}
     \put(148,586.2){\line(0,1){4.3}}
     \put(148,586.2){\line(0,-1){4.3}}
     \multiput(145,585.3)(3,0){5}{\line(1,0){0.5}}
     \put(150.5,577){$\diamondsuit$}
     \put(153,578){\line(0,1){7}}
     \put(153,578){\line(0,-1){7}}

\end{picture}
\end{center}
\caption[gkjf]{Masses of baryons containing one $b$ quark from
lattice QCD. 
Circles use NRQCD for the $b$ quark in the quenched 
approximation, the box uses NRQCD on configurations with 
$n_f=2$ flavours of dynamical fermions. Triangles (\cite{alexcharm})
use Wilson fermions, and diamonds the SW action (\cite{ukqcdcharm})
extrapolating from the region of the charm quark, again in the 
quenched approximation. 
Experimental results are given by horizontal lines (\cite{pdg97},
\cite{delphi95}). (\cite{arifareview}). }
\label{hlbaryons}
\end{figure}
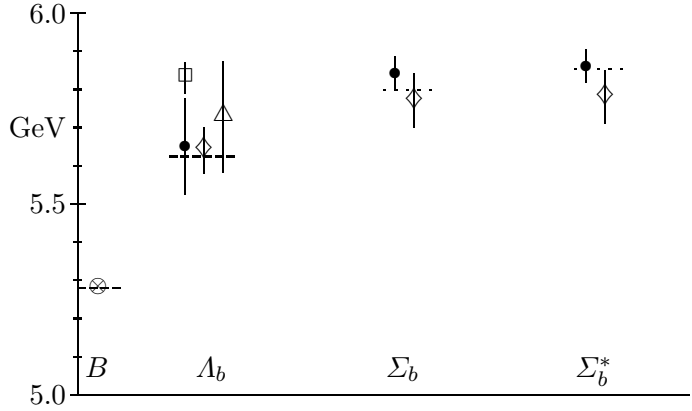

This creates a problem with the bare $b$ quark mass, $m_b$, since it was fixed in 
bottomonium using $a^{-1}$ from that system. It should be fixed again in 
heavy-light systems using the kinetic mass of, say, the $B$. This is difficult to extract 
accurately because $B$s are lighter than $\Upsilon$s. $E(p) - E(0)$ is larger for the 
$B$ than the $\Upsilon$ so the noise in the meson correlation 
function at finite momentum $p$ (set by $E(0)$) is worse. 
An alternative is to calculate the usual 
energy at zero momentum, $E_B(0)$, and apply the energy 
shift per quark in lattice units calculated for heavyonium 
to get $m_Ba$ (\cite{arifa}, \cite{collins-spect}).

These problems mean that the heavy-light 
spectrum cannot be as accurately calculated as that of heavyonium. 
Once dynamical fermions are included sufficiently well to 
mimic the real world there can only be one value of 
$a^{-1}$ and $m_{b/c}$. We are a long way from this 
point at present, however. It is not even possible 
to perform consistent $n_f$ extrapolations 
(to $n_f$ = 3?) of the heavy-light spectrum 
from results at $n_f$ = 0 and 2 (\cite{collins-spect} 
and in preparation). For heavyonium differences 
in methods of fixing $a^{-1}$ disappeared on this 
extrapolation but this is not currently true for 
heavy-light mesons and shows the presence of systematic errors. 

Another difficulty with the heavy-light spectrum is that of fixing the light quark mass. This is a problem shared with light hadron calculations (\cite{weingarten}, \cite{montvay}). The $B$ and $D$ calculations must be done with several 
different light quark masses far from the physical $u/d$ 
masses and the results extrapolated to the chiral limit. This 
inevitably causes an increase in statistical and systematic errors. 
For the $B_s$ and $D_s$, it is possible to interpolate to the $s$ quark mass 
although there are ambiguities in fixing that, again possibly arising from the quenched 
approximation (see, for example, \cite{rajan}). 
 
Figure \ref{hlmesons} shows the $b$-light meson spectrum 
using NRQCD for the $b$ quark, fixing $m_b$ from the $B$ 
mass and $a^{-1}$ from $M_{\rho}$ (from a recent review 
by \cite{arifareview}). The overall agreement with experiment 
is good. The $B^{*} - B$ splitting is too small, however, 
both on quenched and on partially unquenched configurations. 
As before, this may be a quenching effect and/or it may arise 
from radiative corrections to $c_4$ beyond tadpole-improvement
(see the discussion for heavyonium). The problems 
with fixing $m_s$ are clear. The $P$ states still have rather 
large error bars but the ordering, $B_0^{*}, B_1, B_2^{*}$ 
is becoming clear in the lattice results (in disagreement 
with some expectations (\cite{isg97})). The spin 1 $P$ 
states cannot be clearly separated as yet. Experimental 
results on the $P$ states are likewise uncertain.  
Results at a different value of the lattice spacing 
are compared in \cite{jhein}. 

\newcommand{\xorg}{0}
\newcommand{\yorg}{0}
\newcommand{\state}[6]{
   \put(\xorg,0){ \put(#6,-10){\makebox(0,0)[t]{${#1}$}} 		}
   \put(\xorg,\yorg){ \put(#6,#3){\circle*{10}}				}
   \put(\xorg,\yorg){ \put(#6,#3){\line(0,1){#4}}			}
   \put(\xorg,\yorg){ \put(#6,#3){\line(0,-1){#5}}			}
   \put(\xorg,\yorg){ \multiput(#6,#2)(20,0){2}{\line(1,0){10}}		}
   \put(\xorg,\yorg){ \multiput(#6,#2)(-20,0){3}{\line(1,0){10}}	}
                     }
\newcommand{\spectrum}[7]{
   \put(\xorg,0){ \put(0,0){\line(0,1){#1}}				}
   \put(\xorg,\yorg){ \multiput(-20,#2)(0,500){2}{\line(1,0){40}}	}
   \put(\xorg,0){ \multiput(-10,0)(0,100){8}{\line(1,0){20}}		}
   \put(\xorg,\yorg){ \put(-50,#2){\makebox(0,0)[r]{#3}}		}
   \put(\xorg,\yorg){ \put(-50,#4){\makebox(0,0)[r]{#5}}		}
   \put(\xorg,\yorg){ \put(-50,#6){\makebox(0,0)[r]{#7}}		}

                        }

\renewcommand{\yorg}{-1800}
\begin{figure}[hbtp]
\begin{center}
\setlength{\unitlength}{.0025in}
\begin{picture}(900,800)(-70,-10)

\spectrum{800}{2000}{2.0}{2500}{2.5}{2600}{GeV}

\state{D}{1870}{1872}{25}{25}{50}
\state{D^*}{2007}{1990}{14}{14}{125}
\state{D_{^3P_0}}{10000}{2440}{100}{100}{225}
\state{D_1}{2422}{2420}{70}{70}{325}

\state{D_s}{1968}{1990}{14}{14}{550}
\state{D_s^*}{2110}{2103}{15}{15}{625}
\state{D_{s^3P_0}}{10000}{2470}{70}{70}{750}
\state{D_{s1}}{2536}{2500}{50}{50}{860}

\end{picture}

\caption[jglkgj]{
The $D$ meson spectrum from lattice QCD using a 
tadpole-improved SW action in the quenched 
approximation. 
Masses are fixed relative to the spin average of the
$D_s$ and the $D_s^*$ 
(\cite{boyled}). Horizontal
dashed lines mark the experimental results (\cite{pdg}).} 
\label{cmesons}
\end{center}
\end{figure}
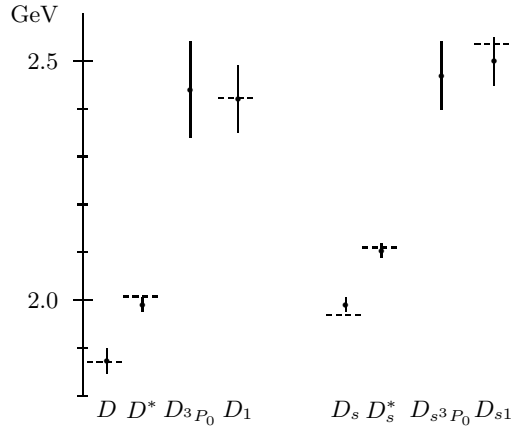

Figure \ref{hlbaryons} shows the $b$-light baryon 
spectrum using NRQCD for the $b$ quark (\cite{arifareview}).
Agreement with experiment is again reasonably good, 
although the $\Lambda_b$ baryon is apparently too 
heavy on the partially unquenched configurations. 
The baryons are probably rather susceptible to finite volume 
effects, and further work is definitely needed on bigger volumes.
The $\Sigma^{*} - \Sigma$ splitting is too 
small in the quenched approximation, which does not 
seem surprising by now. 

Results in the static limit for mesons and 
baryons, for the states that still 
exist there, are similar to those from NRQCD but have been 
in the past
less accurate - see \cite{peisa}, \cite{ukqcdstatic}, \cite{eichten-fb},
\cite{alexstatic} and \cite{duncan-lat92}.
A comparison is made in Figure \ref{hlbaryons} to results 
using heavy Wilson quarks for the $b$. This will be 
discussed further below. 

Arguments earlier showing that the heavy quarks are more 
non-relativistic in heavy-light than heavy-heavy does mean 
that NRQCD should work better for the $D$ than for the 
$\psi$ but currently results are only available for 
$S$-states at one value of the lattice spacing (\cite{jhein}).
For $c$-light mesons and baryons the SW action (or 
other heavy Wilson action) is probably to be preferred
even in this case. 
Figure \ref{cmesons}
shows a recent $D$ spectrum using the tadpole-improved 
SW action for the $c$ quark presented here by 
Peter Boyle (\cite{boyled}). The agreement with experiment 
is encouraging but it has not been possible to extract all the 
$P$ fine structure as yet, and errors bars there are still 
rather large. Uncertainties in how to fix $a^{-1}$ and 
$m_c$ are the same here as for the $b$ case above. In the 
Figure $a^{-1}$ is taken from light hadron spectroscopy. 
Results have been compared at two values of the lattice spacing 
(\cite{boyle}).
No $D$ spectrum is available from unquenched configurations 
as yet. 

Figure \ref{cbaryons} shows the $c$-light baryon spectrum 
using the SW action for the $c$ quark but this time not 
tadpole-improved (\cite{ukqcdcharm}). 
Agreement with experiment for the $m_Q$-independent 
splittings is reasonable but the hyperfine splittings are
much too small. At least a part of this comes from the lack of tadpole-
improvement since this directly affects the `$\vec{\sigma}_Q
\cdot \vec{B}$' term in this formalism. 

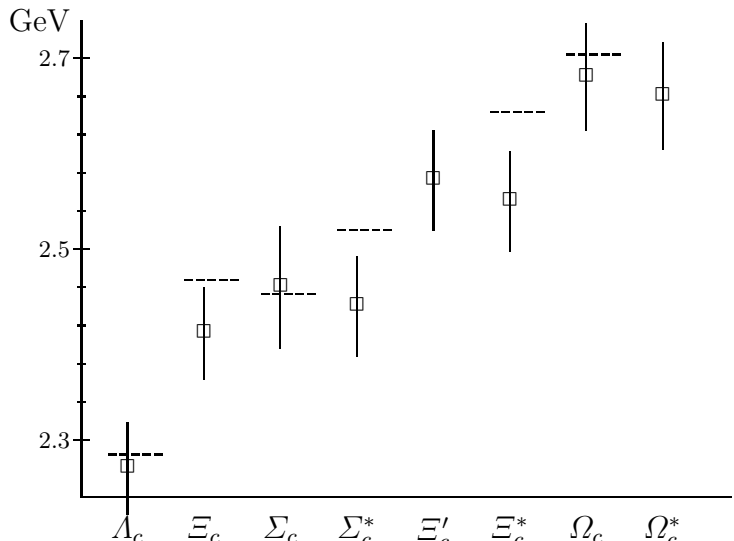
\begin{figure}

\begin{center}
\setlength{\unitlength}{.02in}
\begin{picture}(170,130)(10,930)
\put(15,935){\line(0,1){125}}
\multiput(13,950)(0,50){3}{\line(1,0){4}}
\multiput(14,950)(0,10){10}{\line(1,0){2}}
\put(12,950){\makebox(0,0)[r]{2.3}}
\put(12,1000){\makebox(0,0)[r]{2.5}}
\put(12,1050){\makebox(0,0)[r]{2.7}}
\put(12,1060){\makebox(0,0)[r]{\Large{GeV}}}
\put(15,935){\line(1,0){170}}

\put(27,930){\makebox(0,0)[t]{\Large{$\Lambda_{c}$}}}
\multiput(22,946.3)(3,0){5}{\line(1,0){2}}
\put(27,942.5){\makebox(0,0){$\Box$}}
\put(27,942.5){\line(0,1){12}}
\put(27,942.5){\line(0,-1){12}}

\put(47,930){\makebox(0,0)[t]{\Large{$\Xi_{c}$}}}
\multiput(42,992)(3,0){5}{\line(1,0){2}}
\put(47,978){\makebox(0,0){$\Box$}}
\put(47,978){\line(0,1){12}}
\put(47,978){\line(0,-1){12}}

\put(67,930){\makebox(0,0)[t]{\Large{$\Sigma_{c}$}}}
\multiput(62,988.3)(3,0){5}{\line(1,0){2}}
\put(67,990){\makebox(0,0){$\Box$}}
\put(67,990){\line(0,1){16}}
\put(67,990){\line(0,-1){16}}

\put(87,930){\makebox(0,0)[t]{\Large{$\Sigma_{c}^*$}}}
\multiput(82,1005)(3,0){5}{\line(1,0){2}}
\put(87,985){\makebox(0,0){$\Box$}}
\put(87,985){\line(0,1){13}}
\put(87,985){\line(0,-1){13}}

\put(107,930){\makebox(0,0)[t]{\Large{$\Xi_{c}'$}}}
\put(107,1018){\makebox(0,0){$\Box$}}
\put(107,1018){\line(0,1){13}}
\put(107,1018){\line(0,-1){13}}

\put(127,930){\makebox(0,0)[t]{\Large{$\Xi_{c}^*$}}}
\multiput(122,1036)(3,0){5}{\line(1,0){2}}
\put(127,1012.5){\makebox(0,0){$\Box$}}
\put(127,1012.5){\line(0,1){13}}
\put(127,1012.5){\line(0,-1){13}}

\put(147,930){\makebox(0,0)[t]{\Large{$\Omega_{c}$}}}
\multiput(142,1051)(3,0){5}{\line(1,0){2}}
\put(147,1045){\makebox(0,0){$\Box$}}
\put(147,1045){\line(0,1){14}}
\put(147,1045){\line(0,-1){14}}

\put(167,930){\makebox(0,0)[t]{\Large{$\Omega_{c}^*$}}}
\put(167,1040){\makebox(0,0){$\Box$}}
\put(167,1040){\line(0,1){14}}
\put(167,1040){\line(0,-1){14}}

\end{picture}
\caption[kgj]{The spectrum of baryons containing one $c$ quark
obtained on the lattice using the SW action for the $c$ quark
in the quenched approximation (\cite{ukqcdcharm}). The horizontal
dashed lines give experimental results (\cite{pdg97}). }
\label{cbaryons}
\end{center}
\end{figure}

It is tempting to try extrapolating to the $b$ quark from 
the $c$ quark using the results from the SW action and 
HQS arguments to set the $m_Q$-dependence of splittings. 
This is probably fine for splittings which have little
or no $m_Q$ dependence, in which case extrapolation is not 
really necessary. We have seen that there are several 
splittings for which the leading behaviour is a constant and 
for which there seems almost no sub-leading dependence on 
$m_Q$. For the splittings that have strong $m_Q$ dependence
it is much more difficult to pick up this dependence from 
the small $m_Q$ side than from the large.   
Figure \ref{hlbaryons} compares results from NRQCD 
with extrapolated results from unimproved Wilson 
quarks (\cite{alexcharm}) and SW quarks (\cite{ukqcdcharm}). 
In the latter case the low value obtained for the 
hyperfine $\Sigma^* - \Sigma$ splitting becomes worse
on extrapolation. 
In principle the SW action is safer
for heavy-light mesons than for heavy-heavy as 
was discussed in section 2.4. and so
calculations at the $b$ itself can and should be done 
with this method (\cite{simone}).  

Finally lattice QCD calculations do not have to restrict 
themselves to he physical quark masses but can explore the
whole heavy quark region. This enables a fit to the dependence
on $m_Q$ (or on the pseudoscalar meson mass, say) of a range
of splittings. The coefficients of this dependence are then 
non-perturbative parameters of a heavy quark expansion which 
can be made use of in other heavy quark relations. The values of 
the coefficients can be compared to expectations of powers of
$\Lambda_{QCD}$ and to QCD sum rule results (\cite{collins-spect},
\cite{collins-lat96}, \cite{martinelli}). 

\section{Conclusions}

There has been a lot of progress in heavy hadron spectroscopy 
using the techniques of lattice QCD in recent years, 
converting the qualitative understanding of potential 
models and Heavy Quark Symmetry into clear numerical results
that test QCD. 

Further work is still needed to bring down systematic errors. 
Upsilon spectroscopy is the most accurate at present. Here, 
more calculations need to be done with a non-relativistic 
action which includes next-to-leading spin-dependent terms 
and radiative corrections to leading terms. Finite volume 
effects must be studied for radially excited states. More accuracy 
is needed on dynamical configurations with several different 
values of $n_f$ to allow for a clear extrapolation of fine 
structure to the real world. A prediction for the $\Upsilon -
\eta_b$ mass at the 10\% level is a realisable goal with 
current calculations.  

The charmonium spectrum is more complete experimentally but
more work is needed on the lattice using 
heavy Wilson actions 
to reduce statistical and systematic errors (e.g. from $D^4$ 
terms). Calculations on configurations with dynamical fermions 
are required for extrapolations to compare to experiment.  

In the heavy-light sector statistical and systematic errors 
are inevitably larger and these must be reduced if we are 
to get a clear picture of the fine structure and radial
excitations from the lattice that are now being seen experimentally. 
Analyses of scaling as the lattice spacing is changed 
and finite volume studies for these systems are still at an 
early stage. 
In the next few years a clearer picture will emerge of the 
effect of the quenched approximation on `softer' momentum 
systems such as light and heavy-light hadrons and 
ambiguities of scale setting and quark mass fixing should 
be removed. 

Finally, as noted at the beginning, we are also interested in 
matrix elements for radiative and weak decays of 
heavy hadrons, particularly those which are important for 
the experimental $B$ physics programme. Calculations 
of these are being done on the 
lattice also, using the techniques described here for the 
spectrum. These calculations are much harder and 
accurate spectrum results will 
be a prerequisite for accurate matrix elements.

{\bf Acknowledgements} 

I thank the organisers for a very enjoyable school and the following 
for help in preparing these lectures: Arifa Ali Khan, Gunnar Bali, 
Peter Boyle, Sara Collins, Joachim Hein, Henning Hoeber, Peter
Lepage, Paul McCallum, Colin Morningstar, Junko Shigemitsu, John 
Sloan and Achim Spitz. 
A lot of the work described here was supported by PPARC and 
NATO under grant CRG 941259. I am grateful to the Institute
for Theoretical Physics, UCSB, for hospitality and to 
the Leverhulme Trust and the Fulbright Commission for funding 
while these lectures were being written up.

\end{document}